%% file: furrer_waldmann_rmp_main.tex
\documentclass[rmp,aps,nofootinbib,twocolumn,showpacs]{revtex4}

\usepackage{graphics}
\usepackage{epsfig}

\begin{document}
\title{Magnetic Cluster Excitations}

\author{Albert Furrer}
\email{albert.furrer@psi.ch}
\affiliation{Laboratory for Neutron Scattering, Paul Scherrer Institut, CH-5232 Villigen
PSI, Switzerland}
\author{Oliver Waldmann}
\email{oliver.waldmann@physik.uni-freiburg.de}
\affiliation{Physikalisches Institut, Universit\"at Freiburg, D-79104
Freiburg, Germany}

\begin{abstract}
Magnetic clusters, i.e., assemblies of a finite number (between two or three and several hundred) of interacting spin
centers which are magnetically decoupled from their environment, can be found in many materials ranging from inorganic
compounds, magnetic molecules, artificial metal structures formed on surfaces to metalloproteins. The magnetic
excitation spectra in them are determined by the nature of the spin centers, the nature of the magnetic interactions,
and the particular arrangement of the mutual interaction paths between the spin centers. Small clusters of up to four
magnetic ions are ideal model systems to examine the fundamental magnetic interactions which are usually dominated by
Heisenberg exchange, but often complemented by anisotropic and/or higher-order interactions. In large magnetic clusters
which may potentially deal with a dozen or more spin centers, the possibility of novel many-body quantum states and
quantum phenomena are in focus. In this review the necessary theoretical concepts and experimental techniques to study
the magnetic cluster excitations and the resulting characteristic magnetic properties are introduced, followed by
examples of small clusters demonstrating the enormous amount of detailed physical information which can be retrieved.
The current understanding of the excitations and their physical interpretation in the molecular nanomagnets which
represent large magnetic clusters is then presented, with an own section devoted to the subclass of the single-molecule
magnets which are distinguished by displaying quantum tunneling of the magnetization. Finally, some quantum many-body
states are summarized which evolve in magnetic insulators characterized by built-in or field-induced magnetic clusters.
The review concludes addressing future perspectives in the field of magnetic cluster excitations.
\end{abstract}

\date{Version 02. Oct. 2012}


\maketitle
\tableofcontents

\input{chapter1_introduction_v2012_10_02}
\input{chapter2_basics_v2012_10_02}
\input{chapter3_smallclusters_v2012_10_02}
\input{chapter4_largeclusters_v2012_10_02}
\input{chapter5_smms_v2012_10_02}
\input{chapter6_quantumspins_v2012_10_02}
\input{chapter7_conclusions_v2012_10_02}

\section*{Acknowledgments}
We are grateful to Hans U. G\"udel, Christian R\"uegg, and J\"urgen Schnack for having carefully read parts of the
manuscript and for their useful comments. OW thanks the Deutsche Forschungsgemeinschaft for partial financial support.

\bibliographystyle{apsrmp}
\bibliography{furrer_waldmann_rmp_main}

\end{document}

%% file: chapter1_introduction_v2012_10_02.tex

\section{Introduction}
\label{sec:intro}

Magnetic clusters are defined as systems which consist of a finite number $N$ of interacting spins that are
magnetically isolated from the environment, where the number $N$ may be as small as two or can be several hundreds.
Larger entities of thousands to a few millions of atoms, which are usually referred to as nanoscale or ultrafine
particles, are excluded from the present review. Magnetic clusters either occur naturally in pure compounds, where the
magnetic isolation of the clusters is provided by nonmagnetic ligands, or are formed artificially, e.g., in solid
solutions of magnetic and nonmagnetic compounds. Prototypes of pure compounds are molecular nanomagnets in which a
polynuclear magnetic metal core is embedded in a diamagnetic ligand matrix. Examples associated with cooperative
systems are diluted magnetic compounds, in which the magnetic ions are randomly distributed, so that different types of
magnetic clusters ($N$-mers, $N =1, 2, 3,\ldots$) are simultaneously present.

Many of the characteristic physical properties of magnetic clusters are determined by the magnetic excitation
spectra which reflect the nature of the fundamental magnetic interactions between the spins in a cluster. The
latter can be quantitatively accounted for by a spin Hamiltonian in which a bilinear Heisenberg-type exchange
interaction is usually the dominant term, often complemented by additional terms describing the anisotropic
and/or higher-order interactions. As long as the number $N$ of spins is reasonably small, exact analytical
solutions of the spin Hamiltonian can be obtained. Accordingly, small magnetic clusters are ideal model systems
to explore experimentally the limitations of the theoretical models. However, the synthesis strategies to
produce molecular magnetic materials have advanced enormously in recent years, making available bounded
molecular magnetic clusters with the number $N$ of magnetic centers varying from one to several dozens; the
record currently stands at $N = 84$ in the torus-like molecule Mn$_{84}$ \cite{Chr04-mn84}. In this new class of
magnetic materials, now commonly called molecular nanomagnets \cite{Gat06-book}, clusters with more than four
metal ions are rather the rule than the exception, shifting the main scientific challenge away from that of
studying the nature of the basic interactions towards that of exploring the possible consequences of having such
interactions in a lattice of exchange-coupled spin centers.

In the past decade, magnetic cluster systems have become a topic of increasing interest and relevance in condensed
matter science. They are not only interesting in themselves for determining the origin and the size of the fundamental
magnetic interactions, but they have also important applications in both technology and science. Molecular nanomagnets
are currently considered among the most promising candidates as the smallest nanomagnetic units capable of storing and
processing quantum information \cite{Tro05-qubit}. In particular, magnetic molecules with large spin ground states and
negative axial anisotropy or single-molecule magnets were found to exhibit outstanding properties such as slow
relaxation of the magnetization and stepped magnetic hysteresis curves due to quantum tunneling of the magnetization at
low temperatures \cite{Chr00-smmreview,Gat03-review,Chu98-QTMbook}.

Another emerging field concerns transition-metal perovskites in which magnetic polarons evolve upon hole doping and
behave like magnetic nanoparticles embedded in a nonmagnetic matrix \cite{Phe06}. Furthermore, we mention the
remarkable observations of quantum phase transitions, Bose-Einstein condensation and field-induced three-dimensional
magnetic ordering in weakly interacting antiferromagnetic dimer compounds \cite{Sac99,Gia08} as well as the attractive
phenomenon of spin-Peierls dimerization in both organic and inorganic compounds \cite{Bra75}. Very recently, magnetic
cluster systems could also be fabricated directly on surfaces using scanning microscope techniques \cite{Hir06}. Even
in biology, magnetic metal clusters are important subunits. Polynuclear iron clusters are contained in proteins like
adrenodoxin and ferredoxin, which are involved in the photosynthetic process to convert light energy into chemical
energy based on electron transfer mechanisms \cite{Gri92}. All these systems have a common property, namely the
presence of magnetic clusters, whose characterization is therefore a key issue in both theoretical and experimental
investigations.

An important class of magnetic cluster systems concerns magnetic nanoparticles containing a few hundreds of atoms
produced by e.g. sputtering, inert-gas condensation techniques, direct ball milling or microemulsion-based syntheses.
In principle, the spin dynamics exhibits many features similar to the molecular nanomagnets, such as superparamagnetic
relaxation and quantized spin-wave states. Pioneering neutron scattering experiments in this area have been performed
for Fe nanoparticles \cite{Hen94-intro} and for nanocrystallites of hematite \cite{Kuh06-intro,Han97-intro}. In
practice, the interpretation of experimental data on magnetic nanoparticles is made difficult by line broadening
effects due to variations in size and form, and the presence of a variety of surface spin states. Any discussion of the
magnetic properties needs considering these aspects, and for that reason the magnetic nanoparticles are excluded from
the present review, in particular as excellent reviews of the theoretical and experimental aspects exist
\cite{Hen93-intro,Kod99-intro}.

Our main objective in this review is to introduce the theoretical concepts and the experimental techniques
applied to the study of magnetic cluster excitations as well as to give a snapshot of recent developments
achieved for the most important classes of materials whose properties are largely governed by the presence of
magnetic clusters. The outline of this review is as follows: Sec.~\ref{sec:basics} starts with a summary of the
basic terms of the underlying spin Hamiltonians and with a short description of the most powerful experimental
techniques, followed by representative examples on small magnetic clusters (dimers, trimers, and tetramers) in
Sec.~\ref{sec:sc} to demonstrate the enormous amount of detailed physical information resulting from the study
of magnetic cluster excitations. Clusters with a small number $N$ of coupled magnetic ions are excellent choices
for this purpose, as the basic interactions are all present in them and can be treated exactly, hence a
straightforward comparison between theory and experiment is possible with little ambiguity. This is no longer
the case for the emerging field of large magnetic clusters discussed in Sec.~\ref{sec:lc}, in which the number
$N$ of magnetic centers can be as large as several dozens, so that the interpretation and analysis of the
experimental results require the use of sophisticated tools. In addition, novel physical aspects such as complex
many-body quantum states come into play as a consequence of the large cluster size. An important subclass of the
large magnetic clusters will be addressed in Sec.~\ref{sec:smm}, namely the single-molecule magnets in which at
low temperatures magnetic hysteresis or slow magnetic relaxation and quantum tunneling of the magnetization can
be observed. Many methodologies applied to the large magnetic clusters have their origin in the field of quantum
spin systems which are summarized in Sec.~\ref{sec:qss} for cases where the presence of magnetic clusters is the
most important ingredient to understand their quantum spin properties. We conclude with a brief outlook in
Sec.~\ref{sec:conclusion}. Given the abundant literature on the topic of magnetic cluster excitations, the
present review is necessarily bound to be incomplete in terms of both materials covered and references cited.
The experimental results were chosen according to their didactical suitability to illustrate the concepts,
including both historical data from the pioneering time and data of today's research.

%% file: chapter2_basics_v2012_10_02.tex

\section{Basics}
\label{sec:basics}

\subsection{Spin Hamiltonian}
\label{sec:basics-spinham}

The form of the appropriate spin Hamiltonian of magnetic cluster systems depends on two essentially independent terms:
the nature of the interacting systems in the absence of interactions, and the physical nature of the mechanisms
responsible for the interactions \cite{Wol71}. The total Hamiltonian describing interacting spins can generally be
written as
\begin{eqnarray}
\label{eq:basics-wolf}
\hat{H} = \sum_{i} \hat{H}^{(0)}_i + \sum_{ij} \hat{H}_{ij} + \sum_{i} \hat{H}^{(1)}_i,
\end{eqnarray}
where $\hat{H}^{(0)}_i \gg \hat{H}_{ij}$ and the terms $\hat{H}^{(1)}_i$ may or may not be comparable with
$\hat{H}_{ij}$. The classification of different cases depends on the eigenfunctions of $\hat{H}^{(0)}_{i}$
which provide a basis for the description of $\hat{H}_{ij}$ and $\hat{H}^{(1)}_{i}$. Since
$\hat{H}^{(0)}_{i}$ is by definition large, we generally consider only the ground state at a time, and this
may be characterized as one of following types:

\textbf{Type S}: The system has negligible orbital admixtures. The ionic spins $\hat{\bf{s}}_i$ are good quantum
numbers to define the spin Hamiltonian.

\textbf{Type Q}: The orbital angular momentum is quenched, thus the spin Hamiltonian can be expressed as for type S.
However, orbital effects like the ligand field have to be considered in $\hat{H}^{(1)}_{i}$.

\textbf{Type L}: There is orbital degeneracy (or near degeneracy) with $2l_i+1$ states. The (weak) spin-orbit coupling
$\hat{H} = \lambda \sum_i \hat{\bf{l}}_i \cdot \hat{\bf{s}}_i$  has to be considered in $\hat{H}^{(1)}_{i}$.

\textbf{Type J}: The spin-orbit coupling is large, thus the total angular momentum $\hat{\bf{j}}_i = \hat{\bf{l}}_i +
\hat{\bf{s}}_i$ is a good quantum number. The spin Hamiltonian is expressed as for type S systems (replacing
$\hat{\bf{s}}_i$ by $\hat{\bf{j}}_i$).

It is mentioned that type Q and S are also known as systems with "orbitally non-degenerate ground terms" and type L as
systems with "orbitally degenerate ground terms" or "first-order orbital angular momentum". In the following we write
down $\hat{H}_{ij}$ and $\hat{H}^{(1)}_{i}$ for types S and Q which constitute the large majority of magnetic clusters
studied so far.

A widely used approach to describe the spin interactions $\hat{H}_{ij}$ is the Heisenberg-Dirac-Van Vleck (HDVV)
Hamiltonian \cite{Hei26,Dir26,Vle32},
\begin{eqnarray}
\label{eq:basics-HDVV}
\hat{H} = -2 \sum_{i<j} J_{ij} \hat{\bf{s}}_i \cdot \hat{\bf{s}}_j,
\end{eqnarray}
where $\hat{\bf{s}}_i$ is the spin operator of the $i$th ion in the cluster and $J_{ij}$ the exchange parameter which
couples the magnetic ions at sites $i$ and $j$. In the literature the HDVV Hamiltonian is often described as $- \sum
J_{ij} \hat{\bf{s}}_i \cdot \hat{\bf{s}}_j$ or $+\sum J_{ij} \hat{\bf{s}}_i \cdot \hat{\bf{s}}_j$, in contrast to the
convention adopted here. Hence the exchange parameters given in the present work are always adjusted to be in agreement
with Eq.~(\ref{eq:basics-HDVV}). $\hat H$ commutes with the total spin $\hat{\bf{S}} = \sum_i \hat{\bf{s}}_i$, thus $S$
and $M$ are good quantum numbers and the eigen functions can be written as $|\tau S M\rangle$, where $-S \leq M \leq S$
and $\tau$ stands for any other quantum numbers required for distinguishing the spin multiplets unambiguously. Often
labels are omitted for convenience. Any anisotropic term added to the HDVV Hamiltonian Eq.~(\ref{eq:basics-HDVV}) lifts
the $M$-degeneracy of the spin states $|S M\rangle$.

The exchange coupling may not always be isotropic, thus we extend Eq.~(\ref{eq:basics-HDVV}) to include exchange
anisotropy,
\begin{eqnarray}
\label{eq:basics-anisoexchange}
\hat{H} = -2 \sum_{i<j} \left( J^{xx}_{ij} \hat{s}_{ix} \hat{s}_{jx} + J^{yy}_{ij} \hat{s}_{iy} \hat{s}_{jy} + J^{zz}_{ij} \hat{s}_{iz} \hat{s}_{jz} \right).
\end{eqnarray}
A special case of anisotropic spin-spin coupling is provided by the dipole-dipole interaction which is always present
in addition to the exchange interaction Eq.~(\ref{eq:basics-HDVV}),
\begin{eqnarray}
\label{eq:basics-dipoledipole}
\hat{H} = \sum_{ij} \frac{g^2\mu_B^2}{R^3_{ij}}
\left[
\hat{\bf{s}}_{i}\cdot \hat{\bf{s}}_{j}
-
3 \frac{(\hat{\bf{s}}_{i}\cdot {\bf{R}}_{ij})(\hat{\bf{s}}_{j}\cdot {\bf{R}}_{ij})}{R^2_{ij}}
\right],
\end{eqnarray}
where $g$ is the Land\'e splitting factor, $\mu_B$ the Bohr magneton, and ${\bf R}_{ij} = {\bf R}_i - {\bf
R}_j$ the vector defining the distance between the spins at sites $i$ and $j$ located at ${\bf R}_i$ and
${\bf R}_j$. In most magnetic clusters the typical distance between metal ions is 3.0-3.5~{\AA} yielding $g^2
\mu^2_B / R^3_{ij} \approx 0.1$~K. The dipole-dipole interaction is hence normally much smaller than the HDVV
interaction and can often be disregarded, though in some cases it contributes appreciably to the overall
magnetic anisotropy \cite{Abb01}. Another type of exchange anisotropy is described by the
Dzyaloshinski-Moriya interaction \cite{Dzy58,Mor60} which, however, vanishes for the case of inversion
symmetry,
\begin{eqnarray}
\label{eq:basics-DM}
\hat{H} = \sum_{i<j} {\bf{d}}_{ij} \cdot \left( \hat{\bf{s}}_i \times \hat{\bf{s}}_j \right),
\end{eqnarray}
where ${\bf{d}}_{ij}$ is known as the Dzyaloshinski-Moriya vector.

The HDVV Hamiltonian Eq.~(\ref{eq:basics-HDVV}) is based on the bilinear spin permutation operator \cite{Her66}
\begin{eqnarray}
\label{eq:basics-permut}
\hat{P}_{ij} = \frac{1}{2}\left( 1 + \hat{\bf{s}}_i \cdot \hat{\bf{s}}_j \right).
\end{eqnarray}
A more complete Hamiltonian takes permutations of more than two spins into account. The relevant terms up to second
order (biquadratic terms) are defined by
\begin{eqnarray}
\label{eq:basics-permut2}
 \hat{P}^2_{ij} &=& \frac{1}{4}\left[ 1 + 2\hat{\bf{s}}_i \cdot \hat{\bf{s}}_j + \left(\hat{\bf{s}}_i \cdot \hat{\bf{s}}_j \right)^2  \right],
 \\
 \hat{P}_{ij}\hat{P}_{jk} &=& \frac{1}{4}\left[ 1 + \hat{\bf{s}}_i \cdot \hat{\bf{s}}_j + \hat{\bf{s}}_j \cdot \hat{\bf{s}}_k +
 \left(\hat{\bf{s}}_i \cdot \hat{\bf{s}}_j \right)\left(\hat{\bf{s}}_j \cdot \hat{\bf{s}}_k \right)  \right],
 \quad\\
 \hat{P}_{ij}\hat{P}_{kl} &=& \frac{1}{4}\left[ 1 + \hat{\bf{s}}_i \cdot \hat{\bf{s}}_j + \hat{\bf{s}}_k \cdot \hat{\bf{s}}_l +
 \left(\hat{\bf{s}}_i \cdot \hat{\bf{s}}_j \right)\left(\hat{\bf{s}}_k \cdot \hat{\bf{s}}_l \right)  \right],
\end{eqnarray}
which refer to two-spin, three-spin, and four-spin interactions, respectively.

The Hamiltonian $\hat{H}^{(1)}_{i}$ has to be introduced in essentially two cases. For $s_i \neq 1/2$ systems,
single-ion anisotropy must be considered, which for the case of axial anisotropy reads
\begin{eqnarray}
\label{eq:basics-Dterm}
\hat{H} = \sum_{i} D_i \left[ \hat{s}^2_{iz} - \frac{1}{3} s_i(s_i+1) \right]
\end{eqnarray}
and for planar anisotropy we have
\begin{eqnarray}
\label{eq:basics-Eterm}
\hat{H} = \sum_{i} E_i \left( \hat{s}^2_{ix} - \hat{s}^2_{iy} \right),
\end{eqnarray}
where often $D_i = D$ and/or $E_i = E$ for all sites. In cases with $s_i \geq 2$ also higher-order anisotropy terms
\begin{eqnarray}
\label{eq:basics-O4term}
  \hat{H} &=& \sum_{i} B^0_{4i} \hat{ O }^0_4(s_i) + B^2_{4i} \hat{ O }^2_4(s_i) + B^4_{4i} \hat{ O }^4_4(s_i)
\end{eqnarray}
may have to be included, where $\hat{ O }^m_n(s_i)$ are Stevens operator equivalents built up by fourth-order spin
operators \cite{Hut64}. Finally, the action of an external magnetic field ${\bf B}$ is described by
\begin{eqnarray}
\label{eq:basics-Zeeman}
 \hat{H} = g \mu_B  \sum_{i} {\bf B} \cdot \hat{\bf s}_i = g \mu_B {\bf B} \cdot \hat{\bf S}.
\end{eqnarray}

For type S and Q systems, the HDVV Hamiltonian dominates usually over the anisotropic terms in the total spin
Hamiltonian, and a first-order perturbation treatment of the anisotropy is an excellent starting point (strong-exchange
limit) \cite{Ben90-eprbook}. The energy spectrum is then structured into spin multiplets with a definite value of $S$
for each of them, and the eigen functions are well described by the $|\tau S M\rangle$ spin functions. The energies of
the spin multiplets are governed by the HDVV Hamiltonian (exchange splitting), and each spin multiplet is further split
by the magnetic anisotropy (anisotropy splitting or zero-field splitting, ZFS). The possible transitions may be
distinguished into inter-multiplet ($\Delta S \neq 0$) and intra-multiplet ($\Delta S = 0$). Unless stated otherwise,
the strong-exchange limit is always assumed.

\subsection{Experimental Techniques}
\label{sec:basics-exps}

The spin interactions discussed in the preceding Section give rise to discrete energy levels and wave functions which
can be determined by a variety of experimental methods. The most powerful techniques are certainly spectroscopic
methods such as inelastic neutron scattering (INS) and optical spectroscopies, which allow a direct determination of
the spin states. Some information about the spin states can also be obtained by resonance techniques, e.g., by
paramagnetic resonance experiments (EPR). Information on the spin states is also contained intrinsically in the
thermodynamic magnetic properties, however, an extraction of reliable parameters is not always possible due to the
integral nature of these properties.

In the following the two spectroscopic methods mainly applied to the study of magnetic cluster systems are briefly
introduced. These include INS and optical spectroscopies which both have their merits and should be considered as
complementary methods. Optical spectroscopies, on the one hand, can be applied to very small samples of the order of
10~$\mu$m$^3$; they provide highly resolved spectra so that small line shifts and splittings can be detected, and they
cover a large energy range so that inter-multiplet transitions can easily be observed. Neutron scattering, on the other
hand, is not restricted to particular points in reciprocal space, i.e., interactions between the spins can be observed
through the wave vector dependence, the peak intensities can easily be interpreted on the basis of the wave functions
of the spin states, and data can be taken over a wide temperature range which is important when studying linewidth
phenomena. INS as the most widely used spectroscopic technique is described below in detail, followed by short
descriptions of optical spectroscopies and EPR techniques as well as by a summary of the thermodynamic magnetic
properties.

\subsubsection{Inelastic Neutron Scattering}
\label{sec:basics-ins}

The principal aim of an INS experiment is the determination of the probability that a neutron which is
incident on the sample with wave vector $\bf k$ is scattered into the state with wave vector ${\bf k'}$. The
intensity of the scattered neutrons is thus measured as a function of the momentum transfer
\begin{eqnarray}
\label{eq:basics-ins-Q}
 \hbar {\bf Q} = \hbar ( {\bf k} - {\bf k'}),
\end{eqnarray}
where $\bf Q$ is known as the scattering vector, and the corresponding energy transfer is given by
\begin{eqnarray}
\label{eq:basics-ins-E}
 \hbar \omega = \frac{\hbar^2}{2m} \left( {\bf k}^2 - {\bf k'}^2 \right),
\end{eqnarray}
where $m$ is the mass of the neutron. Eqs.~(\ref{eq:basics-ins-Q}) and (\ref{eq:basics-ins-E}) describe the momentum
and energy conservation of the neutron scattering process, respectively. For $|\bf k| = |{\bf k'}|$ we have from
Eq.~(\ref{eq:basics-ins-E}) $\hbar \omega =0$, i.e., elastic scattering. For inelastic scattering, $\bf Q$ can be
decomposed according to $\bf Q = {\bf G} + {\bf q}$, with a reciprocal lattice vector ${\bf G}$ and a wave vector $\bf
q$. INS experiments thus allow us to measure the magnetic excitation energy at any predetermined point in reciprocal
space, most conveniently by triple-axis crystal spectrometry \cite{Bro55}. In extended systems this yields the
dispersion relation $\hbar \omega({\bf q})$. In magnetic clusters the excitations are dispersion-less but the
scattering intensity shows a characteristic dependence on momentum transfer ($Q$ dependence, \emph{vide infra}). For
INS experiments on polycrystalline samples various types of time-of-flight (TOF) spectrometers are usually more
appropriate \cite{Fur09-book}.

The neutron scattering probability for magnetic cluster excitations can be derived from the master formula
for magnetic scattering \cite{Lov87}:
\begin{eqnarray}
\label{eq:basics-ins-sigma}
 \frac{d^2\sigma}{d\Omega d\omega} = C({\bf Q}) \sum_{\alpha \beta}
 \left( \delta_{\alpha \beta} - \frac{Q_\alpha Q_\beta}{Q^2} \right)
 S_{\alpha \beta}({\bf Q},\omega),
\end{eqnarray}
where
\begin{eqnarray}
\label{eq:basics-ins-C}
 C({\bf Q}) = (\gamma r_0)^2 \frac{k'}{k} F^2({\bf Q}) e^{-2W({\bf Q})}
\end{eqnarray}
and $S_{\alpha \beta}({\bf Q},\omega)$ is the magnetic scattering function
\begin{eqnarray}
\label{eq:basics-ins-S}
 S_{\alpha \beta}({\bf Q},\omega) &=& \sum_{ij} e^{i{\bf Q}\cdot {\bf R}_{ij}}
 \sum_{\lambda \lambda'} p_\lambda \langle \lambda | \hat{s}_{i\alpha} | \lambda' \rangle
\nonumber \\
 &&\times  \langle \lambda' | \hat{s}_{j\beta} | \lambda \rangle \delta( \hbar \omega + E_\lambda -
 E_{\lambda'}).
\end{eqnarray}
Herein is $\gamma = -1.91$, $r_0$ = 0.282$\times$10$^{-12}$~cm the classical electron radius [$(\gamma r_0)^2 =
0.29$~barn], $F({\bf Q})$ the dimensionless magnetic form factor defined as the Fourier transform of the normalized
spin density associated with the magnetic ions, $\exp[-2W({\bf Q})]$ the Debye-Waller factor, and $\alpha,\beta
=x,y,z$. $|\lambda\rangle$ denotes the initial state of the scatterer, with energy $E_\lambda$ and thermal population
factor $p_\lambda$ [Eq.~(\ref{eq:basics-Zp})], and $|\lambda'\rangle$ its final state with energy $E_{\lambda'} $.

The essential factor in the cross section is the magnetic scattering function $S_{\alpha \beta}({\bf Q},\omega)$
which will be discussed in more detail below. There are two further factors which govern the cross section for
magnetic neutron scattering in a characteristic way: Firstly, the magnetic form factor $F({\bf Q})$ which
usually falls off with increasing modulus of the scattering vector $\bf Q$. Secondly, the polarization factor $(
\delta_{\alpha \beta} - Q_\alpha Q_\beta /Q^2)$ tells us that neutrons can only couple to magnetic moments or
spin fluctuations perpendicular to $\bf Q$ which unambiguously allows to distinguish between different
polarizations (transverse and longitudinal) of spin excitations.

The magnetic scattering function $S_{\alpha \beta}({\bf Q},\omega)$ contains two important terms: Firstly,
the structure factor $\exp(i {\bf Q}\cdot {\bf R}_{ij})$ which directly reflects the geometry of the cluster;
secondly, the matrix elements $\langle \lambda |\hat{s}_{i\alpha}| \lambda'\rangle$ which determine the
strength of the transition $| \lambda \rangle\rightarrow | \lambda'\rangle$ as well as corresponding
selection rules.

For magnetic clusters we describe the eigen state $|\lambda\rangle$ by $|\tau S M\rangle$. The matrix
elements can then be calculated by introducing irreducible tensor operators (ITOs) $\hat{T}^{1}_q(s_i)$ of
rank 1, which are related to the spin operators $\hat{s}_{i\alpha}$:
\begin{eqnarray}
\label{eq:basics-ins-T1}
 \hat{T}^{1}_0(s_i) = \hat{s}_{i z},
 \quad
 \hat{T}^{1}_{\pm 1}(s_i) = \mp \frac{1}{\sqrt 2} \left( \hat{s}_{i x} \pm \hat{s}_{i y} \right).
\end{eqnarray}
In the HDVV model the states $|S M\rangle$ are degenerate with respect to the magnetic quantum number $M$, so that
Eq.~(\ref{eq:basics-ins-S}) has to be summed over both $M$ and $M'$. Using the Wigner-Eckart theorem
\begin{eqnarray}
\label{eq:basics-ins-WE}
 \langle S M|\hat{T}^{1}_q(s_i)|S' M'\rangle &=& (-1)^{S-M} \left(\begin{array}{ccc}
S&1&S' \\ -M&q&M' \end{array} \right) \nonumber \\ && \times
 \langle S||\hat{T}^{1}(s_i)||S'\rangle
\end{eqnarray}
we find
\begin{eqnarray}
\label{eq:basics-ins-red}
 \sum_{M M'} \langle S M|\hat{T}^{1}_q(s_i)|S' M'\rangle \langle S' M'|\hat{T}^{1}_{q'}(s_j)|S M\rangle
\nonumber \\
 = \frac{1}{3} \langle S||\hat{T}^{1}(s_i)||S'\rangle \langle S'||\hat{T}^{1}(s_j)||S\rangle.
\end{eqnarray}
The two-row bracket $(...)$ in Eq.~(\ref{eq:basics-ins-WE}) is a Wigner-3$j$ symbol \cite{Rot59}. It vanishes unless
\begin{eqnarray}
\label{eq:basics-ins-selectionrules}
\Delta S \equiv S' - S = 0, \pm 1,
\\
\Delta M \equiv M' - M = 0, \pm 1,
\end{eqnarray}
which establish the selection rule for INS in spin clusters. Thus, INS experiments allow us to detect not only
splittings of individual spin multiplets ($\Delta S=0$), similar to EPR experiments (Sec.~\ref{sec:basics-epr}), but
also splittings produced by magnetic interactions ($\Delta S= \pm1$). The evaluation of the reduced matrix elements on
the right-hand side of Eq.~(\ref{eq:basics-ins-red}) depends on the details or many-body structure of the spin
functions $|\tau S M\rangle$.

Equations~(\ref{eq:basics-ins-S})-(\ref{eq:basics-ins-red}) strictly apply to magnetic cluster systems of type S and Q.
A theoretical treatment of the scattering by L- and J-type ions was given in \onlinecite{Joh66}. However, the
calculation is complicated, and we simply quote the result for $Q \rightarrow 0$. In this case the cross section
measures the magnetization, $\hat{ \bf \mu}_i= -\mu_B( \hat{\bf l}_i + 2 \hat{\bf s}_{i})$, i.e., a combination of spin
and orbital moments that does not allow their separation. This clearly contrasts to magnetic scattering by X-rays. For
INS an approximate result can be obtained for modest values of $Q$. We replace the spin operator $\hat{\bf s}_{i}$ by
\begin{eqnarray}
\label{eq:basics-ins-j}
\hat{\bf s}_{i} = \frac{1}{2} g_i \hat{\bf j}_{i}
\end{eqnarray}
where
\begin{eqnarray}
\label{eq:basics-ins-gj}
g_i = 1 + \frac{j_i(j_i+1) - l_i(l_i+1) + s_i(s_i+1)}{ 2 j_i(j_1 +1)}
\end{eqnarray}
is the Land\'e splitting factor.

If $\omega$ is a positive quantity in the scattering function $S_{\alpha \beta}({\bf Q},\omega)$, the neutron
loses energy in the scattering process and the system is excited from the initial state $\lambda$ which has
energy $\hbar \omega$ less than the final state $\lambda'$. Consider now the function $S_{\alpha \beta}({\bf
Q},-\omega)$ where $\omega$ is the same positive quantity. This represents a process in which the neutron
gains energy. The transitions of the system are between the same states as for the previous process, but now
$\lambda'$ is the initial state and $\lambda$ is the final state. The probability of the system being
initially in the higher state is smaller by the factor $\exp(-\hbar \omega / k_B T)$ as compared to its
probability of being in the lower energy state, hence
\begin{eqnarray}
\label{eq:basics-ins-SS}
S_{\alpha \beta}({\bf Q},-\omega) = \exp(-\frac{\hbar \omega }{ k_B T}) S_{\alpha \beta}({\bf Q},\omega),
\end{eqnarray}
which is known as the principle of detailed balance. Eq.~(\ref{eq:basics-ins-SS}) has to be fulfilled in
experimental data taken in both energy-gain and energy-loss configurations which correspond to the so-called
Stokes and anti-Stokes processes, respectively.

Using the integral representation of the $\delta$ function the scattering function $S_{\alpha \beta}({\bf Q},\omega)$,
Eq.~(\ref{eq:basics-ins-S}), transforms into a physically transparent form:
\begin{eqnarray}
\label{eq:basics-ins-Sint}
S_{\alpha \beta}({\bf Q},\omega) &=&
\frac{1}{2\pi\hbar}
\sum_{ij} e^{i{\bf Q}\cdot {\bf R}_{ij}}
\nonumber \\ &&\times
\int^{+\infty}_{-\infty} \langle\langle \hat{s}_{i\alpha} (0) \hat{s}_{j\beta}(t) \rangle\rangle e^{-i\omega t} dt,
\end{eqnarray}
where $\langle\langle \hat{s}_{i\alpha} (0) \hat{s}_{j\beta}(t) \rangle\rangle$ is the thermal average of
time-dependent spin operators, or the van Hove pair correlation function \cite{VaH54} for spins. A neutron
scattering experiment measures the Fourier transform of the pair correlation function in space and time,
which is clearly what is needed to describe a magnetic system on an atomic scale.

The van Hove representation of the cross section in terms of pair correlation functions is related to the
fluctuation-dissipation theorem \cite{Lov87}:
\begin{eqnarray}
\label{eq:basics-ins-fluct}
S_{\alpha \beta}({\bf Q},\omega) &=&
\frac{\hbar}{\pi}
\left[ 1- \exp( -\frac{\hbar \omega }{ k_B T}) \right] \text{Im} \chi_{\alpha \beta}({\bf Q},\omega).
\nonumber \\
\end{eqnarray}
Physically speaking, the neutron may be considered as a magnetic probe which effectively establishes a
frequency- and wave vector-dependent magnetic field ${\bf B}({\bf Q},\omega )$ in the sample, and detects its
response, the magnetization ${\bf M} ({\bf Q},\omega )$, to this field given by
\begin{eqnarray}
\label{eq:basics-ins-m} M_\alpha ({\bf Q},\omega ) &=& \sum_\beta \chi_{\alpha\beta} ({\bf Q},\omega )
B_\beta({\bf Q},\omega ),
\end{eqnarray}
where $\chi_{\alpha\beta} ({\bf Q},\omega )$ is the generalized magnetic susceptibility tensor. This is really
the outstanding property of the neutron in a magnetic scattering measurement, and no other experimental
technique is able to provide such detailed microscopic information about magnetic compounds.

For polycrystalline material Eq.~(\ref{eq:basics-ins-sigma}) has to be averaged in ${\bf Q}$ space, which in zero
magnetic field can be performed analytically \cite{Wal03-insqdependence}:
\begin{eqnarray}
\label{eq:basics-ins-sigma-powder}
 \frac{d^2\sigma}{d\Omega d\omega} &=& C(Q)
 \sum_{ \lambda \lambda'}  p_\lambda \sum_{ i j}
  [
  \frac{2}{3} j_0( Q R_{ij} ) \tilde{\bf s}_i \cdot \tilde{\bf s}_j
  +
  j_2( Q R_{ij} )
\nonumber \\ && \times
  \sum_q T^{2*}_q({\bf R}_{ij}) T^{2}_q(\tilde{\bf s}_i \tilde{\bf s}_j)
  ]
 \delta( \hbar \omega + E_\lambda - E_{\lambda'})
  .\qquad
\end{eqnarray}
$j_0(x)$ and $j_2(x)$ are the spherical Bessel functions of 0$^{th}$ and 2$^{nd}$ order, and $\tilde{\bf s}_i = (
\langle \lambda | \hat{s}_{ix} | \lambda' \rangle, \langle \lambda | \hat{s}_{iy} | \lambda' \rangle, \langle \lambda |
\hat{s}_{iz} | \lambda' \rangle )$. For an isotropic spin cluster described by only the HDVV Hamiltonian the 2$^{nd}$
order term vanishes:
\begin{eqnarray}
\label{eq:basics-ins-sigma-powder0}
 \frac{d^2\sigma}{d\Omega d\omega} &=&  C(Q) \frac{2}{3}
  \sum_{ \lambda \lambda'} p_\lambda \sum_{ i j}
 \frac{ \sin(Q R_{ij})  }{ Q R_{ij} }
  \langle S || \hat{T}^{1}(s_i)|| S' \rangle
\nonumber \\ && \times
  \langle S' || \hat{T}^{1}(s_j)|| S \rangle
 \delta( \hbar \omega + E_\lambda - E_{\lambda'})
  .
\end{eqnarray}
The Bessel function $\sin(Q R_{ij})/(Q R_{ij})$ is responsible for a characteristic oscillatory $Q$ dependence of the
INS intensity, which is often very helpful in analysis \cite{Fur77,Wal03-insqdependence}. Also, a useful rule of thumb
is inferred: For $Q \rightarrow 0$ the scattering intensity of $\Delta S = \pm 1$ transitions drops to zero, while for
$\Delta S = 0$ transitions it becomes maximal.

Analytical results for the INS cross section were derived for some cases, i.e., for dimers, trimers, and tetramers
\cite{Fur79-inshabil,Har05-insdimer,Gud79-instetramer} and a pentamer and hexamer
\cite{Har09-instrimer,Har11-inspentamer}. Explicit expressions for Eq.~(\ref{eq:basics-ins-sigma-powder}) can be found
in, e.g., \onlinecite{Wal05-smix}.

In practical applications of INS to magnetic cluster compounds the huge incoherent neutron scattering contribution of
hydrogen can easily prevent observation of the magnetic cluster excitations. Removing or reducing the hydrogen content
by e.g. deuteration or fluorination is of course the best solution, but this is often prohibitive, in particular in
molecular clusters. Fortunately, at transfer energies from ca. 0.1 to 3~meV a window exists with comparatively small
hydrogen scattering. This is relevant because otherwise most studies on the cluster excitations in the molecular
nanomagnets for instance would not have been possible.

Another point to be considered is the non-magnetic scattering from the lattice. Besides the "standard tricks"
for identifying the nature of INS features, such as inspecting the temperature and $Q$ dependencies, a
Bose-correction analysis is often helpful. Here, INS data recorded at sufficiently high temperature are
scaled by the Bose factor and then compared to the data at lower temperatures. At high temperatures, where a
large part of the energy spectrum is accessed and the magnetic scattering intensity spread out over
essentially all frequencies, the measured spectrum may reflect the lattice scattering, whose temperature
dependence is governed by the Bose factor $[1 - \exp(\hbar \omega /k_B T)]^{-1}$ (neutron-energy loss).
Accordingly, the Bose-scaled high-temperature data can estimate the lattice contribution at lower
temperatures. Often this works well, especially in large magnetic clusters with a dense higher-lying energy
spectrum, and allows an unambiguous identification of magnetic peaks \cite{Dre10-csfe8ins3,Och08-cr6cr7}.

\subsubsection{Optical Spectroscopies}

\begin{figure}
\includegraphics[width=4.5cm]{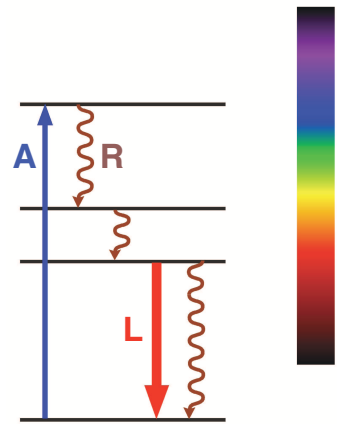}
  \caption{\label{fig:basics-opical} (Color online)
  Sketch of the processes relevant for optical spectroscopies. A: absorption; L: luminescence;
R: radiationless transition.
}
\end{figure}

Optical spectroscopies cover a large range of wavelengths of light. Individual spectrometers are specialized devices
that focus on particular parts of the electromagnetic spectrum produced by lamps, lasers or synchrotron sources. They
therefore exist in a wide variety of types for different applications \cite{Tka06}. One major type of optical
spectroscopy is absorption spectroscopy, where the absorbance of a system is determined by measuring the photons which
pass through (transmittance spectrum). Another important type is emission or luminescence spectroscopy. When a system
is excited by an outside energy source such as light, it eventually returns back to the ground state by releasing the
excess energy either as radiation-less transitions or in the form of photons as illustrated in
Fig.~\ref{fig:basics-opical}.

A variant of the absorption and luminescence spectroscopies, associated with initial transitions to discrete excited
energy states, is the Raman spectroscopy \cite{Lar11} where the system is excited to a virtual energy state and then
quickly relaxes back to a ground-state level. Unlike a luminescence process, Raman scattering involves no transfer of
electron population to the intermediate state. Several variations of Raman spectroscopy have been developed in order to
enhance the sensitivity [e.g., surface-enhanced Raman spectroscopy \cite{Lom08} and resonance Raman spectroscopy
\cite{Cha76}] as well as to improve the spatial resolution [Raman microscopy \cite{Tur96}].

Optical spectroscopies are governed by the energy and momentum conservation laws as in neutron spectroscopy, see
Eqs.~(\ref{eq:basics-ins-Q}) and (\ref{eq:basics-ins-E}). However, as the photon wave vector is about 10$^3$ times
smaller than a typical reciprocal lattice vector, only excitations close to the center of the Brillouin zone are
observed. The calculation of intensities of the observed transitions is a non-trivial task. This is in contrast to
neutron spectroscopy where the intensities of spin excitations are directly proportional to the square of the magnetic
dipole matrix elements, Eq.~(\ref{eq:basics-ins-S}). Since optical spectroscopies often involve intermediate states
which are not known, approximate models have to be employed for the calculation of transition matrix elements
\cite{Lov96}.

The polarization of light has great importance particularly when anisotropic systems are studied. The specific
polarization of both the exciting and the emitted light can be exploited to obtain extra information concerning
the line identification from the observed energy spectra. More specifically, electronic states with transition
dipole moments perpendicular to the electric field orientation will not be excited.

\subsubsection{Electron Paramagnetic Resonance}
\label{sec:basics-epr}

In EPR spectroscopy the absorption of a radio-frequency (rf) magnetic field ${\bf B}^{rf}$ by a magnetic
system is measured \cite{Abr86}. Absorption can occur whenever the energy $h\nu$ of the radiation matches the
energy difference of two eigen states $|\lambda\rangle$ and $|\lambda'\rangle$,
\begin{eqnarray}
\label{eq:basics-epr}
  E_{\lambda'} - E_{\lambda} = \pm h\nu,
\end{eqnarray}
and the absorbed power $P$ is calculated in linear response theory to
\begin{eqnarray}
\label{eq:basics-eprpower}
 P &=&  C(\omega) \sum_{ \alpha \beta}   B_\alpha^{rf}  B_\beta^{rf}
  \sum_{ i j} \sum_{ \lambda \lambda'} p_\lambda
  \langle \lambda | \hat{s}_{i\alpha} | \lambda' \rangle
  \langle \lambda' | \hat{s}_{j\beta} | \lambda \rangle
\nonumber \\ && \times
 \delta( \hbar \omega + E_\lambda - E_{\lambda'})
  ,
\end{eqnarray}
with $C(\omega) = \omega [ 1- \exp( \hbar \omega/k_B T) ](g^2 \mu_B^2)/(8 \hbar \pi)$. For magnetic cluster systems
with eigen states $|\tau S M\rangle$, Eq.~(\ref{eq:basics-eprpower}) can be further evaluated and the EPR selection
rules
\begin{eqnarray}
\label{eq:basics-eprselectionrules}
  \Delta S = 0,
  \\
  \Delta M = \pm 1,
\end{eqnarray}
established from $\sum_i \langle \lambda | \hat{s}_{i\alpha} | \lambda' \rangle = \langle \tau S M |
\hat{S}_\alpha | \tau' S' M'\rangle$.

It follows that EPR spectroscopy is a very direct method to determine anisotropies of the $g$ factor by aligning the
external magnetic field ${\bf B}$ along different directions. Similarly, anisotropies of the form defined by, e.g.,
Eqs.~(\ref{eq:basics-Dterm}) and (\ref{eq:basics-Eterm}), which lift the degeneracy of a particular spin multiplet, can
also be determined from the positions of the lines in the EPR spectra. On the other hand, the exchange splittings or
parameters $J_{ij}$ are not directly attainable, but they can be estimated from the temperature variation of the signal
intensities which follow the Boltzmann populations of the energy levels involved, or in some fortunate cases through
the $S$-mixing mechanism \cite{Wil06-ni4smix}. Finally we point to the distinctive hyperfine structure superimposed on
an EPR spectrum for systems with non-zero nuclear spin quantum numbers.

It is instructive to compare Eq.~(\ref{eq:basics-eprpower}) to the corresponding INS formula
Eq.~(\ref{eq:basics-ins-sigma}) with Eq.~(\ref{eq:basics-ins-S}). The main difference lies in the structure
factor $\exp({\bf Q}\cdot{\bf R}_{ij})$ which is 1 in case of EPR corresponding to ${\bf Q} = 0$ in INS.
Therefore, EPR affords the detection of exactly those magnetic transitions which have INS intensity at ${\bf
Q} \rightarrow 0$, which in the HDVV model are the $\Delta S = 0$ transitions. Physically speaking, in
contrast to a neutron the applied radio frequency establishes a frequency-dependent but spatially homogeneous
magnetic field, ${\bf B}({\bf Q}, \omega )$ with ${\bf Q} = 0$.

Modern EPR spectroscopy techniques permit a large combination of frequency and magnetic field values extending up to
the THz regime and 25~T, respectively. In principle, EPR spectra can be generated by either varying the frequency $\nu$
while holding the magnetic field constant, or doing the reverse. In commercial EPR instruments it is the frequency
which is kept fixed, and typical frequencies are X-band (10~GHz) and Q-band (35~GHz), but W-band (95~GHz) is also
available. However, the EPR techniques have progressed enormously, and multi-frequency high-field EPR and frequency
domain magnetic resonance spectroscopy (FDMRS) experiments are routinely undertaken in various laboratories. A recent
development are THz EPR experiments using radiation from synchrotron sources. For reviews see
\cite{Gat06-smmeprreview,Sla03-fdmrsreview}.

\subsubsection{Thermodynamic Magnetic Properties}
\label{sec:basics-thermodynamic}

The thermodynamic magnetic properties depend explicitly upon both the energies $E_\lambda$ and the eigen
functions $|\lambda\rangle$ of the spin excitations. Based on general expressions of statistical mechanics
for the Gibbs free energy $F$ and internal energy $U$,
\begin{eqnarray}
\label{eq:basics-F}
 F &=& - k_B T \ln Z,
 \\
\label{eq:basics-U}
 U &=& F - T \left( \frac{\partial F}{\partial T} \right)_V,
\end{eqnarray}
we obtain, with the Zeeman term as in Eq.~(\ref{eq:basics-Zeeman}), the magnetization $M_\alpha$, magnetic
susceptibility $\chi_{\alpha \alpha}$, entropy $S$, and Schottky heat capacity $c_V$:
\begin{eqnarray}
 \label{eq:basics-MchiS}
 M_\alpha &=& - \frac{\partial F}{\partial B_\alpha}
 = -g \mu_B \sum_\lambda p_\lambda \langle\lambda| \hat{S}_\alpha |\lambda\rangle ,
\\
 \label{eq:basics-chi}
  \chi_{\alpha\alpha} &=& \frac{\partial M_\alpha}{\partial B_\alpha}
 = \frac{ \left( g \mu_B\right)^2 }{k_B T } \\&& \times \left[
  \sum_\lambda p_\lambda \langle\lambda| \hat{S}_\alpha^2 |\lambda\rangle
 -
  \left( \sum_\lambda p_\lambda \langle\lambda| \hat{S}_\alpha |\lambda\rangle \right)^2
 \right]
, \nonumber
\\
 S &=& - \left( \frac{\partial F}{\partial T} \right)_V
 = k_B \left( \ln Z + \frac{ \sum_\lambda p_\lambda E_\lambda}{k_B T} \right)
,
\\
 c_V &=& \left( \frac{\partial U}{\partial T} \right)_V
 \nonumber \\ &=&
 \frac{1}{k_B T^2} \left[ \sum_\lambda  p_\lambda E^2_\lambda
     -  \left( \sum_\lambda  p_\lambda E_\lambda \right)^2 \right].
\end{eqnarray}
Here, $Z$ and $p_\lambda$ are the partition function and Boltzmann population factor, respectively:
\begin{eqnarray}
\label{eq:basics-Zp} Z = \sum_\lambda \exp(-\frac{E_\lambda}{k_B T}), \quad p_\lambda = \frac{1}{Z}
\exp(-\frac{E_\lambda}{k_B T}).
\end{eqnarray}
For a system with magnetic anisotropy also the magnetic torque $\bf \tau$ appears as a useful thermodynamic quantity:
\begin{eqnarray}
\label{eq:basics-torque}
 \tau = \frac{\partial F}{\partial \theta}, \quad  {\bf \tau} = {\bf M} \times {\bf B},
\end{eqnarray}
where $\theta$ denotes the rotation angle around the torque axis [often the definition $\tau = -\partial F /
\partial \phi$ is found, our convention is consistent with the usual parametrization of magnetic field,
e.g., ${\bf B} = B(\sin\theta,0,\cos\theta)$].

For $T \rightarrow 0$, the free energy reduces to the ground-state energy $E_0$ and the magnetization to the field
derivative $M_\alpha = - \partial E_0 / \partial B_\alpha$. As function of field the ground state often undergoes level
crossings at characteristic fields, which can be detected at very low temperatures as steps in the field-dependent
magnetization (torque) curves. The characteristic fields allow insight into the magnetic excitation spectrum in a
cluster, and low-temperature high-field magnetization (torque) measurements represent an important experimental
technique \cite{Sha02-msteps-review}, though the level crossing can also be detected by other techniques, e.g., proton
nuclear magnetic resonance \cite{Jul99}. In order to check the reliability of the model parameters derived from
spectroscopic data, it is however generally useful to compare the calculated thermodynamic magnetic properties to
corresponding experimental data.

%% file: chapter3_smallclusters_v2012_10_02.tex

\section{Small magnetic clusters}
\label{sec:sc}

The aim of this section is to demonstrate how the various interactions introduced in Sec.~\ref{sec:basics-spinham}
manifest themselves for different experimental techniques. The presented examples will be restricted to small clusters
built up by $N \leq 4$ coupled magnetic ions, for which the underlying models can be treated exactly, since only a
small number of interactions are present and cooperative effects do not occur, so that a straightforward comparison
between theory and experiment is possible with little ambiguity. The examples cover magnetic clusters which naturally
occur in pure compounds as well as clusters which are artificially formed in solid solutions of magnetic and
nonmagnetic compounds. Ideal examples of pure compounds are molecular transition-metal complexes, in which a
polynuclear metal core is embedded in a diamagnetic ligand matrix. Information more directly associated with
cooperative systems results from diluted magnetic compounds, in which the magnetic ions are randomly distributed, so
that different types of clusters ($N$-mers, $N =1, 2, 3, ...$) are simultaneously present. Among the myriads of small
magnetic cluster systems studied up to the present we will choose as a representative of the pure compounds the dimeric
chromium system [(NH$_3$)$_5$CrOHCr(NH$_3$)$_5$]Cl$_5$$\cdot$H$_2$O, which is the first magnetic cluster system
investigated by INS. The class of magnetically diluted systems was pioneered in INS experiments carried out for the
compound KMn$_x$Zn$_{1-x}$F$_3$ \cite{Sve78} and will be exemplified here by the compound CsMn$_x$Mg$_{1-x}$Br$_3$. The
chapter ends with further insights on particular physical aspects resulting from magnetic cluster excitations on some
other compounds.

\subsection{The dimeric chromium compound [(NH$_3$)$_5$CrOHCr(NH$_3$)$_5$]Cl$_5$$\cdot$H$_2$O}
\label{sec:sc-cr2}

\subsubsection{Energy levels}
\label{sec:sc-energylevels}

The simplest magnetic cluster system is the dimer (two coupled spins $\hat{\bf s}_1$ and $\hat{\bf s}_2$) for which the
HDVV Hamiltonian Eq.~(\ref{eq:basics-HDVV}) simplifies to
\begin{eqnarray}
\label{eq:sc-dimerH}
\hat{H} = -2 J \hat{\bf{s}}_1 \cdot \hat{\bf{s}}_2.
\end{eqnarray}
Assuming identical magnetic ions ($s_1=s_2=s$) the eigenvalues of Eq.~(\ref{eq:sc-dimerH}) are
\begin{eqnarray}
\label{eq:sc-dimerE}
E(S) = -J \left[ S(S+1) - 2s (s+1) \right],
\end{eqnarray}
with $0 \leq S \leq 2s$. The energy splittings defined by Eq.~(\ref{eq:sc-dimerE}) satisfy the Land\'e
interval rule
\begin{eqnarray}
\label{eq:sc-Lande}
 E(S) - E(S-1) = -2JS.
\end{eqnarray}
For Cr$^{3+}$ dimers with $s =3/2$ the separation between the ground-state levels will be $2J$, $4J$, and $6J$ ($S=0$
to $S=3$), with the state $S=0$ being the lowest in case of AFM exchange $J<0$. Observed deviations from the Land\'e
interval rule are often attributed to the presence of biquadratic exchange,
\begin{eqnarray}
\label{eq:sc-dimerbi}
\hat{H} = - K \left( \hat{\bf{s}}_1 \cdot \hat{\bf{s}}_2 \right)^2.
\end{eqnarray}
Combining Eqs.~(\ref{eq:sc-dimerH}) and (\ref{eq:sc-dimerbi}) yields the modified eigenvalues
\begin{eqnarray}
\label{eq:sc-dimerEbi}
E(S) &=& -J \eta - \frac{1}{4} K \eta^2,
\\
\eta &=&  S(S+1) - 2s (s+1).
\end{eqnarray}

\subsubsection{Structural and magnetic characterization}

\begin{figure}
\includegraphics[width=6.5cm]{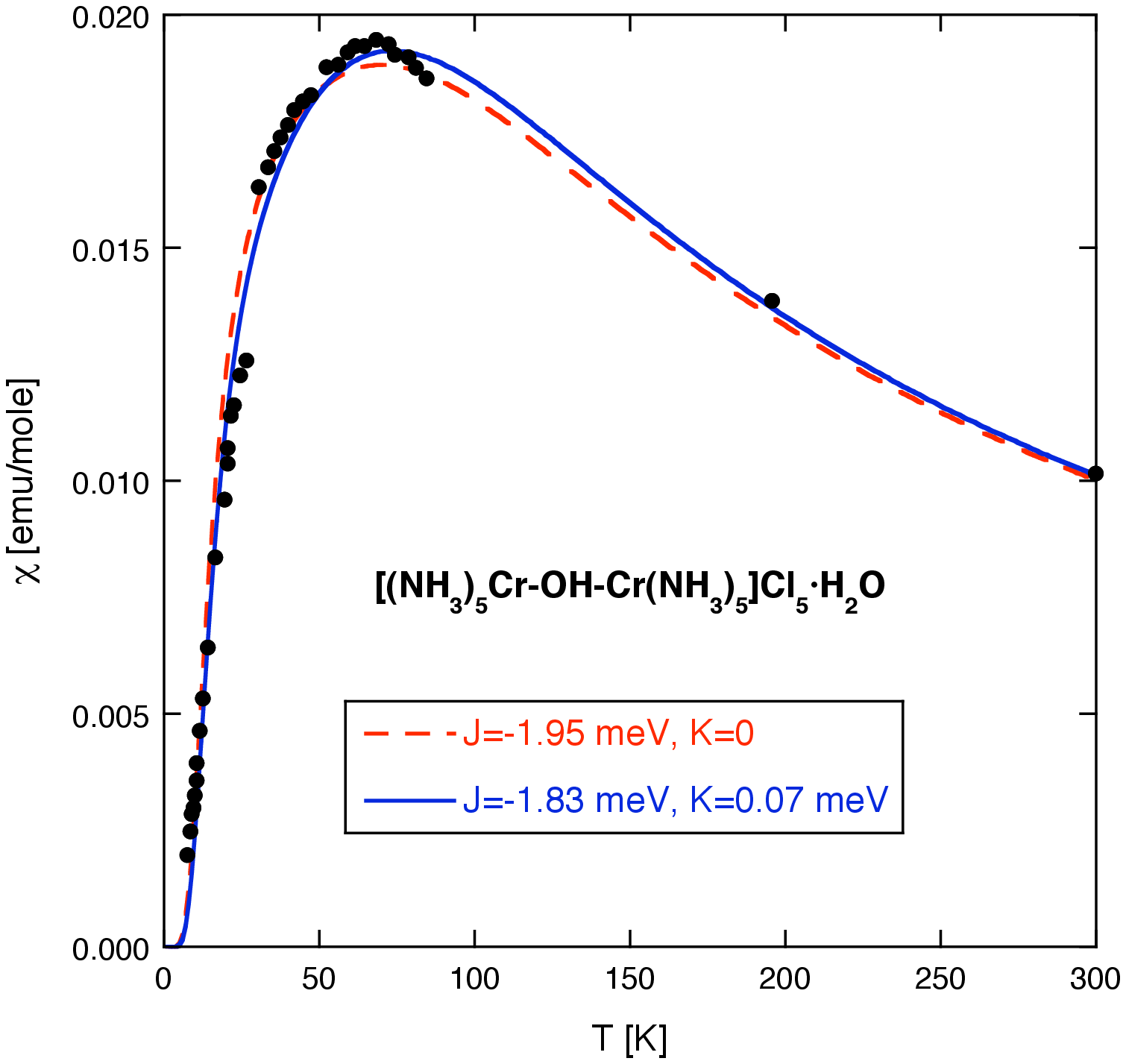}
  \caption{\label{fig:sc-crocrchi} (Color online)
Temperature dependence of the magnetic susceptibilty of [(NH$_3$)$_5$CrOHCr(NH$_3$)$_5$]Cl$_5$$\cdot$H$_2$O.
 Adapted from \onlinecite{Vea73}.
}
\end{figure}

The compound [(NH$_3$)$_5$CrOHCr(NH$_3$)$_5$]Cl$_5$$\cdot$H$_2$O was characterized by X-ray diffraction, EPR, and
magnetic susceptibility measurements \cite{Vea73}. The material crystallizes in the tetragonal space group P4$_2$/mnm
with four formula units in a cell of dimensions $a = 16.259(7)$~{\AA} and $c=7.411(7)$~{\AA}. The two Cr$^{3+}$ ions
are coupled by superexchange via a Cr-O-Cr bridge with a bridging angle of 165.6(9)$^\circ$ and a Cr-O distance of
1.94(1)~{\AA}. The analysis of the X-ray data requires two inequivalent positions of Cr$^{3+}$ dimers in the unit cell.
EPR measurements gave a negligibly small upper limit of $D \leq 0.002$~meV for the single-ion anisotropy defined by
Eq.~(\ref{eq:basics-Dterm}). The magnetic susceptiblity was measured for a polycrystalline sample as shown in
Fig.~\ref{fig:sc-crocrchi}. The data were analyzed according to Eq.~(\ref{eq:basics-chi}) with $g=1.99$ resulting from
the EPR experiments. The agreement between the observed and calculated data is slightly improved when in addition to
the Heisenberg exchange Eq.~(\ref{eq:sc-dimerH}) a biquadratic term Eq.~(\ref{eq:sc-dimerbi}) is included.

\subsubsection{Optical spectroscopies}

\begin{figure}
\includegraphics[width=6.5cm]{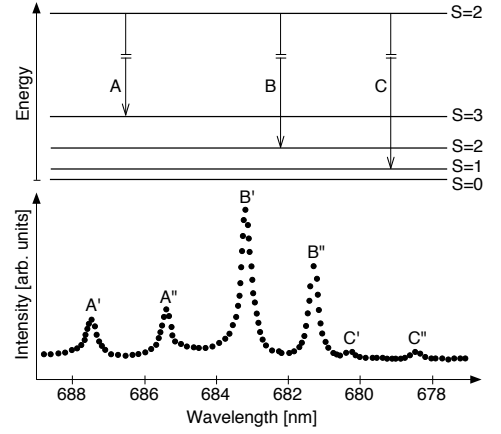}
  \caption{\label{fig:sc-crocropt}
  Polarized luminescence spectrum of [(NH$_3$)$_5$CrOHCr(NH$_3$)$_5$]-Cl$_5$$\cdot$H$_2$O taken at $T=7$~K.
  The corresponding transition diagram is shown at the top.
 Adapted from \onlinecite{Fer73}.
}
\end{figure}

Optical spectroscopies were applied to single crystals of [(NH$_3$)$_5$CrOHCr(NH$_3$)$_5$]Cl$_5$$\cdot$H$_2$O
\cite{Fer73}. Both polarized absorption and polarized luminescence spectra provided well resolved lines from which the
ground-state level scheme could be directly determined. Figure~\ref{fig:sc-crocropt} shows a representative polarized
luminescence spectrum with two sets of transitions (A',B',C') and (A",B",C") reflecting the presence of two
inequivalent dimer sites. The emission starts from an excited state with $S=2$. The appearance of three lines for each
of the transitions indicates that the selection rule $\Delta S=0$ is not exact, but transitions also occur for $\Delta
S= \pm1$ (with much smaller intensities) due to the spin-orbit interaction. The luminescence spectrum accurately
determines the separations between the ground-state levels $S=1, 2,$ and 3, and the separation between $S=0$ and $S=1$
was taken from the absorption spectrum. The ground-state level scheme slightly deviates from the Land\'e interval rule,
so that the data analysis was based on Eq.~(\ref{eq:sc-dimerEbi}). The resulting bilinear and biquadratic exchange
parameters $J$ and $K$ are listed in Table~\ref{tab:sc-tab1}. The luminescence spectrum is strongly temperature
dependent, and it is completely quenched at room temperature.

\subsubsection{Inelastic neutron scattering}

For the analysis of the neutron data we adjust the magnetic scattering function Eq.~(\ref{eq:basics-ins-S})
to the dimer case. We start from the reduced matrix elements introduced in Eq.~(\ref{eq:basics-ins-red}).
Since $\hat{T}^{1}(s_i)$ operates only on the $i$th ion of the coupled system, the reduced matrix elements
can be further simplified:
\begin{eqnarray}
\label{eq:sc-red}
 \langle S||\hat{T}^{1}(s_1)||S'\rangle
&=& (-1)^{2s+S+1} \sqrt{(2S+1)(2S'+1)} \nonumber
  \\&&\times
  \left\{\begin{array}{ccc} S&S'&1 \\ s&s&s \end{array} \right\} \langle s|||\hat{T}^{1}(s)|||s\rangle,
\\
\langle S||\hat{T}^{1}(s_2)||S'\rangle &=& (-1)^{S'-S} \langle S||\hat{T}^{1}(s_1)||S'\rangle,
\\
\langle s|||\hat{T}^{1}(s)|||s\rangle &=& \sqrt{s(s+1)(2s+1)}.
\end{eqnarray}
The two-row bracket $\{...\}$ in Eq.~(\ref{eq:sc-red}) is a Wigner-$6j$ symbol \cite{Rot59} which vanishes
unless $\Delta S= 0, \pm1$. From Eqs.~(\ref{eq:basics-ins-WE}) and (\ref{eq:sc-red}) the INS selection rules
$ \Delta M = 0, \pm1$ and $\Delta S = 0, \pm1$ are recovered. By making use of the symmetry properties of the
reduced matrix elements defined by Eq.~(\ref{eq:sc-red}), we find the following cross section for the dimer
transition $|S\rangle \rightarrow |S'\rangle$:
\begin{eqnarray}
\label{eq:sc-ins-sigma-dimer}
 \frac{d^2\sigma}{d\Omega d\omega} &=& C({\bf Q})
 \frac{ \exp[-\frac{ E(S) }{ k_B T }] }{Z} \sum_\alpha \left[ 1 -  \left(\frac{Q_\alpha}{Q} \right)^2 \right]
  \nonumber \\ &&\times
  \frac{2}{3}  \left[ 1 + (-1)^{\Delta S} \cos({\bf Q} \cdot {\bf R} ) \right] \langle S||\hat{T}^{1}(s_1)||S'\rangle^2
  \nonumber \\ &&\times
  \delta[\hbar \omega + E(S) - E(S')],
\end{eqnarray}
where ${\bf R} = {\bf R}_1-{\bf R}_2$ is the vector defining the intra-dimer separation. The structure factor $[ 1 +
(-1)^{\Delta S} \cos({\bf Q} \cdot {\bf R} )]$ is a powerful means to unambiguously identify dimer excitations from
other scattering contributions due to its characteristic oscillating behavior.

For a polycrystalline material Eq.~(\ref{eq:sc-ins-sigma-dimer}) has to be averaged in $\bf Q$ space:
\begin{eqnarray}
\label{eq:sc-ins-sigma-dimerpoly}
 \frac{d^2\sigma}{d\Omega d\omega} &=& C(Q)
 \frac{ \exp[-\frac{ E(S) }{ k_B T }] }{Z} \frac{4}{3}  \left[ 1 + (-1)^{\Delta S} \frac{ \sin(QR)}{QR} \right]
 \nonumber \\ &&\times
 \langle S||\hat{T}^{1}(s_1)||S'\rangle^2 \delta[\hbar \omega + E(S) - E(S')].
 \nonumber \\ &&
\end{eqnarray}
The polarization and structure factors combine to the interference factor $[ 1 + (-1)^{\Delta S}
\sin(QR)/(QR)]$, which produces a damped oscillatory $Q$ dependence of the intensities.

\begin{figure}
\includegraphics[width=8cm]{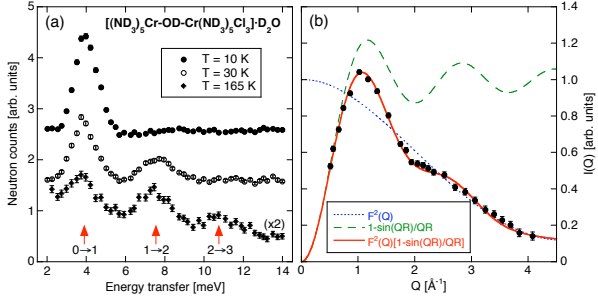}
  \caption{\label{fig:sc-crocrins} (Color online)
(a) Energy spectra of neutrons scattered from deuterated [(NH$_3$)$_5$CrOHCr(NH$_3$)$_5$]Cl$_5$$\cdot$H$_2$O. (b) $Q$
dependence of the intensity of the $|0\rangle \rightarrow |1\rangle$ transition observed at $T=4.2$~K shown in panel (a).
 Adapted from (a) \onlinecite{Gud81} and (b) \onlinecite{Fur77}.
}
\end{figure}

Figure~\ref{fig:sc-crocrins}(a) shows the temperature dependence of neutrons scattered from a polycrystalline
sample of deuterated [(NH$_3$)$_5$CrOHCr(NH$_3$)$_5$]Cl$_5$$\cdot$H$_2$O \cite{Fur77,Gud81}, which
demonstrates the successive appearance of the excited-state transitions with increasing temperature. The data
confirm the ground-state splitting pattern sketched on top of Fig.~\ref{fig:sc-crocropt}. The resulting
parameters based on Eq.~(\ref{eq:sc-dimerEbi}) are listed in Table~\ref{tab:sc-tab1}. The oscillatory
behavior of the intensities vs the modulus of the scattering vector $\bf Q$ predicted by
Eq.~(\ref{eq:sc-ins-sigma-dimerpoly}) is nicely verified as shown in Fig.~\ref{fig:sc-crocrins}(b).

\subsubsection{Comparison of different experimental techniques}

Table~\ref{tab:sc-tab1} lists the results obtained by different experimental techniques presented in the preceding
subsections. From the EPR measurements the anisotropic magnetic effects associated with the compound
[(NH$_3$)$_5$CrOHCr(NH$_3$)$_5$]Cl$_5$$\cdot$H$_2$O were established to be negligibly small. This was verified in
subsequent light and neutron spectroscopic investigations, which did not give evidence for any anisotropy-induced line
splittings. Due to the excellent energy resolution, light spectroscopies provide at low temperatures rather precise
spin coupling parameters, but information on their temperature dependence is severely hampered because of signal
quenching. This is not the case for inelastic neutron scattering, which gives evidence for a strong temperature
dependence of the bilinear exchange parameter $J$ of the order of 15\% upon heating from 7~K to room temperature. The
analysis of the magnetic susceptibility data thus results in a temperature-averaged parameter $J$, and it overestimates
the biquadratic coupling parameter $K$ by a factor of 2.

\begin{table}
\caption{\label{tab:sc-tab1} Coupling parameters $J$ and $K$ of
[(NH$_3$)$_5$CrOHCr(NH$_3$)$_5$]Cl$_5$$\cdot$H$_2$O determined by different experimental techniques.}
\begin{ruledtabular}
\begin{tabular}{cccc}
Technique & $T$ [K] &  $J$ [meV] & $K$ [meV] \\
\hline
mag. suscept.\footnotemark[1]  & 7-300 & -1.83 & 0.07  \\
light spec.\footnotemark[2] & 7 & -1.91(1) & 0.02(1)  \\
INS\footnotemark[3] & 30 & -1.88(5) & 0.03(2)  \\
INS\footnotemark[3] & 165 & -1.83(8) & 0.02(4) \\
INS\footnotemark[3] & 293 & -1.74(14) & 0.03(7)  \\
\end{tabular}
\end{ruledtabular}
 \footnotetext[1]{Ref.~\onlinecite{Vea73}.}
 \footnotetext[2]{Ref.~\onlinecite{Fer73}.}
 \footnotetext[3]{Ref.~\onlinecite{Gud81}.}
\end{table}

\subsection{Manganese $N$-mers in CsMn$_x$Mg$_{1-x}$Br$_3$}
\label{sec:sc-cmmb}

\subsubsection{Structural and magnetic characterization}

Solid solutions of composition CsMn$_x$Mg$_{1-x}$Br$_3$ are ideal model systems for various reasons. Both CsMnBr$_3$
and CsMgBr$_3$ crystallize in the hexagonal space group P6$_3$/mmc, and their unit cell parameters are almost equal:
$a=b=7.609(15)$~{\AA}, $c=6.52(5)$~{\AA} for CsMnBr$_3$ \cite{Goo72} and $a=b=7.610(2)$~{\AA}, $c=6.502(2)$~{\AA} for
CsMgBr$_3$ \cite{McP80}. The structure consists of chains of face-sharing MBr$_6$ octahedra parallel to the $c$ axis,
where M is Mn$^{2+}$ ($s = 5/2$) or Mg$^{2+}$ (diamagnetic). Spin-wave experiments gave evidence for a pronounced
one-dimensional magnetic behavior with the intra-chain exchange interaction exceeding the inter-chain exchange
interaction by three orders of magnitude \cite{Bre77,Fal87a}. All the Mn$^{2+}$ clusters in the mixed compound
CsMn$_x$Mg$_{1-x}$Br$_3$ are thus linear chain fragments with composition Mn$_N$Br$_{3(N+1)}$ ($N=1,2,3,...$) oriented
parallel to the $c$ axis. The Mn$^{2+}$ clusters are statistically distributed with the probability $p_N(x)$ for
$N$-mer formation given by
\begin{eqnarray}
\label{eq:sc-probability}
 p_N(x) = (1-x) x^{N-1}.
\end{eqnarray}
For Mn$^{2+}$ concentrations $x<0.05$, monomers and dimers dominate, but for $x>0.05$ trimers, tetramers, etc. have to
be considered. For the ground-state splitting pattern of dimers we refer to Sec.~\ref{sec:sc-energylevels}. The energy
levels of linear trimers and tetramers are summarized below.

\subsubsection{Energy levels of linear trimers and tetramers}

The HDVV Hamiltonian of a linear trimer is defined by
\begin{eqnarray}
\label{eq:sc-trimerHlin}
\hat{H} = -2J \left(\hat{\bf{s}}_1 \cdot \hat{\bf{s}}_2+\hat{\bf{s}}_2 \cdot \hat{\bf{s}}_3\right) -2J' \hat{\bf{s}}_1 \cdot \hat{\bf{s}}_3.
\end{eqnarray}
It is convenient to introduce the spin quantum numbers $S_{13}$ and $S$ resulting from the spin coupling
scheme defined by the vector sums $\hat{\bf S}_{13} = \hat{\bf s}_1 + \hat{\bf s}_3$ and $\hat{\bf S} =
\hat{\bf s}_2 + \hat{\bf S}_{13}$ with $0 \leq S_{13} \leq 2 s$ and  $|S_{13} - s| \leq S \leq (S_{13}+s)$,
respectively, assuming $s_1=s_2=s_3=s$. The trimer states are therefore defined by $|S_{13} S M\rangle$, and
their degeneracy is $(2S+1)$. With this choice of spin quantum numbers, the Hamiltonian
Eq.~(\ref{eq:sc-trimerHlin}) is diagonal and the eigenvalues can thus readily be derived as
\begin{eqnarray}
\label{eq:sc-trimerElin}
E(S_{13},S) &=& -J \left[ S(S+1) - S_{13}(S_{13}+1) - s (s+1) \right]
\nonumber \\&&
- J' \left[ S_{13}(S_{13}+1) - 2s (s+1) \right].
\end{eqnarray}
The HDVV Hamiltonian of a linear tetramer is given by
\begin{eqnarray}
\label{eq:sc-tetraHlin}
\hat{H} &=& -2J \left(\hat{\bf{s}}_1 \cdot \hat{\bf{s}}_2 + \hat{\bf{s}}_2 \cdot \hat{\bf{s}}_3 + \hat{\bf{s}}_3 \cdot \hat{\bf{s}}_4 \right)
\nonumber \\&&
 -2J' \left( \hat{\bf{s}}_1 \cdot \hat{\bf{s}}_3 + \hat{\bf{s}}_2 \cdot \hat{\bf{s}}_4 \right)
- 2J'' \hat{\bf{s}}_1 \cdot \hat{\bf{s}}_4.
\end{eqnarray}
To solve Eq.~(\ref{eq:sc-tetraHlin}), the total spin $\hat{\bf S} =\hat{\bf s}_1 + \hat{\bf s}_2 + \hat{\bf
s}_3 + \hat{\bf s}_4$ is still a good quantum number, but for a complete characterization of the tetramer
states additional intermediate spin quantum numbers are needed, e.g., $\hat{\bf S}_{12} = \hat{\bf s}_1 +
\hat{\bf s}_2$ and $\hat{\bf S}_{34} = \hat{\bf s}_3 + \hat{\bf s}_4$ with $0 \leq S_{12} \leq 2 s$ and $0
\leq S_{34} \leq 2 s$, respectively. The total spin is then defined by $|S_{12} - S_{34}| \leq S \leq
(S_{12}+S_{34})$, and the basis states are the wave functions $|S_{12} S_{34} S M\rangle$. There is no spin
coupling scheme which results in a diagonal Hamiltonian matrix, so that the eigenvalues of
Eq.~(\ref{eq:sc-tetraHlin}) have to be calculated numerically or by spin-operator techniques \cite{Jud63}.

\subsubsection{Electron paramagnetic resonance}

\begin{figure}
\includegraphics[width=6.5cm]{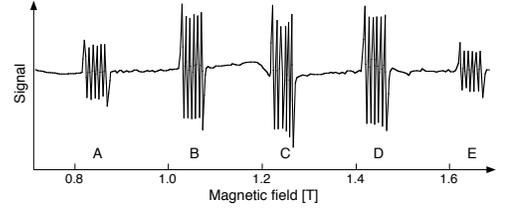}
  \caption{\label{fig:sc-cmmbepr}
  Q-band EPR spectrum of Mn$^{2+}$ ions in CsMgBr$_3$ at $T=77$~K.
  Adapted from \onlinecite{McP74}.
}
\end{figure}

\begin{figure}
\includegraphics[width=6.5cm]{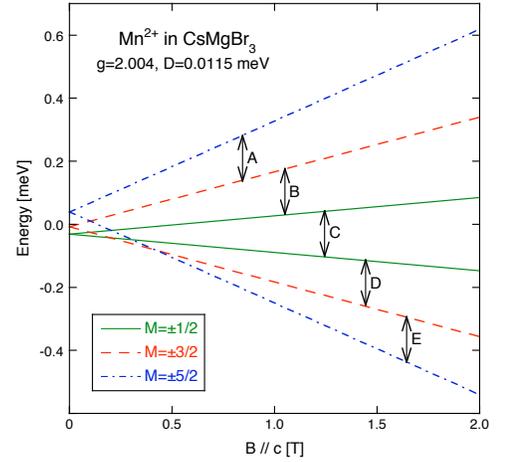}
  \caption{\label{fig:sc-cmmbepr2} (Color online)
Field dependence of the ground-state levels of Mn$^{2+}$ ions in CsMgBr$_3$. The calculations are based on
Eqs.~(\ref{eq:basics-Dterm}) and (\ref{eq:basics-Zeeman}), with $D>0$.
 The double arrows mark the resonances  A to E observed in the EPR spectrum of
Fig.~\ref{fig:sc-cmmbepr}. }
\end{figure}

Single crystals of CsMgBr$_3$ doped with Mn$^{2+}$ ions ($s=5/2$) were studied by EPR measurements at Q- and X-band
frequencies with the magnetic field parallel and perpendicular to the $c$ axis \cite{McP74}. The EPR spectrum displayed
in Fig.~\ref{fig:sc-cmmbepr} shows the hyperfine and fine structures expected for Mn$^{2+}$ ions in an axial
environment. The Q-band frequency of $\nu =35$~GHz produces five resonances A to E whenever the spacing of adjacent
Zeeman-split energy levels corresponds to $\Delta E=0.145$~meV [see Eq.~(\ref{eq:basics-epr})] as illustrated in
Fig.~\ref{fig:sc-cmmbepr2}. Each resonance is characterized by six oscillations due to the hyperfine interaction, since
the nuclear spin of manganese is $I=5/2$. The positions of the five resonances do not occur at equidistant spacings,
which indicates the presence of a non-zero single-ion anisotropy. From the positions of the resonances the Land\'e
splitting factor $g=2.004(1)$ and the axial anisotropy parameter $|D|= 0.0115(2)$~meV were obtained. Note that the sign
of $D$ cannot be determined from EPR experiments at elevated temperatures.

\subsubsection{Optical spectroscopies}

\begin{figure}
\includegraphics[width=8cm]{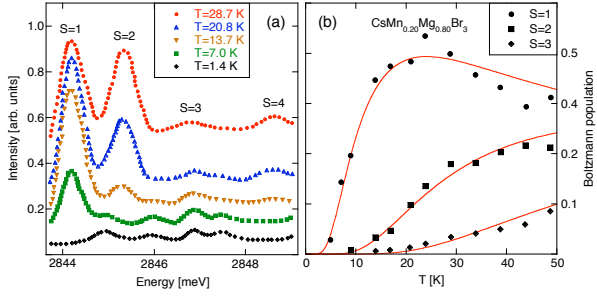}
  \caption{\label{fig:sc-cmmbopt} (Color online)
(a) Temperature dependence of absorption spectra resulting from a spin-flip process
observed for CsMn$_{0.20}$Mg$_{0.80}$Br$_3$. The successive
population of the Mn$^{2+}$ dimer states is marked by $S=1$ to $S=4$. (b) Observed intensities of the bands
$S=1$ to $S=3$ as a function of temperature. The lines correspond to the Boltzmann populations calculated
from Eq.~(\ref{eq:sc-dimerE}) with $J=-0.88$~meV .
 Adapted from \onlinecite{McC84}.
}
\end{figure}

Single crystals of CsMn$_x$Mg$_{1-x}$Br$_3$ ($0.04 \leq x \leq 0.20$) were investigated by optical spectroscopies
\cite{McC84}. In particular, Mn$^{2+}$ pair excitations were observed in the absorption spectra as shown in
Fig.~\ref{fig:sc-cmmbopt}. The weak absorptions at $T=1.4$~K cannot be assigned with certainty; they are either
single-ion absorptions or due to Mn$^{2+}$ clusters with $N>3$. With increasing temperature additional bands appear due
to the successive population of the cluster states $S=1$ to $S=4$. The temperature dependence of the intensities is
best described by using the HDVV Hamiltonian Eq.~(\ref{eq:sc-dimerE}) with $J=-0.88$~meV as illustrated in
Fig.~\ref{fig:sc-cmmbopt}.

\subsubsection{Inelastic Neutron Scattering}
\label{sec:IIIB5}

\begin{figure}
\includegraphics[width=6.5cm]{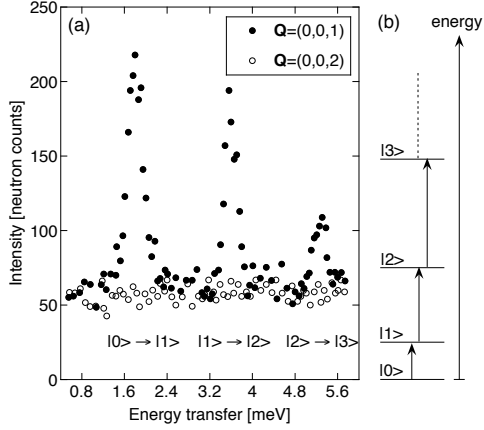}
  \caption{\label{fig:sc-cmmbins}
(a) Energy spectra of neutrons scattered from Mn$^{2+}$ pairs in CsMn$_{0.28}$Mg$_{0.72}$Br$_3$ at $T=30$~K. (b)
Energy-level sequence of an antiferromagnetically coupled spin pair.
 Adapted from \onlinecite{Fal84}.
}
\end{figure}

INS experiments performed on a single crystal of CsMn$_{0.28}$Mg$_{0.72}$Br$_3$ gave evidence for well defined
Mn$^{2+}$ dimer transitions as shown in Fig.~\ref{fig:sc-cmmbins} \cite{Fal84}. The observed intensities are in
excellent agreement with the predictions by the structure factor Eq.~(\ref{eq:sc-ins-sigma-dimer}); with ${\bf R}
=(0,0,1/2)$ the intensity has a maximum for ${\bf Q}=(0,0,1)$ and vanishes for ${\bf Q} =(0,0,2)$. The energies of the
transitions $|0\rangle \rightarrow |1\rangle$, $|1\rangle \rightarrow |2\rangle$, $|2\rangle \rightarrow |3\rangle$,
and $|3\rangle \rightarrow |4\rangle$ turned out to be 1.80(1), 3.60(1), 5.27(2), and 6.74(3)~meV, which considerably
deviate from the Land\'e interval rule, so that the data analysis was based on Eq.~(\ref{eq:sc-dimerEbi}). The
resulting parameters are $J = -838(5)$~$\mu$eV and $K = 8.8(8)$~$\mu$eV.

\begin{table}
\caption{\label{tab:sc-tab2} Resulting parameters from least-squares fits to the observed Mn trimer transitions in
CsMn$_{0.28}$Mg$_{0.72}$Br$_3$ \cite{Fal86}.}
\begin{ruledtabular}
\begin{tabular}{cccccc}
Model & $J$ [$\mu$eV] &  $J'$ [$\mu$eV] & $K$ [$\mu$eV] & $L$ [$\mu$eV] & $\chi^2$ \\
\hline
a $(K=L=0)$            & -870(12) & -8(14)  &  0  & 0  & 5.22\\
b $(K \neq 0, L=0)$    & -786(7)  & -12(10)  & 14(2)  &  0  & 3.13 \\
c $(K\neq 0, L\neq 0)$ & -777(6)  & -11(9)  & 8(1)  &  6(1) &  1.67\\
\end{tabular}
\end{ruledtabular}
\end{table}

Later INS experiments gave evidence for well defined Mn$^{3+}$ trimer and tetramer transitions \cite{Fal87b}. For the
evaluation of the differential neutron cross section we refer to references in Sec.~\ref{sec:basics-ins}. For a trimer
the selection rules of the transition $|S_{13} S M\rangle \rightarrow |S_{13}' S' M'\rangle$ are derived as
\begin{eqnarray}
\label{eq:sc-trimerlin-selctionrules}
 \Delta S &=& 0, \pm1,
\\
 \Delta S_{13} &=& 0, \pm1,
\\
 \Delta M &=& 0, \pm1.
\end{eqnarray}

\begin{figure}
\includegraphics[width=6.5cm]{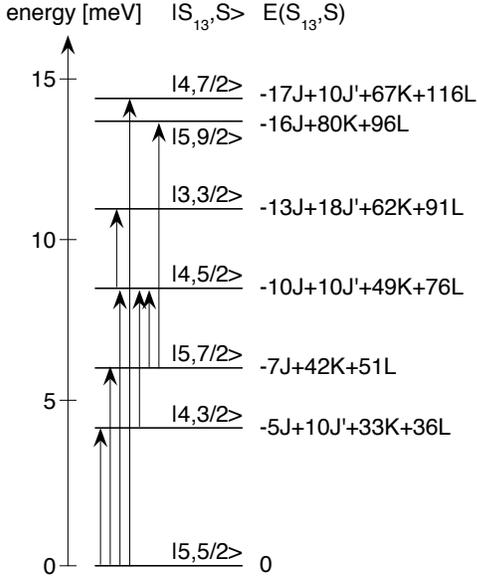}
  \caption{\label{fig:sc-cmmbspk}
Energy level splittings of Mn trimers in CsMn$_{0.28}$Mg$_{0.72}$Br$_3$. The arrows denote the
observed transitions.
 Adapted from \onlinecite{Fal86}.
}
\end{figure}

The smallest magnetic systems to identify three-spin interactions are spin trimers. The bilinear Hamiltonian
Eq.~(\ref{eq:sc-trimerHlin}) has to be extended in the following way:
\begin{eqnarray}
\label{eq:sc-trimerHlinBi}
\hat{H} &=& -2J \left( \hat{\bf{s}}_1 \cdot \hat{\bf{s}}_2+\hat{\bf{s}}_2 \cdot \hat{\bf{s}}_3 \right)
-2J' \hat{\bf{s}}_1 \cdot \hat{\bf{s}}_3
\nonumber \\ &&
 -K \left[ \left(\hat{\bf{s}}_1 \cdot \hat{\bf{s}}_2\right)^2 + \left(\hat{\bf{s}}_2 \cdot \hat{\bf{s}}_3\right)^2 \right]
 -K' \left(\hat{\bf{s}}_1 \cdot \hat{\bf{s}}_3\right)^2
\nonumber \\ &&
 -L \left[ \left(\hat{\bf{s}}_1 \cdot \hat{\bf{s}}_2\right)\left(\hat{\bf{s}}_2 \cdot \hat{\bf{s}}_3\right)
 + \left(\hat{\bf{s}}_3 \cdot \hat{\bf{s}}_2\right)\left(\hat{\bf{s}}_2 \cdot \hat{\bf{s}}_1\right)
 \right]
.
\end{eqnarray}
$K$, $K'$, and $L$ denote biquadratic two-spin and three-spin exchange parameters, respectively, which give rise to
off-diagonal matrix elements, so that Eq.~(\ref{eq:sc-trimerHlinBi}) was diagonalized in first-order perturbation
theory. The biquadratic $K'$ term is neglected, since $|K'| \ll |K|$. The low-energy part of the eigenvalues
$E(S_{13},S)$ is illustrated in Fig.~\ref{fig:sc-cmmbspk} for the case of Mn trimer excitations in
CsMn$_{0.28}$Mg$_{0.72}$Br$_3$, which were identified in INS experiments according to the characteristic dependence of
the cross section Eq.~(\ref{eq:basics-ins-sigma}) upon $\bf Q$ and $T$ \cite{Fal86}. The observed transitions are
marked by arrows in Fig.~\ref{fig:sc-cmmbspk}. Least-squares fits based on Eq.~(\ref{eq:sc-trimerHlinBi}) with
different parameter selection gave the results listed in Table~\ref{tab:sc-tab2}. The model including only bilinear
exchange interactions failed as expected. The model including the bilinear and biquadratic terms of the Hamiltonian
Eq.~(\ref{eq:sc-trimerHlinBi}) resulted in an improved standard deviation $\chi^2$, but only the least-squares fit
including the three-spin interaction was able to reproduce the observed transitions satisfactorily.

\begin{figure}
\includegraphics[width=6.5cm]{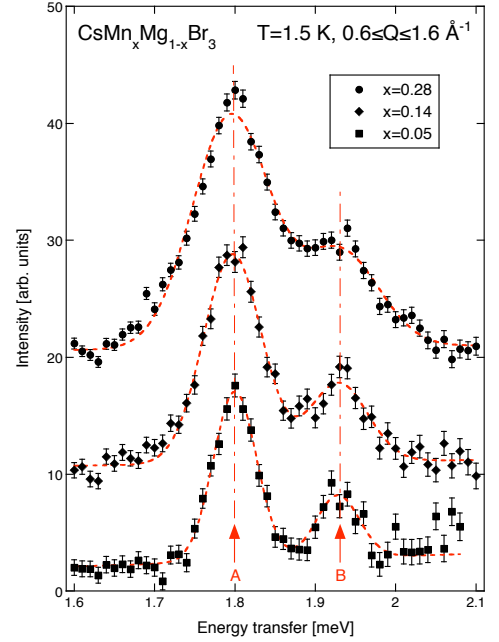}
  \caption{\label{fig:sc-cmmbxins} (Color online)
Energy spectra of neutrons scattered from CsMn$_x$Mg$_{1-x}$Br$_3$. The energy resolution
amounts to 55~$\mu$eV. For clarity, the data for $x=0.14$ and 0.28 are shifted by 10 and 20 intensity units,
respectively. The lines refer to Gaussian peak fits with equal line widths for both transitions A and B.
 Adapted from \onlinecite{Fur11}.
}
\end{figure}

Recent INS experiments performed with increased instrumental energy resolution gave evidence for anisotropy-induced
splittings of Mn$^{2+}$ dimer and tetramer transitions \cite{Fur11}. This is demonstrated in Fig.~\ref{fig:sc-cmmbxins}
for the dimer $|S=0\rangle \rightarrow |S=1\rangle$ transition. There are two well defined lines A and B which
according to the approximate intensity ratio 2:1 can be attributed to the $|S=0,M=0\rangle \rightarrow
|S=1,M=\pm1\rangle$ and $|S=0,M=0\rangle \rightarrow |S=1,M=0\rangle$ transitions, respectively. A similar
anisotropy-induced splitting was observed for the lowest tetramer $|S=0\rangle \rightarrow |S=1\rangle$ transition as
well. The dimer and tetramer data could be rationalized by the combined action of a single-ion anisotropy parameter
$D=0.0183(16)$~meV defined by Eq.~(\ref{eq:basics-Dterm}) and of two-ion anisotropic coupling parameters
$J=J_{xx}=J_{yy}=-0.852(3)$~meV and $J_{zz}/J=0.997(1)$ defined by Eq.~(\ref{eq:basics-anisoexchange}). The two-ion
anisotropy is most likely due to the anisotropic part of the dipole-dipole interaction
Eq.~(\ref{eq:basics-dipoledipole}). The exchange coupling $J$ is sufficiently strong to keep the spins $\hat{\bf s}_i$
antiferromagnetically aligned at low temperatures $T \leq |J|/k_B$, but their direction with respect to ${\bf R} ||
{\bf c}$ is free to rotate. Therefore, the second term of Eq.~(\ref{eq:basics-dipoledipole}) has to be averaged in
space:
\begin{eqnarray}
\label{eq:sc-dipoledipoleaveraged}
\hat{H} = \sum_{ij} \frac{g^2\mu_B^2}{R^3_{ij}}
\left[
\hat{\bf{s}}_{i}\cdot \hat{\bf{s}}_{j}
-
3 \frac{(\hat{\bf{s}}_{i}\cdot {\bf{R}}_{ij})(\hat{\bf{s}}_{j}\cdot {\bf{R}}_{ij})}{ \pi^2 R^2_{ij}}
\right],
\end{eqnarray}
The dipole-dipole anisotropy calculated from Eq.~(\ref{eq:sc-dipoledipoleaveraged}) is $J_{zz}/J=0.997$, in agreement
with the experimental findings.

\subsubsection{Comparison of different experimental techniques}
\label{sec:IIIB6}

Table~\ref{tab:sc-tab3} lists the results obtained by different experimental techniques presented in the preceding
subsections. The sign of the axial anisotropy parameter $D$ could unambiguously be determined by neutron spectroscopy,
in contrast to the EPR experiments. The parameters $D$ and $J$ exhibit a pronounced temperature dependence probably due
to the lattice expansion with increasing $T$, whereas the parameter $K$ remains constant. Since the analysis of the
optical data was based on a model with $K=0$, the resulting exchange parameter $J$ cannot be compared with the results
of the INS experiments. It was shown in \onlinecite{Fal84,Str04} that the presence of biquadratic exchange ($K \neq 0$)
is caused to a major extent by the mechanism of exchange striction \cite{Kit60}.

\begin{table}
\caption{\label{tab:sc-tab3} Axial anisotropy parameter $D$ and spin coupling parameters $J$ and $K$ of Mn$^{2+}$
dimers in CsMn$_x$Mg$_{1-x}$Br$_3$ determined by different experimental techniques.}
\begin{ruledtabular}
\begin{tabular}{ccccc}
Technique & $T$ [K] &  $D$ [meV] & $J$ [meV] & $K$ [meV]  \\
\hline
EPR\footnotemark[1]  & 77 & $\pm$0.0115(2) & - & -  \\
Optical\footnotemark[2] & 13 & - & -0.88 & -  \\
INS\footnotemark[3] & 30 & - & -0.838(5) & 0.0088(8)  \\
INS\footnotemark[4] & 50 & - & -0.823(1) & 0.0087(2)  \\
INS\footnotemark[5] & 1.5 & 0.0183(16) & -0.852(3) & 0.0086(2)  \\
\end{tabular}
\end{ruledtabular}
 \footnotetext[1]{Ref.~\onlinecite{McP74}.}
 \footnotetext[2]{Ref.~\onlinecite{McC84}.}
 \footnotetext[3]{Ref.~\onlinecite{Fal84}.}
 \footnotetext[4]{Ref.~\onlinecite{Str04}.}
 \footnotetext[5]{Ref.~\onlinecite{Fur11}.}
\end{table}

In extended antiferromagnets the observation of the spin-wave dispersion by single-crystal INS experiments is usually
the most common approach to determine exchange parameters. By applying the spin-wave formalism to $\hat H$ from
Eq.~(\ref{eq:sc-trimerHlinBi}), which includes higher-order exchange terms, we find
\begin{eqnarray}
\label{eq:sc-swtE}
\hbar \omega(q) &=& 4 s |J_{eff}| \sin(q c),
\\
J_{eff} &=& - | \sqrt{J(J-4J')} + \frac{5s}{2} (K+2L) |,
\end{eqnarray}
where $q$ is the wave number of the spin wave propagating along the $c$ axis. From spin-wave experiments performed for
the one-dimensional antiferromagnet CsMnBr$_3$ ($s=5/2$) the exchange coupling was determined to be $J_{eff}=
-0.89(2)$~meV \cite{Bre77,Fal87a}. The analysis of the spin-wave dispersion just yields an effective exchange parameter
$J_{eff}$, but the individual sizes of the bilinear and biquadratic exchange parameters cannot be determined. This is
in contrast to experiments on small magnetic clusters as discussed in the preceding sections. A numerical comparison of
the $J_{eff}$ values obtained from the three models listed in Table~\ref{tab:sc-tab2} is interesting. Using
Eq.~(\ref{eq:sc-trimerHlinBi}) and $s=5/2$ we find $J_{eff}= -0.89(2)$, -0.90(3) and -0.92(3)~meV for models a, b and
c, respectively. The three values are identical within experimental error, and they excellently agree with $J_{eff}$
determined from spin-wave experiments. They also agree with $J_{eff}= -0.88$~meV derived from the optical
spectroscopies applied to the Mn$^{2+}$ dimer excitations, see Table~\ref{tab:sc-tab3}. $J_{eff}$ is obviously
independent of the geometric size of the coupled magnetic ions and can therefore be regarded as a measure of the
magnetic energy per Mn$^{2+}$ ion in the AFM state of CsMnBr$_3$.

\subsection{Further insights from magnetic cluster excitations}
\label{sec:sc-further}

\subsubsection{Exchange parameters from high-field magnetization steps}

\begin{figure}
\includegraphics[width=6.5cm]{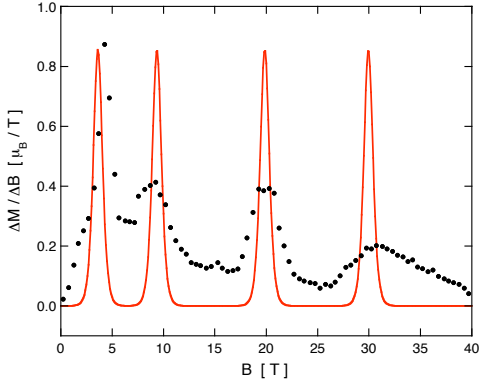}
  \caption{\label{fig:sc-ni4mag} (Color online)
High-field differential magnetization of \{Ni$_4$Mo$_{12}$\}. The circles represent the
experimental data taken at 0.44~K, and the line a calculation based on the
Hamiltonians Eqs.~(\ref{eq:sc-ni4H}), (\ref{eq:basics-Dterm}) and (\ref{eq:basics-Eterm}) with the model parameters
listed in the text.
 Adapted from \onlinecite{Sch06}.
}
\end{figure}

Step-like features in high-field magnetization data result from level crossings associated with the ground state of
magnetic clusters and thereby provide information about the exchange parameters. This is demonstrated here for the
tetrameric nickel compound [Mo$_{12}$O$_{28}$(OH)$_{12}$\{Ni(H$_2$O)$_3$\}$_4$]$\cdot$13H$_2$O, henceforth abbreviated
as \{Ni$_4$Mo$_{12}$\}. The magnetization is enhanced from zero up to the saturation value of 8~$\mu_B$ in steps of
2~$\mu_B$ at the fields 4.5, 8.9, 20.1, and 32~T as illustrated by the differential magnetization data in
Fig.~\ref{fig:sc-ni4mag} \cite{Sch06}.

\begin{figure}
\includegraphics[width=6.5cm]{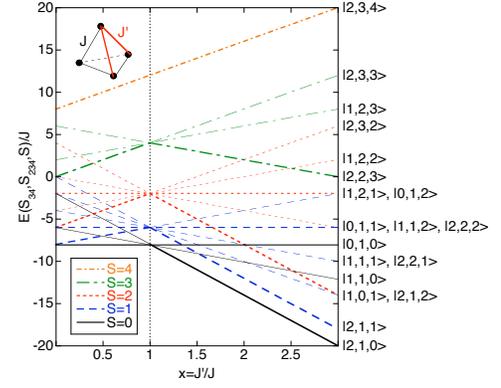}
  \caption{\label{fig:sc-ni4spk} (Color online)
Energy levels of $S=1$ tetramers calculated from Eq.~(\ref{eq:sc-ni4E}) with $J<0$. The lowest states of a
given $S$ value are marked by bold lines. The $|S_{34} S_{234} S\rangle$ states are identified at the right
hand side. The insert shows the coupling parameters in a slightly distorted tetrahedron as realized for the
compound \{Ni$_4$Mo$_{12}$\}.
}
\end{figure}

The four antiferromagnetically coupled Ni$^{2+}$ ions ($s=1$) in \{Ni$_4$Mo$_{12}$\} are arranged in a
slightly distorted tetrahedron, i.e., the Ni(2)-Ni(3) and Ni(2)-Ni(4) distances are slightly shorter than the
other four Ni-Ni distances as shown in the insert of Fig.~\ref{fig:sc-ni4spk}, thus the spin Hamiltonian is
described by
\begin{eqnarray}
\label{eq:sc-ni4H}
  \hat{H} &=& -2J \left(\hat{\bf{s}}_1 \cdot \hat{\bf{s}}_2 + \hat{\bf{s}}_1 \cdot
\hat{\bf{s}}_3 + \hat{\bf{s}}_1 \cdot \hat{\bf{s}}_4 + \hat{\bf{s}}_3 \cdot \hat{\bf{s}}_4 \right)
 \nonumber\\&&
 -2J' \left( \hat{\bf{s}}_2 \cdot \hat{\bf{s}}_3 + \hat{\bf{s}}_2 \cdot \hat{\bf{s}}_4 \right),
\end{eqnarray}
which can be brought to diagonal form by choosing the spin quantum numbers according to the vector couplings
$\hat{\bf S}_{34}= \hat{\bf s}_3 + \hat{\bf s}_4$, $\hat{\bf S}_{234}= \hat{\bf s}_2 + \hat{\bf S}_{34}$, and
$\hat{\bf S}= \hat{\bf s}_1 + \hat{\bf S}_{234}$ with $0 \leq S_{34} \leq 2s$, $|S_{34} - s| \leq S_{234}
\leq (S_{34}+s)$, and $|S_{234}-s| \leq S \leq (S_{234}+s)$, respectively. The eigenvalues of
Eq.~\ref{eq:sc-ni4H} for $s_1=s_2=s_3=s_4=s$ are then given by
\begin{eqnarray}
\label{eq:sc-ni4E}
 E( S_{34},S_{234},S) &=& -J [ S(S+1)-S_{234}(S_{234}+1)
 \nonumber\\&&
  -S_{34}(S_{34}+1) ]
  -J' [ S_{234}(S_{234}+1)
 \nonumber\\&&
  - S_{34}(S_{34}+1) -s(s+1) ].
\end{eqnarray}

Figure~\ref{fig:sc-ni4spk} displays the energy levels $E( S_{34},S_{234},S)$ normalized to $J$ (assuming AFM coupling
$J<0$) as a function of the ratio $x=J'/J$. For $x=1$ the energy levels are degenerate with respect to the total spin
$S$, and the energy splittings follow the Land\'e rule Eq.~(\ref{eq:sc-Lande}). By applying a magnetic field the ground
state changes in steps from $S=0$ to $S=4$ for the field values corresponding to the maxima of the $dM/dH$ data
displayed in Fig.~\ref{fig:sc-ni4mag}.

\begin{figure}
\includegraphics[width=6.5cm]{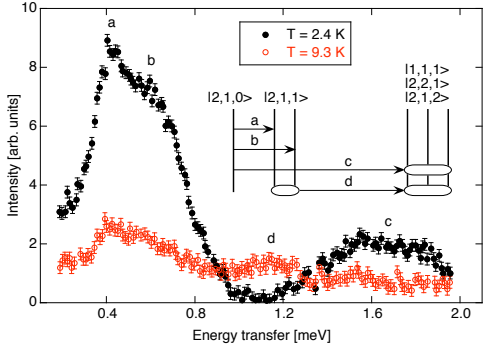}
  \caption{\label{fig:sc-ni4ins} (Color online)
Energy spectra of neutrons scattered from polycrystalline \{Ni$_4$Mo$_{12}$\} with an incident neutron energy
of 3.27~meV. The insert attributes the observed transitions to the low-energy part of the
splitting pattern. Adapted from \onlinecite{Neh10}.
}
\end{figure}

INS spectra measured for a polycrystalline sample of \{Ni$_4$Mo$_{12}$\} are shown in Fig.~\ref{fig:sc-ni4ins}, from
which two ground-state transitions can be identified at ca. 0.5 and 1.7~meV \cite{Neh10}. The former is composed of two
subbands at 0.4 and 0.6~meV attributed to a ZFS from magnetic anisotropy. An excited-state transition appears at 1.2
meV. From Fig.~\ref{fig:sc-ni4spk} we can readily conclude that the singlet $|2,1,0\rangle$ has to be the ground state.
Moreover, the first-excited state has to be the triplet $|2,1,1\rangle$ centered at 0.5 meV, since transitions between
$S=0$ states are not allowed. The observed splitting of the triplet $|2,1,1\rangle$ into two components can be ascribed
to an axial single-ion anisotropy defined by Eq.~(\ref{eq:basics-Dterm}), which has the effect to split the states
$|S_{34} S_{234} S \rangle$ into the states $|S_{34} S_{234} S M\rangle$. A least-squares fit to the energy spectra of
Fig.~\ref{fig:sc-ni4ins} on the basis of the Hamiltonians Eq.~(\ref{eq:sc-ni4H}) and (\ref{eq:basics-Dterm}) converged
to the parameters $J= -0.25(2)$~meV, $J'= -0.53(4)$~meV, $D= 0.22(5)$~meV \cite{Fur10b} which nicely reproduce the
high-field magnetization data, see Fig.~\ref{fig:sc-ni4mag}. The resulting low-energy splitting pattern is sketched in
the insert of Fig.~\ref{fig:sc-ni4ins}.

\subsubsection{Pressure dependence of exchange parameters}
\label{sec:sc-pressure}

\begin{figure}
\includegraphics[width=6.5cm]{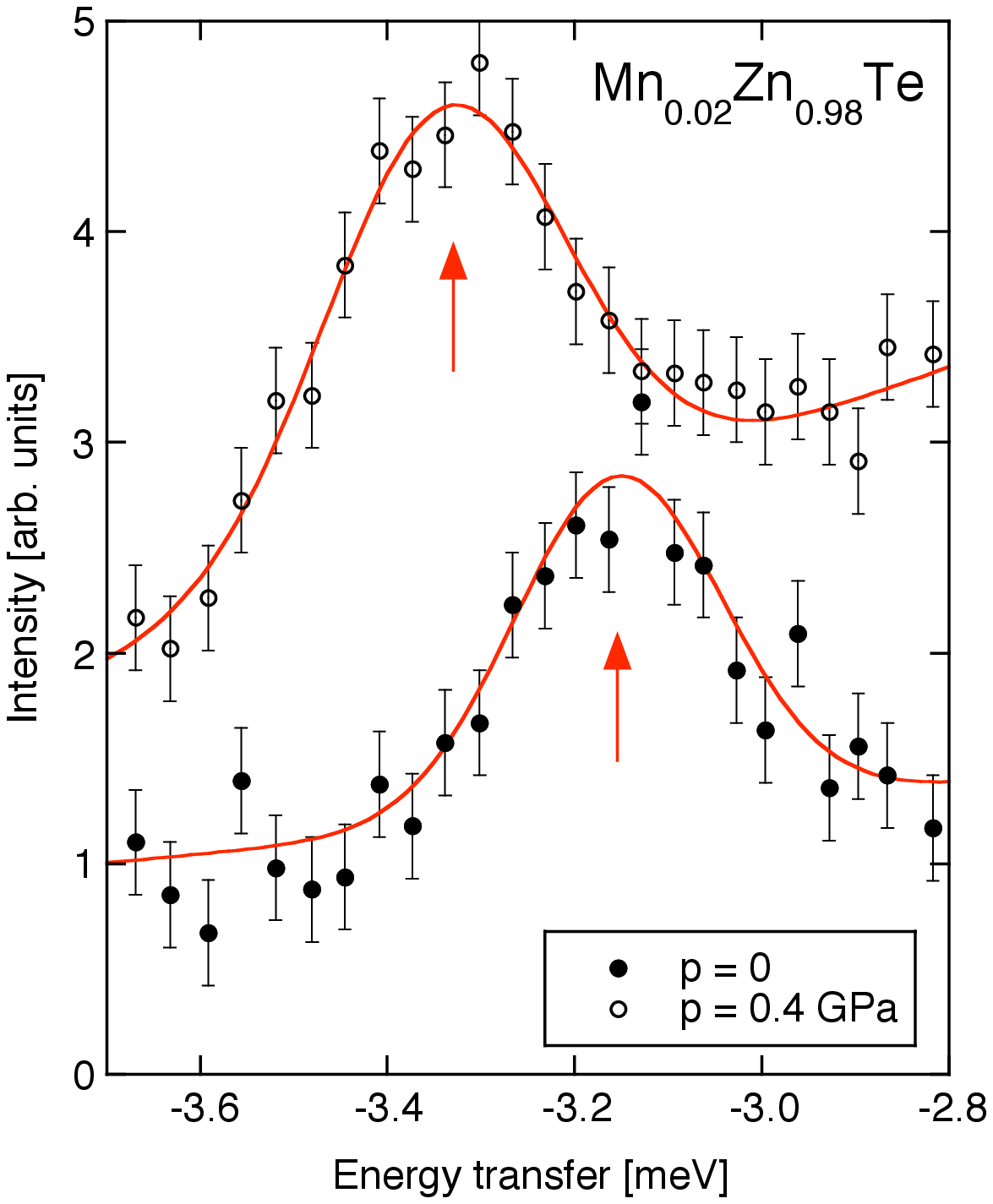}
  \caption{\label{fig:sc-mztp} (Color online)
Pressure dependence of the $|2\rangle \rightarrow |1\rangle$ transition associated with Mn$^{2+}$ dimers in
Mn$_{0.02}$Zn$_{0.98}$Te measured by INS. The lines denote Gaussian fits to the data.
 Adapted from  \onlinecite{Kol06}.
}
\end{figure}

By using external pressure the exchange parameters $J$ can be determined for varying distance $R$ between the
magnetic ions. This is of importance for testing and improving theoretical models of the exchange interaction,
where the distance usually enters in a straightforward manner. A detailed understanding of the exchange
interaction is, for instance, indispensable for the engineering of spintronics devices made of magnetic
semiconductors. In an effort to shed light on this issue the pressure dependence of $J$ was investigated for
antiferromagnetically coupled Mn$^{2+}$ dimers in the semiconducting compound Mn$_{0.02}$Zn$_{0.98}$Te
\cite{Kol06}. The corresponding energy-level scheme is indicated in Fig.~\ref{fig:sc-cmmbins}. INS experiments
performed for pressures of $p=0$ and $p=0.4$~MPa gave evidence for an appreciable pressure-induced upward shift
of the observed dimer excitations as illustrated for the $|2\rangle \rightarrow |1\rangle$ transition with
energy $4|J|$ in Fig.~\ref{fig:sc-mztp}. The pressure-induced change of $J$ amounts to $dJ=-0.040(9)$~meV,
accompanied by a 0.49\% decrease of the intra-dimer distance $R$, resulting in a linear distance dependence
$|dJ/dR|=1.8(4)$~meV/{\AA} for $dR \ll R$. Similar INS experiments performed for CsMn$_{0.28}$Mg$_{0.72}$Br$_3$
gave evidence for a much stronger distance dependence of $J$ with $|dJ/dR|=3.6(3)$~meV/{\AA} \cite{Str04}.

\subsubsection{Doping dependence of exchange parameters}

\begin{figure}
\includegraphics[width=6.5cm]{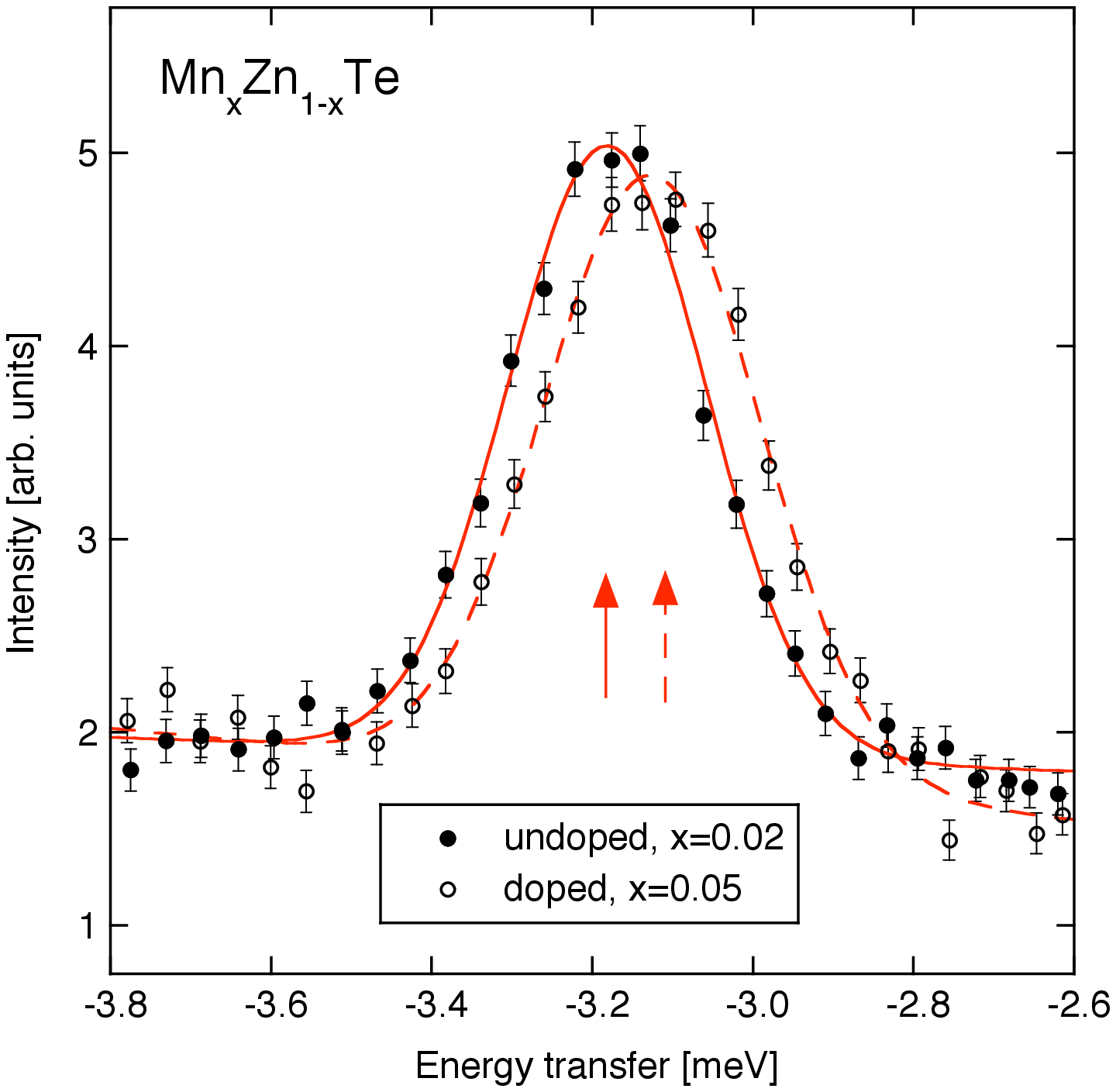}
  \caption{\label{fig:sc-mztdop} (Color online)
Doping dependence of the $|2\rangle \rightarrow |1\rangle$ transition associated with Mn$^{2+}$ dimers in
Mn$_x$Zn$_{1-x}$Te measured by INS. The lines denote Gaussian fits to the data.
 Adapted from  \onlinecite{Kep03}.
}
\end{figure}

It has been shown that a magnetic semiconductor can be converted by hole doping from its intrinsic AFM state to
a ferromagnet \cite{Fer01}. It is still an open question whether the holes are localized or itinerant. The
doping dependence of the exchange parameters may shed light on this issue. Neutron spectroscopic measurements
were performed for single crystals of Mn$_x$Zn$_{1-x}$Te, one with $x=0.05$ and doped with P to a level of
5$\times$10$^{19}$~cm$^{-3}$, and another undoped reference sample with $x=0.02$ \cite{Kep03}. The experiments
were similar as those described in Sec.~\ref{sec:sc-pressure} and gave evidence for a distinct doping-induced
downward shift of the observed Mn$^{2+}$ dimer excitations as illustrated for the $|2\rangle \rightarrow
|1\rangle$ transition with energy $4|J|$ in Fig.~\ref{fig:sc-mztdop}. The doping-induced change of the exchange
energy $J$ amounts to $dJ=0.013(3)$~meV, in reasonable agreement with $dJ=0.010$~meV calculated from the RKKY
interaction which indicates that the ferromagnetic exchange is mediated by weakly localized holes.

\subsubsection{Anisotropic exchange interactions}

\begin{figure}
\includegraphics[width=6.5cm]{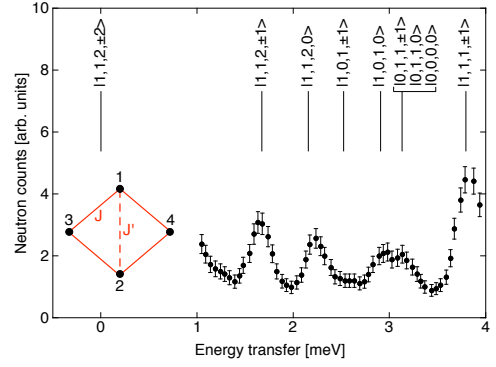}
  \caption{\label{fig:sc-co4} (Color online)
Energy spectrum of neutrons scattered from K$_{10}$[Co$_4$(D$_2$O)$_2$(PW$_9$O$_{34}$)$_2$]$\cdot$20D$_2$O at
$T=1.7$~K. The inserts show a sketch of the tetrameric Co$^{2+}$ unit and the resulting
energy level scheme, labeled with the dominant component $|S_{12} S_{34} S, \pm M\rangle$ of the wave
function. The highest states $|1,1,1,0\rangle$ and $|1,1,0,0\rangle$ located at 5.04 and 5.79~meV,
respectively, are not shown.
 Adapted from \onlinecite{Cle97}.
}
\end{figure}

Exchange anisotropy can generally be expected for type-L and type-J compounds where orbital degeneracy is
present. This is exemplified here for the type-L compound
K$_{10}$[Co$_4$(D$_2$O)$_2$(PW$_9$O$_{34}$)$_2$]$\cdot$20D$_2$O which contains a tetrameric Co$^{2+}$ cluster as
sketched in Fig.~\ref{fig:sc-co4}. The combined action of spin-orbit and crystal-field interactions splits the
$^4$T$_1$ single-ion ground state of the Co$^{2+}$ ions into six anisotropic Kramers doublets \cite{Car86}. By
considering only the lowest single-ion level, an effective spin Hamiltonian of the Co$^{2+}$ tetramer with $s
=1/2$ for all ions can be written as
\begin{eqnarray}
\label{eq:sc-co4H}
  \hat{H} &=& -2J [
         \hat{s}_{1,x} \hat{s}_{3,x} + \hat{s}_{1,x} \hat{s}_{4,x}
       + \hat{s}_{2,x} \hat{s}_{3,x} + \hat{s}_{2,x} \hat{s}_{4,x}
 \nonumber\\&&
       + \hat{s}_{1,y} \hat{s}_{3,y} + \hat{s}_{1,y} \hat{s}_{4,y}
       + \hat{s}_{2,y} \hat{s}_{3,y} + \hat{s}_{2,y} \hat{s}_{4,y}
 \nonumber\\&&
 + \alpha (
         \hat{s}_{1,z} \hat{s}_{3,z} + \hat{s}_{1,z} \hat{s}_{4,z}
       + \hat{s}_{2,z} \hat{s}_{3,z} + \hat{s}_{2,z} \hat{s}_{4,z}
   )
   ]
 \nonumber\\&&
 -2J' (
         \hat{s}_{1,x} \hat{s}_{2,x}
       + \hat{s}_{1,y} \hat{s}_{2,y}
 + \alpha'
         \hat{s}_{1,z} \hat{s}_{2,z}
   ).
\end{eqnarray}
It turns out that for this particular system the eigen functions are approximately given by the spin functions $|S_{12}
S_{34} S M\rangle$ constructed through the spin coupling scheme $\hat{\bf S}_{12} = \hat{\bf s}_1+ \hat{\bf s}_2$,
$\hat{\bf S}_{34} = \hat{\bf s}_3+ \hat{\bf s}_4$, $\hat{\bf S} = \hat{\bf S}_{12}+ \hat{\bf S}_{34}$, with less than
1\% $S$-mixing.

INS experiments gave evidence for well defined transitions associated with the Co$^{2+}$ tetramer as shown in
Fig.~\ref{fig:sc-co4} \cite{Cle97}. The data analysis based on Eq.~(\ref{eq:sc-co4H}) provided the exchange
parameters $J=0.52$~meV, $\alpha =2.4$, $J'=0.11$~meV, and $\alpha'=4.6$, resulting in the energy level scheme
indicated in Fig.~\ref{fig:sc-co4}. Both interactions $J$ and $J'$ are ferromagnetic, thus leading to an $M= \pm
2$ ground state, in agreement with magnetic susceptibility and EPR experiments \cite{Gom92}. The exchange
anisotropy is rather large as expected from the anisotropy of the Land\'e $g$ matrix with components in the
range 2.6-7.0 observed by EPR experiments.

\subsubsection{Higher-order single-ion anisotropies}
\label{sec:sc-higherorders}

Anisotropy-induced ground-state level splittings are essential to understand the step-like magnetic hysteresis curves
and the related relaxation and spin reversal phenomena observed in the single-molecule magnets (see Sec~\ref{sec:smm}).
EPR is the experimental tool of choice to determine ground-state level splittings, but because of the typically large
ZFS parameters $D$ high magnetic fields and/or high frequencies are needed to obtain sufficiently resolved spectra. INS
experiments offer a valuable alternative in zero field which will be demonstrated here for the tetrameric iron compound
Fe$_4$(OCH$_3$)$_6$(dpm)$_6$, or Fe$_4$ in short, which has an $S=5$ ground state and shows slow relaxation of the
magnetization below 1~K \cite{Bar99}. The four iron atoms lie exactly in a plane, with the inner Fe atom being in the
center of an isosceles triangle. EPR experiments \cite{Bou01} have shown that the single-ion anisotropies defined by
Eqs.~(\ref{eq:basics-Dterm}) and (\ref{eq:basics-Eterm}) are not sufficient to reproduce the observed signals in the
ground-state multiplet, but higher-order anisotropy terms as given in Eq.~(\ref{eq:basics-O4term}) are needed. The
anisotropy parameters resulting from the EPR experiments are listed in Table~\ref{tab:sc-fe4}.

\begingroup
\squeezetable
\begin{table}
\caption{\label{tab:sc-fe4} Anisotropy parameters determined for Fe$_4$ (isomer AA) by EPR \cite{Bou01} and
INS \cite{Amo01} experiments.}
\begin{ruledtabular}
\begin{tabular}{cccccc}
 & $D$ [$\mu$eV] & $E$ [$\mu$eV] & $B^0_4$ [$\mu$eV] & $B^2_4$ [$\mu$eV] & $B^4_4$ [$\mu$eV]   \\
\hline
EPR & -25.5(2) & 1.2(4) & -1.4(3)$\times$10$^{-3}$ & -1.0(4)$\times$10$^{-2}$ &  -0.(4)$\times$10$^{-4}$ \\
INS & -25.3(2) & -2.5(2) & -1.5(3)$\times$10$^{-3}$ & - & - \\
\end{tabular}
\end{ruledtabular}
\end{table}
\endgroup

\begin{figure}
\includegraphics[width=6.5cm]{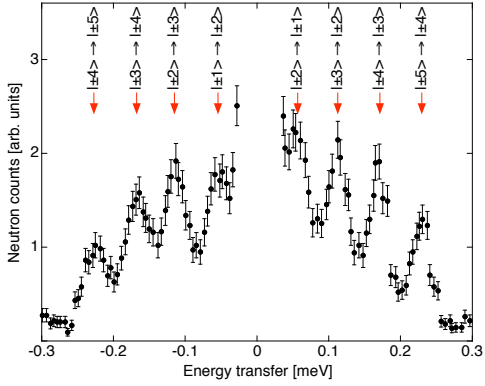}
  \caption{\label{fig:sc-fe4} (Color online)
Energy spectrum of neutrons scattered from Fe$_4$ at $T=6.5$~K . The transitions $ |\pm M\rangle
\rightarrow |\pm M'\rangle$ within the $S=5$ ground-state multiplet are marked for each observed peak.
 Adapted from \onlinecite{Amo01}.
 }
\end{figure}

It can be seen from Table~\ref{tab:sc-fe4} that $D$ is the dominant anisotropy parameter which splits the $S=5$ ground
state into a sequence of five doublets ($M= \pm5, \pm4, \pm3, \pm2, \pm1$) and a singlet ($M=0$) with energies $D M^2$
according to Eq.~(\ref{eq:basics-Dterm}). The other anisotropy parameters produce a slight mixing of the $| M\rangle$
states and/or give rise to small energy shifts. Nevertheless, the relevant selection rule in INS experiments is
retained with good accuracy as $\Delta M= \pm 1$, as demonstrated by the data displayed in Fig.~\ref{fig:sc-fe4}
\cite{Amo01}. The decreasing energy spacing with decreasing energy transfer prevented the transition $ |M=\pm1\rangle
\rightarrow |M= 0\rangle$ to be resolved from the elastic line. The anisotropy parameters resulting from the INS
experiments are listed in Table~\ref{tab:sc-fe4}. The value of $E$ is twice that obtained by the EPR measurements.
Small discrepancies between the parameters determined by INS and EPR are frequently observed and are probably due to
the high magnetic fields used by the latter technique, producing a rather large Zeeman splitting so that a mixing of
the ground and the excited spin multiplets cannot be neglected.

\subsubsection{Dzyaloshinski-Moriya interactions}

\begin{figure}
\includegraphics[width=5cm]{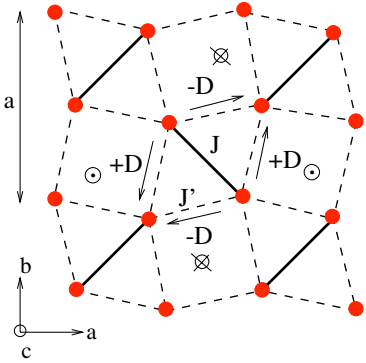}
  \caption{\label{fig:sc-scbostruct} (Color online)
Structural arrangement of the Cu$^{2+}$ dimers (red balls connected with black lines)
in SrCu$_2$(BO$_3$)$_2$ including the Dzyaloshinski-Moriya
interactions whose vectors are perpendicular to the ($a$,$b$)-plane. The arrows show the order
of the spins in the expression ${\bf d}( \hat{\bf s}_i \times \hat{\bf s}_j)$. Adapted from \onlinecite{Cep01}.
 }
\end{figure}

\begin{figure}
\includegraphics[width=8cm]{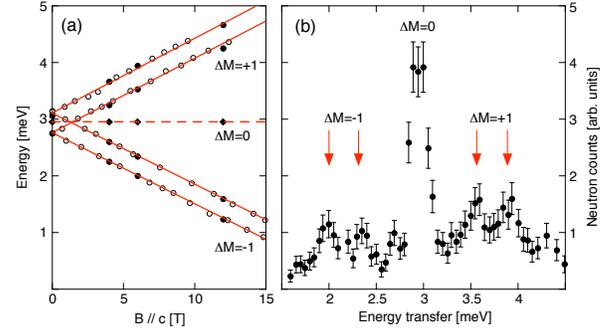}
  \caption{\label{fig:sc-scboins} (Color online)
(a) Magnetic field dependence ($B||c$) of the singlet-triplet excitations observed in SrCu$_2$(BO$_3$)$_2$ by EPR (open
dots) and INS (solid dots) experiments. (b) INS spectrum of SrCu$_2$(BO$_3$)$_2$ at ${\bf Q} =(1,0,0)$ and
$B=6$~T ($B||c$).
 Adapted from \onlinecite{Noj99} and  \onlinecite{Cep01}.
 }
\end{figure}

The compound SrCu$_2$(BO$_3$)$_2$ is a two-dimensional spin-gap system with tetragonal unit cell. It consists of
alternately stacked Sr and CuBO$_3$ planes. The latter are characterized by a regular array of mutually perpendicular
Cu$^{2+}$ dimers ($s=1/2$) as illustrated in Fig.~\ref{fig:sc-scbostruct}. The gap associated with the singlet-triplet
dimer excitations was determined by INS experiments to be 3.0~meV \cite{Kag00}. An almost perfect center of inversion
at the middle of the dimer bonds forbids the Dzyaloshinski-Moriya (DM) interaction [Eq.~(\ref{eq:basics-DM})] between
the two spins of a dimer. However, each dimer is separated from the neighboring dimer by a BO$_3$ unit, for which there
is no center of inversion, so that the DM interaction is allowed between the next-nearest-neighbor (nnn) Cu$^{2+}$
spins as described by the Hamiltonian
\begin{eqnarray}
\label{eq:sc-scboH}
  \hat{H} &=& \sum_{nnn} \pm d {\bf e}_c \cdot \left( \hat{\bf s}_i \times \hat{\bf s}_j \right)
\end{eqnarray}
where the sign depends on the bond (see Fig.~\ref{fig:sc-scbostruct}), and ${\bf e}_c$ is the unit vector in the $c$
direction \cite{Cep01}. The effect of Eq.~(\ref{eq:sc-scboH}) is to split the $\Delta M= \pm 1$ transition associated
with the singlet-triplet splitting into four branches under the application of a magnetic field parallel to the $c$
axis. This prediction was verified by EPR measurements \cite{Noj99} and later confirmed by INS experiments \cite{Cep01}
as shown in Fig.~\ref{fig:sc-scboins}. In addition to the four $\Delta M= \pm 1$ branches, the field-independent
$\Delta M=0$ transition was observed in the INS measurements. The DM parameter resulting from these investigations
turned out to be $d \approx 0.18$~meV, which roughly compares with the estimated value $d = (\Delta g/g) J_{nnn}
\approx 0.5$~meV.

\subsubsection{ A novel tool for local structure determination}

Conventional crystallography is the standard tool for structure determination, and a periodic lattice is a
prerequisite for such studies. However, complex materials are often characterized by local deviations from
perfect periodicity which may be crucial to their properties. The most prominent bulk methods for local
structure determination are x-ray absorption fine structure, nuclear magnetic resonance, and atomic
pair-distribution function analysis. All these methods provide a spatial resolution of typically 0.1~{\AA},
and their performance can hardly be improved. Magnetic cluster excitations are able to push the spatial
resolution beyond the present limits through the dependence of the exchange coupling $J$ on the interatomic
distance $R$, which for most materials is governed by the linear law $dJ/dR = \alpha$ as long as $dR \ll R$.
Modern spectroscopies measure exchange couplings with a precision of $dJ/J \approx 0.01$, thus spatial
resolutions of $dR \approx 0.01$~{\AA} can be achieved as demonstrated below.

\begin{figure}
\includegraphics[width=6.5cm]{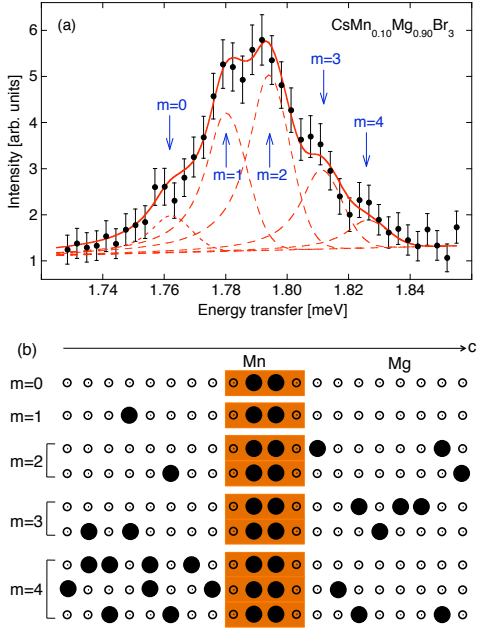}
  \caption{\label{fig:sc-cmmbinsstructure} (Color online)
(a) Energy distribution of the $|S=0,M=0\rangle \rightarrow |S=1,M=1\rangle$ Mn$^{2+}$ dimer transition
observed for CsMn$_{0.1}$Mg$_{0.9}$Br$_3$ at $T=1.6$~K. (b) Sketch of statistical distributions on Mn$^{2+}$
and Mg$^{2+}$ ions along the $c$ axis. $m$ is the number of Mn$^{2+}$ ions replacing Mg$^{2+}$ ions outside
the central Mn$^{2+}$ dimer embedded in the brown area. Adapted from \onlinecite{Fur11b}.
 }
\end{figure}

\begin{table}
\caption{\label{tab:sc-tab5} Analysis of the INS spectrum of CsMn$_{0.1}$Mg$_{0.9}$Br$_3$ given in
Fig.~\ref{fig:sc-cmmbinsstructure}(a). $E_m$ and $I^{obs}_m$ denote the energy transfers and the normalized
intensities of the individual bands $m$, respectively. The intensities $I^{calc}_m$ were calculated according to
a statistical model described in \onlinecite{Fur11b}. Relative error bars are given for the exchange couplings
$J_m$ and for the local Mn-Mn distances $R_m$. }
\begin{ruledtabular}
\begin{tabular}{cccccc}
m & $E_m$ [K] &  $I^{obs}_m$ & $I^{calc}_m$ & $J_m$ [meV] & $R_m$ [{\AA}]  \\
\hline
0 & 1.762(2) & 0.081(22) & 0.119 & -0.8321(10) & 3.2311(3) \\
1 & 1.780(2) & 0.302(25) & 0.289 & -0.8408(10) & 3.2287(3) \\
2 & 1.794(2) & 0.383(22) & 0.310 & -0.8477(10) & 3.2268(3) \\
3 & 1.811(2) & 0.173(22) & 0.193 & -0.8563(10) & 3.2244(3) \\
4 & 1.826(3) & 0.061(18) & 0.089 & -0.8637(14) & 3.2224(4) \\
\end{tabular}
\end{ruledtabular}
\end{table}

We turn to Fig.~\ref{fig:sc-cmmbxins} which displays Mn$^{2+}$ dimer excitations in CsMn$_x$Mg$_{1-x}$Br$_3$.
The total linewidths of the transitions A and B show an $x$-dependent increase beyond the instrumental energy
resolution (FWHM = 55~$\mu$eV) which was further investigated by high-resolution INS experiments with FWHM =
15~$\mu$eV \cite{Fur11b}. As shown in Fig.~\ref{fig:sc-cmmbinsstructure}(a), the energy spectrum observed for
the transition A exhibits marked deviations from a normal Gaussian distribution. It is best described by the
superposition of five individual bands which correspond to specific Mn$^{2+}$ dimer configurations with
particular exchange couplings $J_m$ ($m=0,1,2,3,4$). The linear law $dJ/dR=\alpha$ was established for
CsMn$_x$Mg$_{1-x}$Br$_3$ with $\alpha =3.6$~meV/{\AA} \cite{Str04}, thus each of the five $J_m$ values can be
associated with a particular local distance $R_m$ as listed in Table~\ref{tab:sc-tab5}.

How can the discrete nature of the local Mn-Mn distances be explained? In Fig.~\ref{fig:sc-cmmbinsstructure}(b)
different configurations along the Mn chain structure are sketched, where $m$ is the number of peripheral
Mn$^{2+}$ ions replacing the Mg$^{2+}$ ions. The introduction of additional Mn$^{2+}$ ions exerts some internal
pressure within the chain, since the ionic radii of the Mn$^{2+}$ (high spin) and Mg$^{2+}$ ions are different
with $r_{Mn}=0.83$~{\AA}$> r_{Mg}=0.72$~{\AA} \cite{Sha76}, so that the atomic positions have to rearrange. In
particular, the Mn-Mn bond distance of the central Mn$^{2+}$ dimer is expected to shorten gradually with
increasing number of Mn$^{2+}$ ions as compared to the case $m=0$. For any number $m$ there is a myriad of
structural configurations, resulting in a continuous distribution of local distances $R_m$. This view, however,
is in contrast to the observed energy spectrum displayed in Fig.~\ref{fig:sc-cmmbinsstructure}(a) which is
clearly not continuous. In other words, the bond distance is not smoothly adjusted to its surrounding but locks
in at a few specific values $R_m$. Obviously the realization of discrete local distances $R_m$ is governed by
the number $m$ of peripheral Mn$^{2+}$ ions and not by their specific arrangement in the chain. This surprising
result is due to the one-dimensional character of the compound CsMn$_x$Mg$_{1-x}$Br$_3$ in which the mixed
Mn$_x$Mg$_{1-x}$ chains behave like a system of hard core particles \cite{Kri96}. In conclusion, the use of
high-resolution spectroscopies allows a rather direct determination of local interatomic distances in small
magnetic clusters, in contrast to other techniques which usually have to be combined with simulations.

%% file: chapter4_largeclusters_v2012_10_02.tex

\section{Large magnetic clusters}
\label{sec:lc}

\subsection{Introduction}
\label{sec:lc-intro}

In the previous section, the scientific questions addressed by the small magnetic clusters focused on demonstrating and
elucidating the nature of the various basic interactions between the spin centers in condensed matter systems. However,
in large magnetic clusters, or the molecular nanomagnets in the context of this review, the huge size of the Hilbert
space makes a complete experimental characterization of the magnetic cluster excitations (usually) impossible.
Accordingly, the fine details in the spin interactions such as exchange anisotropy are not detected, and the modeling
of the data can with much success be based on spin models taking into account only the dominant interactions, which in
most cases is Heisenberg exchange. The spectroscopic experiments typically reveal the low-lying excitations or the
low-energy sector of these spin models, and one is hence naturally directed towards questions such as what is the
nature of the ground state and elementary excitations.

The key distinguishing feature as compared to the previous section is the many-body structure of the
wave-functions in the large magnetic clusters, and a main goal could be formulated as what novel quantum
states are realized and which physical concepts allow us to rationalize them. At this point the close ties to
the field of, e.g., quantum spin systems becomes apparent, and indeed methodologies developed there are often
applied to molecular nanomagnets. Most of the examples presented in this section will elaborate on that.

However, also distinguishing novel aspects come into play as a consequence of the fact that the molecular
nanomagnets are not extended. For instance, phase transitions, either classical or quantum, are not possible
in a strict sense. Conceptually most important, however, is that the wave vector ${\bf q}$ ceases to be a
useful quantum number. One can actually expect that exactly those lattice topologies which cannot be expanded
into an extended lattice will exhibit the most interesting novel complex quantum states and magnetic
phenomena. Research into this direction has just started, however, and only preliminary results are available
at the time of the writing of this manuscript. A further important point not emphasized yet is that the spins
of the magnetic centers in molecular nanomagnets are generally rather large, with $s_i = 3/2, 2, 5/2$ being
most often found, in contrast to quantum spin system, where much focus is on spin-1/2 systems. Typical metal
ions would be Cr$^{3+}$ and Mn$^{4+}$ ($s_i = 3/2$), Mn$^{3+}$ ($s_i = 2$), and Mn$^{2+}$ and Fe$^{3+}$($s_i
= 5/2$). The quantum-classical correspondence hence enters naturally in the discussion of the magnetic
excitations in molecular nanomagnets.

In the following those classes of molecules will be discussed for which considerable insight into the magnetic cluster
excitations has been obtained. Important classes such as odd-membered wheels \cite{Yao06-fe9,Hos09-v7,Cad04-cr8ni},
ferromagnetic clusters \cite{Cle99-ni4fmins,Low06-cr10ins,Stu11-mn10}, discs \cite{Hos-review,Koi07-disc}, and others
are not mentioned.

An important subclass of molecular nanomagnets are the single-molecule magnets (SMMs), which have received
the largest attention in the past and could be described as having given birth to the molecular nanomagnets
as a research field. The above questions are also relevant, but the most interesting phenomena in SMMs, such
as quantum tunneling of magnetization, are mainly related to magnetic anisotropy. The SMMs will be discussed
in Sec.~\ref{sec:smm}.

\subsection{Theoretical description}
\label{sec:lc-theory}

\subsubsection{Spin Hamiltonian}
\label{sec:lc-spinH}

As mentioned before, experimental results on large magnetic clusters can often very well be reproduced by
spin Hamiltonians which include only the most dominant terms. For clusters containing only type S and Q metal
ions, on which we focus in this review, these are the HDVV Hamiltonian Eq.~(\ref{eq:basics-HDVV}), the
single-ion anisotropy Eqs.~(\ref{eq:basics-Dterm}) and (\ref{eq:basics-Eterm}), and the Zeeman term
Eq.~(\ref{eq:basics-Zeeman}). However, in molecular nanomagnets the site symmetries of the individual spin
centers in the cluster are very low, if they have symmetry at all. Accordingly, the single-ion anisotropy and
$g$ factors should in general be described as tensors. Also, in molecular nanomagnets often different kinds
of metal centers are incorporated. This gives rise to the spin Hamiltonian
\begin{eqnarray}
\label{eq:lc-microscopic}
\hat{H} &=& -2 \sum_{i<j} J_{ij} \hat{\bf{s}}_i \cdot \hat{\bf{s}}_j
 + \sum_i \hat{\bf{s}}_i \cdot {\bf D}_{i} \cdot \hat{\bf{s}}_i
\nonumber \\ &&
 + \mu_B \sum_i \hat{\bf{s}}_i \cdot {\bf g}_{i} \cdot {\bf B},
\end{eqnarray}
which in the following will be referred to as microscopic Hamiltonian in order to clearly distinguish it from
effective models which also appear. The dipole-dipole interaction Eq.~(\ref{eq:basics-dipoledipole}) has also
to be included, but its effect on energy spectrum and magnetic behavior is very similar to that of the
single-ion anisotropy term and may hence be lumped into the single-ion parameters. Experimental $D$ and $E$
values which were derived with the dipole-dipole interaction explicitly included in
Eq.~(\ref{eq:lc-microscopic}) will be indicated by a superscript "lig".

In contrast to the sites, the molecule itself may exhibit, or closely approximate, a high molecular symmetry. In fact,
clusters with a particular symmetry are appealing for physical studies, and are thus preferred objects of
investigations. The microscopic Hamiltonian simplifies then enormously and includes only few parameters.

Usually the HDVV Hamiltonian dominates over the single-ion anisotropy, and the strong-exchange limit
(Sec.~\ref{sec:basics-spinham}) is an excellent starting point. Although the magnetic anisotropy cannot be ignored in
the analysis of experimental data, the physics of interest in these cases is (usually) related to the Heisenberg
interactions, and the discussion focuses on the corresponding Heisenberg spin models (notable exceptions are the SMMs
discussed in Sec.~\ref{sec:smm}). The magnetic anisotropy may however also be so large that important effects appear
which are not covered by the strong-exchange limit ($S$ mixing)\cite{Liv02-smix,Wal05-smix} or may need completely
different physical concepts for their description. The examples selected below will demonstrate this point. It is added
that from the values of magnetic parameters, such as $J$ and $D$, it is usually not possible to infer \emph{a priori}
whether the strong-exchange limit is obeyed or not. The ratio $D/J$ is generally small in molecular nanomagnets, and
which case is realized needs a case-to-case analysis.

The dimension of the Hilbert spaces encountered in large magnetic clusters poses a major obstacle, similar to that
found in other areas such as quantum spin systems, and the same conceptual ideas are followed to tackle it. Indeed,
essentially any technique developed in the context of quantum spin systems is also of interest for large magnetic
clusters. However, some of them had been of particular use, and are mentioned next.

\subsubsection{Numerical techniques}
\label{sec:lc-numerics}

A most straight forward approach is to numerically solve the spin Hamiltonian for its energies and eigen functions,
which is achieved in two steps, setting up the Hamiltonian matrix and then diagonalizing it.

The major decision in the first step is the choice of the basis set, which could be the product states
$|\{M_i\}\rangle$ (with an obvious shorthand notation) or the spin functions $|\eta S M\rangle$, where $\eta$ denotes
the intermediate spin quantum numbers generated in a particular spin coupling scheme. The product states are most
easily handled in computer code, yield sparse Hamiltonian matrices, and are eigenfunctions of $\hat{S}_z$ which allows
a block factorization for magnetic clusters with uniaxial symmetry. On the other hand, for the spin functions it is
numerically demanding to calculate matrix elements (a number of Wigner symbols need to be calculated) and the matrices
are dense, but they have the intrinsic advantage of being eigenfunctions of the total spin operator $\hat{ \bf S}$
which results in a very effective block-factorization for the HDVV Hamiltonian. Interestingly, nearly all numerical
work in the field of quantum spin systems has been based on product states; spin functions are rarely used. However,
for large magnetic clusters diagonalization using spin functions has been used with much success
\cite{Del93-fe8,Wal99-fe6,Gui04-mn3x3,Bar08}, and efficient ITO based techniques have been developed for calculating
matrix elements \cite{Gat93-ito} and employing spatial symmetries \cite{Wal00-sym,Sch10-sym}.

Complete information on the system is obtained by a full exact diagonalization, and several canned computer codes are
available \cite{Bai00}. The largest dimension of the Hilbert space which can be handled is ca. 100~000 on a super
computer, or about 15~000 on a (32 bit) personal computer. If symmetries are systematically taken into account, quite
large magnetic clusters can be treated on personal computers, e.g., a mixed valent Mn-[3$\times$3] grid molecule with a
Hilbert space dimension of 4~860~000 \cite{Wal06-mixedmn3x3}.

If full exact diagonalization is not possible, one may attempt to obtain the energies and eigen functions in
a subspace. A first approach is to truncate the space of basis function, but the success of the procedure
obviously depends on how well the selected basis states represent the sought-after eigen functions
\cite{Sch09-approx}.

A set of very efficient diagonalization methods is given by the sparse matrix diagonalization techniques, which allow
an exact numerical calculation of a small number ($\sim$100) of selected energies and/or eigen states, typically the
low-lying states \cite{Bai00}. In physics the most prominently used variant is the Lanczos method, while in chemistry
the Jacobi-Davidson algorithm is more often applied. However, for the specific purpose of calculating the
low-temperature observables of large magnetic clusters, the simpler subspace-iteration techniques turn out to be quite
powerful, since they provide both energies and eigen function even in the presence of degeneracies with very comparable
convergence rates.

These techniques employ an iterative process and work best for sparse matrices. Because of the latter it is most
natural to use the product states, though spin functions were applied in few cases \cite{Gui04-mn3x3}. The iterative
process consists of repeatedly applying the Hamiltonian matrix $\bf H$ to a vector $\bf x$,
\begin{eqnarray}
\label{eq:lc-subspace}
{\bf x}_{n+1} = ( {\bf H} - \sigma {\bf 1} ) \cdot {\bf x}_{n},
\end{eqnarray}
where $\sigma$ introduces a shift which allows to optimize the convergence rate, and the starting vector ${\bf x}_0$
may be a random vector. If more than one eigen pair is searched, the iteration is applied to a subspace of vectors $\bf
X$. For a practical algorithm, some further improvements are suggested \cite{Bai00}. The Lanczos and Jacobi-Davidson
algorithms are also based on matrix-vector multiplications ${\bf H} \cdot {\bf x}$, but employ more sophisticated
algorithms to extract the information on the eigen pairs from the generated vectors \cite{Bai00}.

Besides these approaches a number of numerical techniques exist which aim at calculating observables directly without
evaluating eigen pairs explicitly. Among these are e.g. Quantum Monte Carlo, Chebyscheff expansion, dynamic and
finite-temperature Lanczos, transfer matrix and (dynamical) density matrix renormalization group methods. However,
although promising, these methods have not yet been applied systematically to the analysis of large magnetic clusters
as defined in this review, but few applications were reported
\cite{Exl03-fe30dmrg,Eng06,Tor07-qmc,Sch10-finiteTlanczos,Umm12-Fe18}. More efforts in this direction would obviously
be desirable.

\subsubsection{Effective Hamiltonian techniques}
\label{sec:lc-effH}

Another approach to describe the relevant low-energy excitations in a particular spin model is to replace the
microscopic Hamiltonian by an effective spin Hamiltonian (mapping), which acts in a Hilbert space of (significantly)
reduced dimension. It is emphasized that the states in the Hilbert space of the effective Hamiltonian do not have to be
identical to those of the microscopic Hamiltonian. An effective Hamiltonian may be constructed from various procedures,
but the following simple technique had been particularly useful for rationalizing the magnetism in a number of large
magnetic clusters.

The method can be described as a first-order perturbation treatment, as integrating out a degree of freedom or as a
mean-field argument and is guided by physical intuition. It starts with combining a subset of the spins $\hat{\bf s}_i$
in a collective spin $\hat{\bf S}_\mu = \sum_{i \in \mu} \hat{\bf s}_i$, where $\mu$ stands for the set of sites $i$,
and selecting the sector of interest via the value of the quantum number $S_\mu$ associated to $\hat{\bf S}_\mu$, which
is usually the minimal or maximal value. Then it holds
\begin{eqnarray}
\label{eq:lc-effspin}
\hat{\bf s}_i = c_i \hat{\bf S}_\mu,
\end{eqnarray}
with a projection coefficient $c_i$, which depends on $s_i$ and $S_\mu$. For the sector where all spins in the subset
$\mu$ are ferromagnetically aligned and $S_\mu$ assumes its maximal value $S_\mu = \sum_i s_i$ holds $c_i = s_i/S_\mu$.
The effective Hamiltonian is then finally obtained by inserting Eq.~(\ref{eq:lc-effspin}) in the microscopic
Hamiltonian, which removes the spins $\hat{\bf s}_i$ of the subset $\mu$ and replaces them by the collective spin
$\hat{\bf S}_\mu$.

A typical situation where the above scheme could be applied is a cluster with one or few ferromagnetic
exchange couplings which are much larger than the other ones. Then the spins linked by a strong ferromagnetic
coupling can be combined into one collective spin which replaces them in the spin Hamiltonian. Another
situation is an antiferromagnetic bipartite lattice, where the spins on each sublattices $A$ or $B$ are
ferromagnetically aligned with respect to each other in the ground state. This suggests to introduce two
sublattice spins $\hat{\bf S}_A$ and $\hat{\bf S}_B$, and to transform the Hamiltonian according to
Eq.~(\ref{eq:lc-effspin}). The subsections below will provide examples for the practical application of the
scheme.

\subsubsection{Condensed matter techniques}

Sophisticated techniques for calculating properties of extended (quantum) spin systems have been developed.
Some of them can directly be applied to large magnetic clusters as they work for any number $N$ of spin
centers, i.e., the thermodynamic limit is not assumed \emph{a priori} but taken after completion of the
calculation. However, in magnetic clusters translational invariance is not present, and it is hence a
characteristic feature of the application of these techniques to clusters that they have to be adapted to
work in real space and not momentum space as in extended systems, which may bring in novel aspects. This area
is still largely unexplored, but two of the simpler techniques in this class of methods have been applied
with some success. They both elaborate on the observation that in molecular nanomagnets the spins $s_i$ are
relatively large.

Spin-wave theory (SWT), or the set of techniques embraced by this acronym such as linear SWT, interacting SWT, or
modified and finite-size SWT \cite{Rad63,Iva04-swtreview,Tak87-mswt,Hir89-mswt,Zho93-fsswt}, is certainly a most
successful theory in magnetism. All variants have in common that they start from a classical spin configuration in the
ground state, and expand around it by approximating the single-site spin operators $\hat{\bf s}_i$. They are hence
semi-classical in nature, and - technically - can be applied to any type of lattice, and in particular to large
magnetic clusters when formulated in real space \cite{Cep05-swtfe30}. It is necessary to distinguish between SWTs for
ferromagnetic and antiferromagnetic clusters. For ferromagnetic clusters, SWT allows us to calculate the
zero-temperature excitation spectrum exactly for any arbitrary cluster. However, since SWT breaks spin rotational
invariance it is conceptually inappropriate for antiferromagnetic systems with disordered ground states, such as
one-dimensional (1D) systems or magnetic clusters. Nevertheless, it is often found to produce significant results e.g.
for excitation energies \cite{Iva04-swtreview}, but the reliability of the results should carefully be checked
case-by-case. The fundamental questions as regards the applicability of SWT to antiferromagnetic clusters shall not be
addressed in this review, but results of some few case-by-case checks will be reported. For the technical
implementation of SWT for magnetic clusters we refer to \cite{Stu11-mn10,Wal07-swtfe30}.

The large spins in molecular nanomagnets suggest an entirely classical description, where each spin operator
$\hat{\bf s}_i$ is replaced by a classical vector ${\bf s}_i$ of appropriate length, or a unit vector ${\bf
e}_i$ times a prefactor $\tilde{s}_i$, ${\bf s}_i = \tilde{s}_i {\bf e}_i$. A subtlety arises here as regards
the appropriate value of $\tilde{s}_i$, which can be argued to be best taken as $\tilde{s}_i = \sqrt{
s_i(s_i+1) }$ \cite{Joy67-class,Lus97-class} or $\tilde{s}_i = s_i$ \cite{Fis64-class}. With this replacement
the Hamiltonian becomes a classical functional, and the ground-state energy and spin configuration is
obtained by minimization. Also, thermodynamic quantities can be calculated for quite large magnetic clusters
of hundreds of sites, and it is here where the classical approach was mostly applied to molecular nanomagnets
\cite{Sch05-frust,Mue01-classquantum,Yao06-fe9}. The calculation of dynamic quantities is also possible
\cite{Lus97-class,Lus98-class}, but the discrete energies of the excitations in magnetic clusters can of
course not be recovered, though their intensities \cite{Men00-classquantum}. Hence, the classical
calculations are not actually suited to reproduce experimental spectroscopic data, but nevertheless can
provide significant insight into the physics of the excitations in a large magnetic cluster through the
quantum-classical correspondence \cite{Sch05-meta}, which is probably the most important aspect.

\subsection{Even-membered antiferromagnetic molecular wheels}
\label{sec:lc-wheels}

\begin{figure*}
\includegraphics[width=15cm]{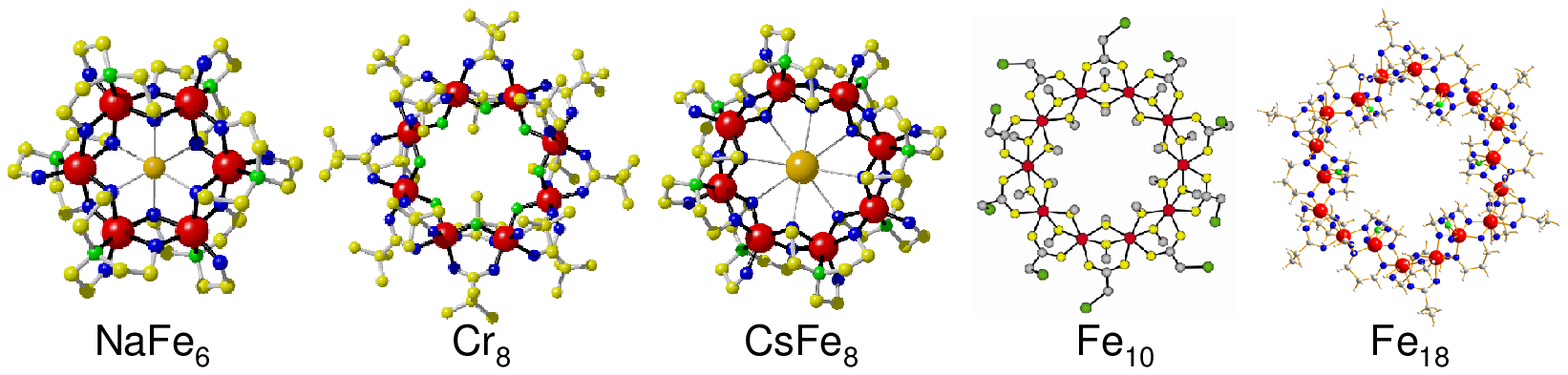}
 \caption{\label{fig:wheels-wheels} (Color online) Molecular structures of the five even-membered antiferromagnetic wheels
 NaFe$_6$: Na[Fe$_6$tea$_6$]Cl,
 Cr$_8$: [Cr$_8$F$_8$Piv$_{16}$],
 CsFe$_8$: Cs[Fe$_8$tea$_8$]Cl,
 Fe$_{10}$: [Fe$_{10}$(OCH$_3$)$_{20}$(O$_2$CCH$_2$Cl)$_{10}$], and
 Fe$_{18}$: [Fe$_{18}$(pd)$_{12}$(pdH)$_{12}$(O$_2$CEt)$_6$(NO$_3$)$_6$].}
\end{figure*}

Molecular wheels are species in which the metal ions form almost perfect ring-like structures. The decanuclear wheel
[Fe$_{10}$(OCH$_3$)$_{20}$(O$_2$CCH$_2$Cl)$_{10}$], or Fe$_{10}$ in short, was the first wheel whose magnetic
properties were studied \cite{Taf94-fe10,Gat94-science}. Since then, because of their aesthetical appeal and
unprecedented magnetism, the class of the molecular wheels has enormously grown and dozens of wheels with nuclearity
ranging from $N=6$ to $N=18$ have been synthesized. In Fig.~\ref{fig:wheels-wheels} the crystal structures of five
representatives are displayed. The physics in the molecular wheels, and their relatives such as the modified wheels
(Sec.~\ref{sec:lc-modwheels}), turned out to be surprisingly rich. The presentation here is hence necessarily limited.

The even-membered antiferromagnetic wheels have been the subject of intense research. In this section,
homo-nuclear wheels will be discussed with $s_i = s$ for all $i$. In view of the high molecular symmetry,
which in hexa-nuclear wheels can be a crystallographic S$_6$ symmetry and in octa-nuclear wheels a C$_4$
symmetry, the magnetism should be expected to be very well described by the generic spin Hamiltonian
\begin{eqnarray}
\label{eq:wheels-H} \hat{H} &=& -2 J \left( \sum_{i=1}^{N-1} \hat{\bf{s}}_i \cdot \hat{\bf{s}}_{i+1} +
\hat{\bf{s}}_N \cdot \hat{\bf{s}}_{1} \right) \nonumber \\ && + D \sum_i \left[ \hat{s}_{iz}^2 -
\frac{1}{3}s_i(s_i+1) \right] + \mu_B g \hat{\bf{S}} \cdot {\bf B}.
\end{eqnarray}
Additional terms describing e.g. a variation of the exchange constants along the wheel, tilted single-site
anisotropy tensors or higher-order terms should in principle also be present, but these are difficult to
resolve in experiment. Sometimes a planar term $E \sum_i( \hat{s}_{ix}^2 - \hat{s}_{iy}^2)$ had to be added
to Eq.~(\ref{eq:wheels-H}) (\emph{vide infra}), and evidence for weak Dzyaloshinski-Moriya interactions were
reported \cite{Cin02-fe6-dm,Lan09-spinpeierls}. However, the latter manifests itself only in very specific
experiments, and is not further considered. The Hamiltonian Eq.~(\ref{eq:wheels-H}) (plus sometimes an $E$
term) generally provides a very good description.

The exchange coupling $2J$ is typically on the order of few meV, and the single-ion anisotropy weak, $|D| <
0.1$~meV. In most cases it is of the easy-axis type ($D<0$), with the exception of Fe$_{18}$, where $D>0$
(\emph{vide infra}). The ratio $D/J$ is small in the antiferromagnetic wheels, but the strong-exchange limit
is not necessarily well obeyed. In fact, the effect of the anisotropy is rather determined by $(N s)^2 |D/J|$
\cite{Chi98-wheel-mqt,Wal02-wheel-qt}, and antiferromagnetic wheels should be classified according to wether
the anisotropy is "weak" or "strong" by the different physics in the two limits.

\begin{figure}[b]
\includegraphics[width=6.5cm]{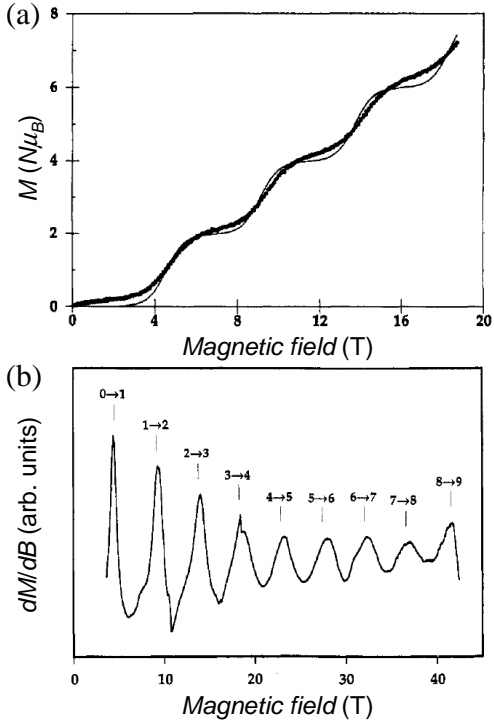}
 \caption{\label{fig:wheels-fe10mag} (a) Magnetization vs field and (b) differential
 magnetization $dM/dB$ at a temperature of 0.6~K of the antiferromagnetic wheel Fe$_{10}$. The data
 show the sequence of magnetization steps, which were be traced here for fields up to 40~T.
  Adapted from \onlinecite{Taf94-fe10}.}
\end{figure}

\begin{figure}
\includegraphics[width=6.5cm]{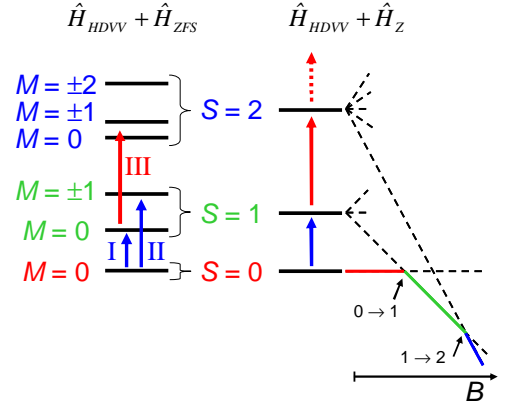}
\caption{\label{fig:wheels-spectrum} (Color online) Sketch of the low-lying energies in even
antiferromagnetic molecular wheels. The Heisenberg interactions give rise to spin multiplets $S = 0, 1, 2,
\ldots$ which split in a magnetic field and lead to level crossings in the ground state, which changes from
$|S=0,M=0\rangle$ to $|1,-1\rangle$, $|1,-1\rangle \rightarrow |2,-2\rangle$, etc., and steps in the
magnetization curve. The magnetic anisotropy splits the spin multiplets in zero field, as indicated to the
left. Some allowed INS transitions are indicated by arrows.}
\end{figure}

The magnetic susceptibility data have conclusively pointed towards a $S = 0$ ground state in the
antiferromagnetic wheels, which is intuitively anticipated from the expected alternating up and down spin
configuration in the classical ground state. The first general insight into the excitation spectrum has been
obtained from low-temperature magnetization curves on the Fe$_{10}$ wheel, which are reproduced in
Fig.~\ref{fig:wheels-fe10mag}. A sequence of magnetization steps is seen as function of field, where at each
step the magnetization changes by 2~$\mu_B$, and which occur in regular field intervals of ca. 4.2~T. This
demonstrates that as function of field the ground state changes from $S = 0$ to $S = 1$ at the first step, $S
= 1$ to $S = 2$ at the second step, and so on, as sketched in Fig.~\ref{fig:wheels-spectrum}. From the fields
of the magnetization steps or ground-state level crossings one can infer the zero-field energies of the
states \cite{Sha02-msteps-review}, with the result
\begin{eqnarray}
\label{eq:wheels-lc}
E(S) = \frac{\Delta}{2} S(S+1).
\end{eqnarray}
The excitation spectrum consists hence of a set of spin multiplets with $S = 0, 1, 2, ...$ whose energies
satisfy the Land\'e rule Eq.~(\ref{eq:sc-Lande}), see also Fig.~\ref{fig:wheels-spectrum}, where $\Delta$ is
the energy gap between the singlet and triplet. Such a spectrum is also generated by a Heisenberg dimer
$\hat{H}_{AB} = \Delta \hat{\bf S}_A \cdot \hat{\bf S}_B$, with appropriate spins $S_A$ and $S_B$. The
Hamiltonian $\hat{H}_{AB}$ appears hence as an effective spin Hamiltonian for the antiferromagnetic
Heisenberg model on a ring. There is however some deeper physics in here, which will be mentioned in the next
subsection.

\begin{figure}
\includegraphics[width=5.5cm]{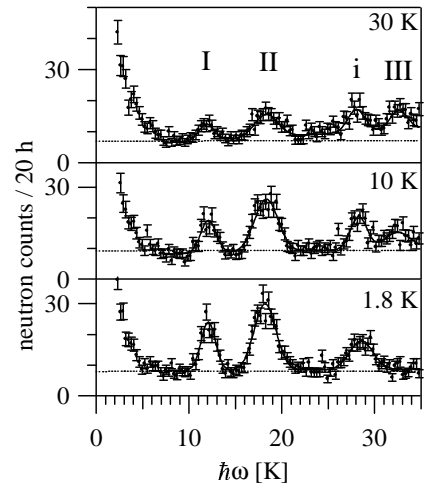}
 \caption{\label{fig:wheels-nafe6ins}
Neutron energy-loss spectrum of NaFe$_6$ at the indicated temperatures ($Q = 1.0$~\AA$^{-1}$). The transition
i could be assigned to an impurity species in the sample. Adapted from \onlinecite{Wal99-fe6}.}
\end{figure}

The role of magnetic anisotropy in antiferromagnetic wheels was elucidated by EPR
\cite{Pil97-esr,Pil01-esr,Pil03-esr,Sla02-cr8} and measurements of the magnetic torque
\cite{Cor99-torque-ac,Cor99-torque-prb,Wal99-fe6}. The first INS experiment on a wheel was performed on the
cluster Na[Fe$_6$\{N(CH$_2$CH$_2$O)$_3$\}$_6$]Cl, or NaFe$_6$ (Fig.~\ref{fig:wheels-wheels})
\cite{Wal99-fe6}. The experimental data are reproduced in Fig.~\ref{fig:wheels-nafe6ins}, and are
characterized by two cold transitions I and II, and a hot transition III. The interpretation of the observed
transitions is straightforward: Peaks I and II correspond to transitions from the $S = 0$ ground state to the
zero-field-split $S=1$ multiplet, and peak III to a transition from the $|S=1,M = 0\rangle$ level to the
$S=2$ multiplet, as indicated in Fig.~\ref{fig:wheels-spectrum}. The INS energies yield directly $2J =
-2.0$~meV and $D = -0.038$~meV, in excellent agreement with magnetic susceptibility and magnetic torque
measurements, which puts NaFe$_6$ in the "weak" anisotropy category.

The following three subsections are organized by the "strength" of the magnetic anisotropy or parameter $(N
s)^2 D/J$. First, the physics in the "weak" anisotropy case, which is that of the $L \& E$-band picture, is
considered followed by the "intermediate" anisotropy case. Finally, wheels with "strong" anisotropy which may
display quantum tunneling of the N\'eel vector are discussed.

\subsubsection{Antiferromagnetic wheels with "weak" anisotropy and the $L \& E$-band picture}
\label{sec:lc-cr8}

Most antiferromagnetic wheels fall into the weak anisotropy category. The anisotropy leads to a zero-field
splitting of the spin multiplets, which is detected in e.g. INS experiments as was shown by the example of
NaFe$_6$, but the physics of the magnetic excitations is not affected by it. Hence, the appropriate model for
the discussion of the physics is that of the antiferromagnetic Heisenberg ring (AFMHR) or
Eq.~(\ref{eq:wheels-H}) with $D=0$.

Initially, the wheels were regarded as models for 1D antiferromagnetic chains with the implication that the
physical concepts developed for them, which are characterized by the strong quantum fluctuations in 1D,
should also apply to AFMHRs. However, the experimental and theoretical work has suggested a very different
picture of the excitations, which is denoted here as $L \& E$-band concept. This concept has its roots in
Anderson's 1952 paper on antiferromagnetic spin waves \cite{And52-swt}, and emerged in the course of several
works \cite{And64,Ber92-qdjs,Can96,Chi98-wheel-mqt,Sch00-rotbands,Wal02-spindyn,Lhu02-review}. The $L \&
E$-band concept is classical in nature, though some subtleties are present, and the meaning of "classical" in
this context is in fact not yet completely understood. It rather seems that in magnetic clusters of the size
discussed here the line between classical and quantum physics is blurred \cite{Kon11-review}. Fundamental
questions hence remain. However, from a practical point of view the $L \& E$-band concept seems to apply
whenever the classical spin structure appears to describe well the ground state, and has allowed
rationalizing the magnetism in a number of different classes of molecules. The molecular wheel
[Cr$_8$F$_8$Piv$_{16}$], or Cr$_8$ (Fig.~\ref{fig:wheels-wheels}), has played an important role in this
context, as it was the first system for which a detailed, and indeed complete, experimental demonstration of
the concept was achieved by exploiting the full power of INS.

\begin{figure}
\includegraphics[width=6.5cm]{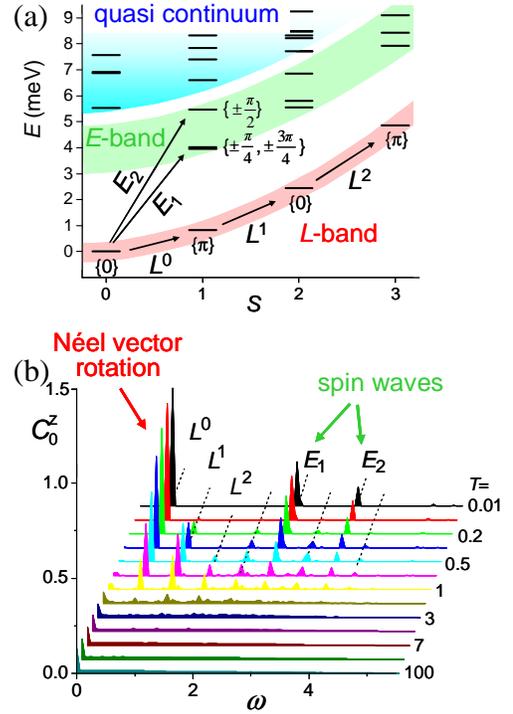}
 \caption{\label{fig:wheels-cr8theory}  (Color online)
(a) Simulated energy spectrum for a $N=8$, $s = 3/2$ AFMHR with $2J = -1.46$~meV as for Cr$_8$. The low-lying
levels form rotational bands. The $L$ band can be rationalized as (quantized) rotation of the N\'eel vector.
The next higher-lying bands are denoted collectively as $E$ band and may be regarded as (discrete)
antiferromagnetic spin waves. Some INS transitions are indicated by arrows with associated labels. The
numbers in brackets indicate the shift quantum number $q$ for some spin multiplets. (b) Simulated spin-spin
auto-correlation function $C^z_0(\omega)$ in units of $2|J|$ for various temperatures. The $L$- and $E$-band transitions
indicated in panel (a) are identified. Adapted from \onlinecite{Wal03-cr8} and \onlinecite{Wal02-spindyn}.}
\end{figure}

Descriptions of the $L \& E$-band concept are available \cite{Wal05-gridreview}; only major aspects are mentioned here.
The key ingredient is the notion of rotational bands, where a rotational band is a sequence of spin multiplets with $S
= S_{min}, S_{min} +1, S_{min}+2, ...$ whose energies increase according to the Land\'e rule $E(S) \propto S(S+1)$. The
analogy of this spectrum with that of a rigid rotator $\hat{H} = \hat{\bf L}^2/(2I)$, where $\hat{\bf L}$ is an angular
momentum and $I$ the moment of inertia, is not coincidental. The $L \& E$-band concept can be summarized as:

(1) In an energy-vs-$S$ plot, the low-energy sector is characterized as a set of rotational modes; the lowest-lying
mode called $L$ band and a number $n_E$ of higher-lying modes called $E$ band. The next higher-lying states are
collectively denoted as quasi continuum.

(2) The classification of the energy states in (1) is justified by a selection rule: Spin transitions from the $L$ band
into the quasi continuum are forbidden. Hence, at low temperatures only INS transitions between states of the $L$ band
or from the $L$ to the $E$ band are allowed.

Importantly, these points have to come together \cite{Wal02-spindyn}. The presence of a rotational mode is by itself
not sufficient to ensure the validity of the $L \& E$-band concept and its consequences. In
Fig.~\ref{fig:wheels-cr8theory} are shown the energy spectrum and spin-spin auto-correlation function $C^z_0(\omega)$
of a $N=8$, $s=3/2$ AFMHR. The structure of the energies detailed in (1) is clearly seen. Also, the correlation
function demonstrates, e.g., that transitions from the $S=0$ ground state into the quasi continuum do not occur. The
above two points completely define the concept, but some further comments are at place:

\textbf{$L$ band:} The $L$ band is intimately related to the occurrence of long-range N\'eel order: For lattices which
can be expanded to infinity, the presence of the $L$ band in the spectrum is a necessary but not sufficient requirement
for an ordered N\'eel ground state in the infinite lattice \cite{Ber92-qdjs}. It is hence also related to the
corresponding antiferromagnetic Goldstone mode, and the $L$-band states have spatial symmetries consistent with the
ordered N\'eel state. An excellent discussion is found in \onlinecite{Lhu02-review}. Qualitatively, the $L$ band
reflects the rotational degree of freedom of the classical antiferromagnetic spin configuration in the ground state, or
(quantized) rotation of the N\'eel vector. The $L$ band is also known as the tower of states or quasi-degenerate joint
states.

\textbf{$E$ band:} The $E$ band is deeply related to spin waves: For lattices which can be expanded to infinity, the
$E$ band evolves into the familiar antiferromagnetic spin waves in the infinite lattice (where we deviate from the
usual notation by not regarding the Goldstone mode or antiferromagnetic Bragg peak as a spin wave, see Ref.~25 in
\onlinecite{Wal07-swtfe30}). In finite clusters, the spin-wave spectrum becomes of course discretized. It then consists
of $n_E$ energies, and each of them corresponds to one of the $n_E$ rotational modes in the $E$ band. Although
originating from extended lattices, the concept of spin waves can be carried over in an obvious manner to any lattice
which obeys the $L \& E$ band concept, not only those characterized by a wave vector \cite{Cep05-swtfe30}.
Qualitatively, the $E$ band reflects the possible higher-energetic internal spin structures which results as
excitations from the classical antiferromagnetic ground-state spin configuration.

In the AFMHR the cyclic symmetry gives rise to a shift quantum number $q$ defined via the shift operator $\hat{T}
|q\rangle = e^{i q} |q\rangle$ with $q = 0, \pm 2\pi/N, ..., \pi$ by which the excitations can be classified. A
complete theory is not existing but phenomenologically the energies of the $L$ and $E$ band for $S \geq 1$ can then be
approximated by
\begin{eqnarray}
\label{eq:wheels-dispersion}
 E(S,q) = \frac{1}{2} \epsilon( \pi ) S(S+1) + \epsilon( q ) - \epsilon( 0 )
\end{eqnarray}
with $q=0, \pi$ for the $L$ band. $\epsilon( q )$ can be regarded as the finite-size version of the spin-wave
dispersion relation \cite{Dre10-csfe8ins3}. Indeed, in the infinite chain $q$ would become the wave vector, and
$\epsilon( q )$ would agree with the familiar spin-wave dispersion \cite{And52-swt}. However, it is emphasized again
that the $L \& E$-band concept is not limited to clusters with cyclic or a similar high symmetry, it can also be
observed in clusters with different point group symmetries or no symmetry at all (\emph{vide infra}).

\begin{figure}
\includegraphics[width=6.5cm]{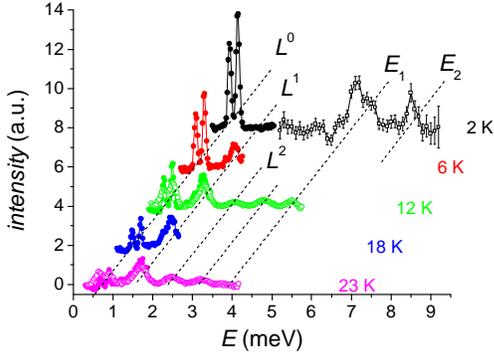}
 \caption{\label{fig:wheels-cr8ins} (Color online)
Neutron energy-loss spectrum of Cr$_8$ at the indicated temperatures. Adapted from \onlinecite{Wal03-cr8}.}
\end{figure}

We turn now to discussing the molecular wheel Cr$_8$. In this molecule, eight Cr$^{3+}$ ions ($s = 3/2$) form
an almost perfect planar octagon. The system crystallizes in space group P42$_1$2, and the molecule nominally
exhibits C$_4$ symmetry. However, disorder in some pivalate ligands and tBu groups is present, suggesting
that the individual molecules are slightly distorted. INS data yielded the exchange parameter $2J =
-1.46(4)$~meV and anisotropy $D = -0.038(5)$~meV \cite{Car03-cr8}, consistent with thermodynamic and EPR
results \cite{Sla02-cr8}. A variation of the exchange and anisotropy parameters along the wheel as allowed by
a C$_4$ symmetry was not detected. Evidence for a weak rhombic term with $E/D \approx 0.11$ was found.

A careful analysis of the INS data provided a detailed picture of the excitations in Cr$_8$ \cite{Wal03-cr8}. The
experimental INS spectra, corrected for non-magnetic scattering, are compiled in Fig.~\ref{fig:wheels-cr8ins}.
Comparison with the theoretical result shown in Fig.~\ref{fig:wheels-cr8theory} will be made. The $L^0$ transition, or
transition from the $S=0$ ground state to the lowest $S=1$ multiplet, is split into two close peaks because of the
zero-field splitting from the anisotropy, exactly as discussed before for NaFe$_6$. The splitting is however small
demonstrating the "weak" anisotropy case in Cr$_8$ and justifying an analysis in terms of the AFMHR model. This is
further corroborated by comparing the experimental INS spectra to the theoretical correlation function, which agree in
any detail, demonstrating the validity of the above point (1).

The INS data allowed also a detailed comparison of the excitation intensities or oscillator strengths $|\langle \lambda
|| \hat{T}^{(1)}(s_i)|| \lambda' \rangle|^2$, and very good agreement between experiment and theory was found.
Furthermore, magnetic scattering intensity at energies higher than that corresponding to transition $E_2$ in
Fig.~\ref{fig:wheels-cr8theory} was not observed, demonstrating the selection rule in point (2).

\begin{figure}
\includegraphics[width=8cm]{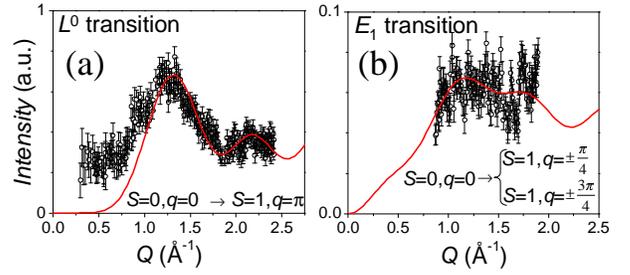}
 \caption{\label{fig:wheels-cr8insQ} (Color online)
$Q$ dependence of the integrated neutron scattering intensity of Cr$_8$ for the peak (a) $L^0$ and (b) $E_1$. Both
peaks arise from transitions from the $S=0$ ground state into the $S=1$ sector,  but the involved final spin multiplets
differ by their shift quantum number $q$, as indicated in the panels. The solid lines represent the theoretical curves.
 Adapted from \onlinecite{Wal03-cr8}.}
\end{figure}

Finally, also the $Q$ dependence of the peak intensities were analyzed. The experimental results for the $L^0$ and
$E_1$ transitions and the theoretical expectations are presented in Fig.~\ref{fig:wheels-cr8insQ}. The good agreement
is obvious. However, most important, the $Q$ dependence provides a fingerprint of the underlying many-body structures
of the quantum states involved in a transition. Hence, the observed different $Q$ dependencies of the $L^0$ and $E_1$
transitions directly demonstrate the different physical nature of the excitations in the $L$ band (rotations of the
N\'eel vector) and the $E$ band (spin waves).

\begin{figure}
\includegraphics[width=6.5cm]{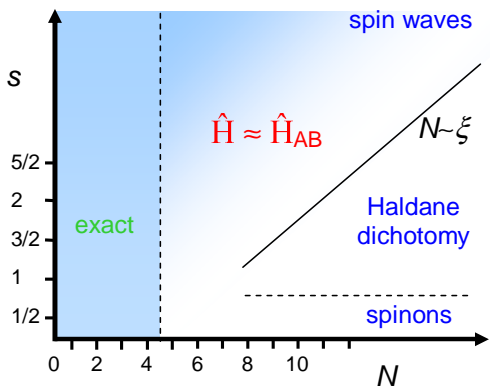}
 \caption{\label{fig:wheels-cr8theory2} (Color online)
Physical properties of the antiferromagnetic Heisenberg ring as a function of its size $N$ and the spin
length $s$. In the blue shaded area quantum fluctuation are weak and the spin structure is essentially
"classical". Here the $L \& E$-band concept becomes valid and $\hat{\bf H}_{AB}$ a good effective spin
Hamiltonian. This "classical" regime is reached when the size $N$ is significantly smaller than a
characteristic correlation length $\xi$, which roughly increases exponentially with $s$. Adapted from
\onlinecite{Wal02-spindyn}.}
\end{figure}

Having demonstrated the $L \& E$-band picture for Cr$_8$, the question arises of how general it is. A conclusive answer
is not yet available, but the results so far suggest Fig.~\ref{fig:wheels-cr8theory2}: The $L \& E$ band concept works
the better the larger $s$ and the smaller $N$ \cite{Eng06,Wal02-spindyn}. The crossover is qualitatively determined by
comparing the size $N$ of the cluster to a length $\xi$ characterizing the decay of the antiferromagnetic correlations
with distance, as indicated in Fig.~\ref{fig:wheels-cr8theory2}. A precise definition of $\xi$ can be subtle
\cite{Hal83a,Aff87}, but in any (disordered) system the correlations decay on a characteristic length scale, which is
what is meant by $\xi$. The spin length enters then through the dependence of $\xi$ on $s$, which, motivated by
Haldane's and other results \cite{Hal83a,Hal83b,Aff87}, is roughly exponential. Since for $s=1/2$, $\xi$ is about 6
sites, it is concluded that large $s=1/2$ clusters are distinguished from those with $s > 1/2$ in that the $L \& E$
band picture is never obeyed in them.

The $L \& E$-band concept has some useful consequences, valid not only for the AFMHR. First, the $L$ band can
well be reproduced by an effective Hamiltonian which may be constructed along the lines outlined in
Sec.~\ref{sec:lc-effH}. This is carried out now for the example of the even antiferromagnetic wheels. Their
lattice is bipartite and introducing two sublattices $A$ and $B$ is natural, where the spins on each
sublattice are ferromagnetically aligned. Then an effective Hamiltonian is obtained by combining the spins on
each sublattice to $\hat{\bf S}_A = \sum_{i\in A} \hat{\bf s}_i$ and $\hat{\bf S}_B = \sum_{i\in B} \hat{\bf
s}_i$, and inserting $\hat{ \bf s}_i = (2/N) \hat{\bf S}_A$ or $\hat{ \bf s}_i = (2/N) \hat{\bf S}_B$,
depending on the spin's sublattice, into the microscopic Hamiltonian, which results in
\begin{eqnarray}
\label{eq:wheels-HAB}
\hat{H}_{AB} = -2j \hat{\bf S}_A \cdot \hat{\bf S}_B,
\end{eqnarray}
with $j = a_1 J$. For $a_1$ one finds $a_1 = a_1^{AB}$, where $a_1^{AB} \equiv 4/N$. The two-sublattice Hamiltonian
$\hat{H}_{AB}$ produces obviously an energy spectrum according to Eq.~\ref{eq:wheels-lc}, immediately explaining the
step-like magnetization curves (Fig.~\ref{fig:wheels-fe10mag}) and $L$ band in the energy spectrum. The approximation
can be improved to yield nearly exact results if $a_1$ is slightly corrected to account for the weak quantum
fluctuations \cite{Wal02-wheel-qt}. For Cr$_8$ one obtains $a_1 = a_1^{qm}$ with $a_1^{qm} = 0.5586$.

Secondly, the nature of the $E$ band suggests applying SWT to the calculation of its energies. This route was explored
in the last years, but the status is not yet clear and more work is needed. However, it seems that despite the
conceptual problems of antiferromagnetic SWT the standard interacting SWTs do, at least for bipartite lattices, produce
reasonable results for the energies of the $E$-band states in the $S=1$ sector. For instance, for the AFMHR the
"dispersion relation" $\epsilon(q)$ in Eq.~(\ref{eq:wheels-dispersion}) are obtained (\emph{vide infra}).

\subsubsection{Antiferromagnetic wheels with "intermediate" anisotropy}
\label{sec:lc-csfe8}

In antiferromagnetic wheels with substantial magnetic anisotropy the ground state and lowest excitation may
better be described by quantum tunneling of the N\'eel vector (Sec.~\ref{sec:lc-nvt}), but for the
next-higher lying levels the $L \& E$-band picture is still appropriate, though $S$ mixing occurs as a novel
feature. In the following the effects of a significant anisotropy on the excitations will be discussed by the
example of CsFe$_8$, as it is one of the best characterized large magnetic clusters and the generic
Hamiltonian Eq.~(\ref{eq:wheels-H}) has been confirmed with high precision.

The chemical formula of CsFe$_8$ is Cs[Fe$_8$tea$_8$]Cl. Eight Fe$^{3+}$ ions ($s = 5/2$) form an almost
perfect octagon with a Cs$^+$ ion at the center (Fig.~\ref{fig:wheels-wheels}). Depending on the synthesis,
the system co-crystallizes with different solvent molecules in space groups P4/n, Pna21, and P21/n
\cite{Saa97-csfe8synth}, and the molecule exhibits ideal or approximate C$_4$ symmetry, but a dependence of
the magnetic excitations on the solvent was experimentally not observed. CsFe$_8$ is a member of a family of
wheels which are distinguished by the templating central alkaline ion, and a magneto-structural correlation
was established \cite{Wal01-csfe8torque,Pil03-esr}. The magnetic excitations in CsFe$_8$ were studied by
low-temperature high-field torque magnetometry, single-crystal high-frequency EPR at Q-band and 190~GHz,
single-crystal nuclear magnetic resonance (NMR) \cite{Sch07-csfe8nmr}, and several INS experiments covering
energies up to 25~meV
\cite{Wal01-csfe8torque,Dre10-csfe8epr,Sch07-csfe8nmr,Wal05-csfe8ins1,Wal06-csfe8ins2,Dre10-csfe8ins3}.

The data analysis could be accomplished by solving the microscopic Hamiltonian Eq.~(\ref{eq:wheels-H}) using the
numerical sparse matrix techniques described in Sec.~\ref{sec:lc-numerics}. However, the Hilbert space dimension is 1
679 616 and the task can become time consuming. Hence, approximate schemes are desired, with the advantage that their
accuracy can always be checked by comparing with the exact results from Eq.~(\ref{eq:wheels-H}) for some few cases. The
lowest-level description built on the strong-exchange limit works well for, e.g., EPR experiments, but involves many
parameters and furthermore misses important effects (for a detailed description of the strong-exchange limit approach
we refer to the book by \onlinecite{Ben90-eprbook}, or to \onlinecite{Dre10-csfe8epr}). However, if only the energy
levels in the $L$-band sector - to use the language of the previous section - are desired, a high-accuracy higher-level
description is provided by the sublattice Hamiltonian approach. Applying the ideas used for the HDVV term
(Sec.~\ref{sec:lc-cr8}) to the anisotropy (and Zeeman) term in the microscopic Hamiltonian yields the sublattice
Hamiltonian
\begin{eqnarray}
\label{eq:wheels-HABfull}
 \hat{H}_{AB} = -2j \hat{\bf S}_A \cdot \hat{\bf S}_B
   + d \left( \hat{S}^2_{A,z} + \hat{S}^2_{B,z} \right)
   + g \mu_B \hat{\bf S} \cdot {\bf B},
\end{eqnarray}
with $j = a_1 J$ and $d = b_1 D$, where $a_1 = a_1^{AB}$ and $b_1 = b_1^{AB}$ [$ b_1^{AB} \equiv (2s-1)/(N s - 1)$].
The approximation can again be improved by adjusting $a_1$ and $b_1$ to account for the weak quantum fluctuations,
which results in essentially exact energies and yields transition intensities to within 10\% accuracy
\cite{Wal02-wheel-qt,Wal06-csfe8ins2}. For CsFe$_8$, $a_1^{qm} = 0.5536$ and $b_1^{qm} = 0.1870$. The energy spectrum
of CsFe$_8$ is nearly identical to that of Cr$_8$ shown in Fig.~\ref{fig:wheels-cr8theory}, and the labeling of states
and transitions are carried over.

\begin{figure}
\includegraphics[width=7.5cm]{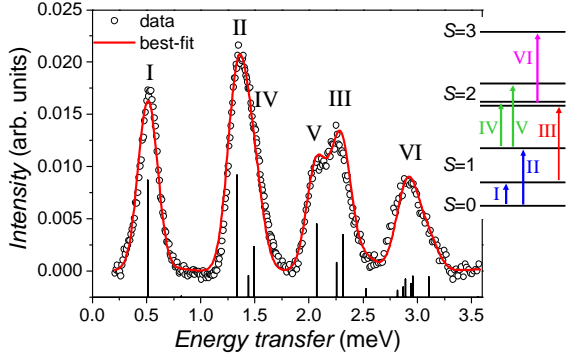}
 \caption{\label{fig:wheels-csfe8ins2} (Color online)
Neutron-energy loss spectrum in CsFe$_8$, after background correction, at $T = 17$~K, and best-fit curve
calculated from Eq.~(\ref{eq:wheels-HABfull}). Adapted from \onlinecite{Wal06-csfe8ins2}. }
\end{figure}

\begin{figure}
\includegraphics[width=6.5cm]{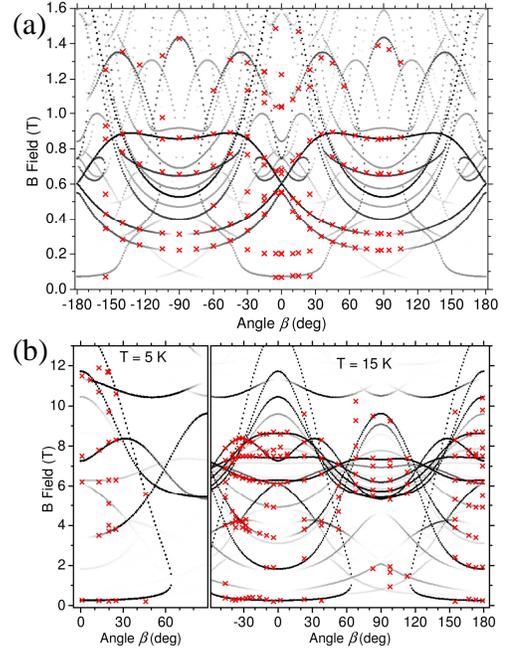}
 \caption{\label{fig:wheels-csfe8epr} (Color online)
Experimental (crosses) and simulated (doted lines) EPR transitions as observed in angle-resolved EPR
experiments on single-crystals of CsFe$_8$ at frequencies of (a) Q band (35~GHz) ($T=25$~K) and (b) 190~GHz
($T=5$ and 15~K). Adapted from \onlinecite{Dre10-csfe8epr}.}
\end{figure}

INS experiments allowed the observation of all $L$-band states up to $S=5$ at an energy of 14.4~meV [transitions
$L^0$-$L^4$ with respect to Fig.~\ref{fig:wheels-cr8theory}]. The data could very accurately be simulated,
Fig.~\ref{fig:wheels-csfe8ins2}. Also, the EPR transitions observed in angle-resolved high-frequency EPR in the field
range of 0 - 12~T could be described with high accuracy, Fig.~\ref{fig:wheels-csfe8epr}. In the EPR experiments the
$L$-band states up to $S=4$ were probed. In the fits to the data only two free parameters, $2J$ and $D$, were involved.
The determined values are given in Table.~\ref{tab:lc-csfeparams} (INS \#2 and HFEPR).

\begin{figure}
\includegraphics[width=8cm]{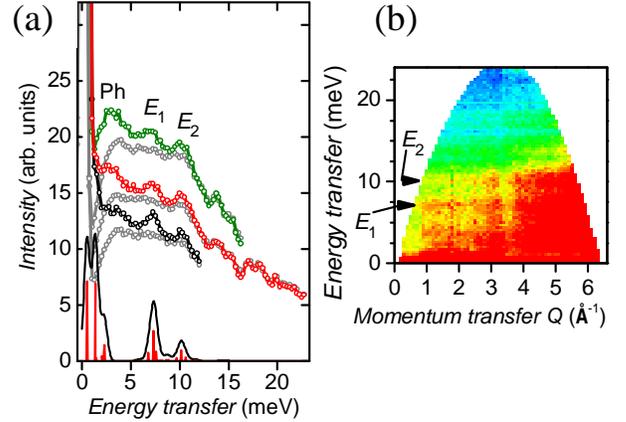}
 \caption{\label{fig:wheels-csfe8ins3} (Color online)
Neutron-energy loss spectrum in CsFe$_8$ ($T = 5$~K). (a) $Q$-sliced INS data, with curves offset for
clarity. Gray curves represent the INS data recorded at $T=58$~K after Bose-correction. The solid line
represents the best-fit simulation of the INS spectrum. (b) $S(Q,\omega)$ plot. Adapted from
\onlinecite{Dre10-csfe8ins3}. }
\end{figure}

The excitation spectrum in CsFe$_8$ was also probed for energies up to 25~meV in a high-energy INS experiment
\cite{Dre10-csfe8ins3}. Results are shown in Fig.~\ref{fig:wheels-csfe8ins3}. At lower energies the transitions within
the $L$ band are again observed, however, at energies of ca. 7.5 and 10~meV two further cold transitions are detected,
which can unambiguously be related to the discrete spin-wave excitations $E_1$ and $E_2$ expected in an octa-nuclear
antiferromagnetic wheel. Above transition $E_2$ no magnetic scattering intensity is observed, confirming the selection
rule associated to the $L \& E$-band picture in Sec.~\ref{sec:lc-cr8}. The determined $2J$ and $D$ values are listed in
Table.~\ref{tab:lc-csfeparams} (INS \#3).

The INS data were recorded on a non-deuterated poly-crystalline sample. This may explain the huge
non-magnetic scattering in Fig.~\ref{fig:wheels-csfe8ins3}, which is typically observed in molecular
nanomagnets above energies of ca. 2-3~meV. For CsFe$_8$ it could however very well be accounted for using a
Bose-correction analysis (Sec.~\ref{sec:basics-ins}), which allowed the unambiguous identification of the
magnetic peaks, see Fig.~\ref{fig:wheels-csfe8ins3}(a).

\begin{table}
\caption{\label{tab:lc-csfeparams} Comparison of the magnetic parameters for CsFe$_8$ obtained by different
experimental techniques.}
\begin{ruledtabular}
\begin{tabular}{cccc}
Technique & $2J$ [meV] & $D$ [meV] & Reference \\
\hline
torque & -1.90(10) & -0.045(3) & \onlinecite{Wal01-csfe8torque} \\
INS \#1& -1.78(4) & -0.048(1) & \onlinecite{Wal05-csfe8ins1} \\
INS \#2& -1.80(2) & -0.050(1) & \onlinecite{Wal06-csfe8ins2} \\
INS \#3& -1.79(5) & -0.050(7) & \onlinecite{Dre10-csfe8ins3} \\
HFEPR & -1.87(25) & -0.0493(1) & \onlinecite{Dre10-csfe8epr}
\end{tabular}
\end{ruledtabular}
\end{table}

Table~\ref{tab:lc-csfeparams} compiles the determined $2J$ and $D$ values, including those obtained by torque
and high-resolution INS (INS \#1) \cite{Wal01-csfe8torque,Wal05-csfe8ins1}. The consistency is excellent, in
particular considering the large range of energy scales probed in the experiments ($\sim$0.01~meV in 35~GHz
EPR, $\sim$10~meV in high-energy INS). Efforts were made to infer the significance of further terms in the
microscopic spin Hamiltonian not included in Eq.~(\ref{eq:wheels-H}). A $J_1-J_2$ modulation of the exchange
constants along the wheel was found to be smaller than 20\%, and the rhombic anisotropy to be negligible, $E
= 0.0000(3)$~meV \cite{Dre10-csfe8ins3,Dre10-csfe8epr}.

The determined $2J$ value in the EPR experiment deserves a comment. The EPR selection rules do not allow a
direct observation of exchange splittings (Sec.~\ref{sec:basics-epr}). Exchange constants may though be
determined indirectly, through the temperature dependence of the EPR resonance intensities, which however in
large clusters is challenging, or through the $S$-mixing mechanism
\cite{Liv02-smix,Wal05-smix,Wil06-ni4smix,Bar07-mn12transverse,Pil09-cr7msmix}, which was the case in
CsFe$_8$. In the strong-exchange limit, the anisotropy splitting produces the "normal" zero-field splitting
pattern, e.g., a $D$ term in the microscopic spin Hamiltonian produces a zero-field splitting of the spin
multiplet, which follows the $M^2$ behavior corresponding to $\hat{S}^2_z$. However, if anisotropy is
stronger, as compared to the exchange $J$, then the pattern is modified and deviations from $M^2$ occur. In
perturbation theory this corresponds to higher-order terms $(\hat{S}^2_z)^n$ with $n>1$ coming in, with
weights $\propto (D/J)^n$. Detecting these shifts in the zero-field splitting pattern allows the indirect
determination of the strength of the exchange. The excellent agreement of the EPR $2J$ value demonstrates
hence that the generic Hamiltonian Eq.~(\ref{eq:wheels-H}) does also describe the subtle $S$-mixing effects
very well in CsFe$_8$.

\begin{figure}
\includegraphics[width=6.5cm]{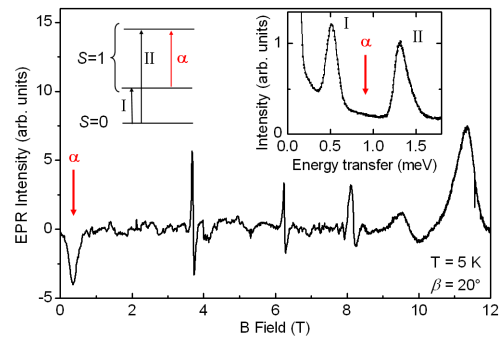}
 \caption{\label{fig:wheels-csfe8inseps} (Color online)
190~GHz EPR spectrum on a single crystal of CsFe$_8$. The intra-multiplet transition $\alpha$ is clearly
observed. The inset to the left depicts the three lowest energy levels and the allowed INS transitions I, II,
and $\alpha$. The inset to the right shows INS data recorded at $T = 9.7$~K in which the transition $\alpha$
should have been observed if it were of appreciable intensity. Adapted from \onlinecite{Dre10-csfe8epr}.}
\end{figure}

The comparison of the INS and EPR experiments reveal another interesting aspect of the excitations in
antiferromagnetic wheels. According to the INS selection rules the transition $|S=1,M=0\rangle
\leftrightarrow |S=1,M=\pm1\rangle$, or $\alpha$ henceforth, is allowed and should be detected at appropriate
temperatures, yet it is not observed in INS experiments, albeit in EPR experiments
(Fig.~\ref{fig:wheels-csfe8inseps}). It turns out that \emph{intra}-multiplet transitions, such as $\alpha$,
are orders of magnitude weaker than \emph{inter}-multiplet transitions because of the particular many-body
structure of the wave functions, which is that of a bipartite lattice of two mesoscopically sized spins on
each sublattice ($S_A = S_B = 10$ in CsFe$_8$) \cite{Wal05-csfe8ins1}. By that reason the
\emph{intra}-multiplet transitions become too weak to be observed by INS. CsFe$_8$ provides a convincing
example, since here the transition $\alpha$ should have easily been detected by INS, see
Fig.~\ref{fig:wheels-csfe8inseps}. Hence, the combined INS and EPR data demonstrate directly a hallmark
feature of antiferromagnetic wheels, namely the mesoscopic antiferromagnetic sublattice structure.

\begin{figure}
\includegraphics[width=6.5cm]{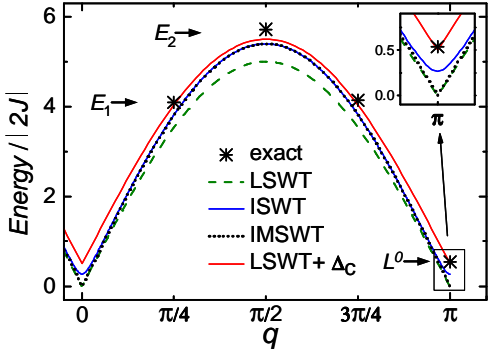}
 \caption{\label{fig:wheels-csfe8swt} (Color online)
Zero-temperature excitation spectrum of the $N=8$, $s=5/2$ AFMHR, as function of the shift quantum number $q$
as calculated using exact numerical diagonalization (stars) and the indicated SWTs (lines)(in the latter $q$
was assumed as continuous for clarity). LSWT = linear SWT, ISWT = interacting SWT, IMSWT =
full-diagonalization interacting modified SWT, LSWT+$\Delta_c$ = linear SWT with a shift. Adapted from
\onlinecite{Dre10-csfe8ins3} to which we also refer for details.}
\end{figure}

Although anisotropy is appreciable in CsFe$_8$, the energies of the spin-wave excitations $E_1$ and $E_2$ are actually
little affected. This, and fundamental interest, motivated an analysis of the elementary excitations of the $N=8$,
$s=5/2$ AFMHR model using different variants of SWT. All models provide predictions for the $E$-band states or the
"dispersion relation" $\epsilon(q)$ in Eq.~(\ref{eq:wheels-dispersion}), but only the last three yield also estimates
for the singlet-triplet gap $\Delta$ ($= a_1 |2J|$) or $L$-band indeed. The findings are compared in
Fig.~\ref{fig:wheels-csfe8swt} to the exact energies. Interestingly, all SWT models reproduce roughly the spin-wave
excitation spectrum, which supports the notion that the $L \& E$-band concept is essentially classical. However,
significant differences exist, and the ISWT and LSWT+$\Delta_c$ models do best for the $E$-band excitations. As regards
the singlet-triplet gap, the LMSWT and IMSWT estimate it almost a factor of two too small. LSWT+$\Delta_c$ comes
closest, to within 7\%. A similar analysis for the larger Fe$_{18}$ wheel ($N= 18$, $s= 5/2$) confirmed the
observations \cite{Umm12-Fe18}.

\subsubsection{Antiferromagnetic wheels with "strong" anisotropy and quantum tunneling of the N\'eel vector}
\label{sec:lc-nvt}

The possibility of quantum tunneling of the N\'eel vector (QTNV) in antiferromagnetic materials attracted
huge interest \cite{Bar90-nvt,Kri90-nvt,Gun95-QTMbook,Gid95-ferritin,Chu98-QTMbook,Shp07}, and initial
attempts to establish QTNV concentrated on ferritin proteins
\cite{Aws92-ferritin,Tej96-tc,Gar96-tc,Gid96-tc}. The prediction that coherent QTNV might also be realized in
antiferromagnetic molecular wheels with strong anisotropy \cite{Chi98-wheel-mqt} stimulated intense research.
In this context the molecules Fe$_{10}$ (Sec.~\ref{sec:lc-wheels}), CsFe$_8$ (Sec.~\ref{sec:lc-csfe8}), and
Fe$_{18}$ appeared as promising candidates (Fig.~\ref{fig:wheels-wheels}). Reviews and follow-up articles on
the QTNV scenario in antiferromagnetic wheels are available
\cite{Leu03-nvtreview,Mei01-wheel-mqt,Kon11-review}.

\begin{figure}
\includegraphics[width=6.5cm]{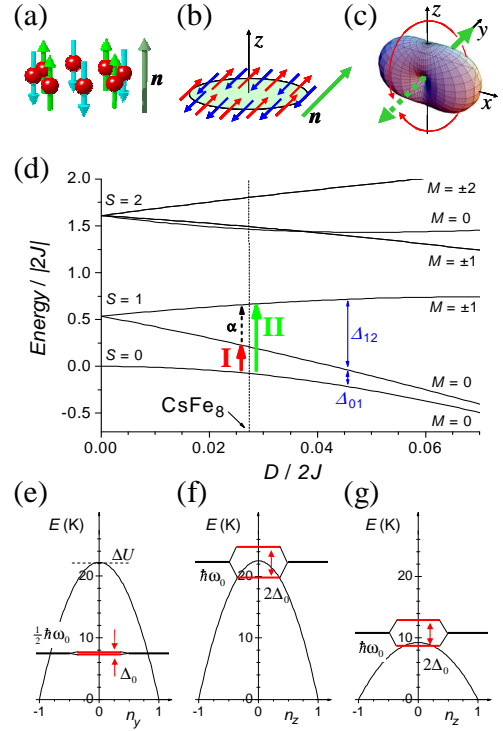}
 \caption{\label{fig:wheels-nvt} (Color online)
(a,b) Classical ground-state spin configuration for a (a) $N=8$, (b) $N=18$ antiferromagnetic wheel with the
N\'eel vector (long green arrow) pointing along (a) $\bf z$, (b) $\bf y$. (c) Shape of the potential $V({\bf
n})$ in the case (B). The two tunneling paths from ${\bf n} = +y$ to $-y$ are indicated. Simulated low-lying
energies of a $N=8$, $s=5/2$ antiferromagnetic wheel vs $D/(2J)$ in zero field ($M$ is then an exact quantum
number). Transitions I and II are observed in INS experiments (Fig.~\ref{fig:wheels-nvtins}). (e-g) Energy
scheme as given by semiclassical theory for (e) Fe$_{18}$, (f) CsFe$_8$, and (g) Fe$_{10}$. Adapted from
\onlinecite{Kon11-review,Wal09-fe18,Wal05-csfe8ins1}. }
\end{figure}

The discussion in the following refers to Eq.~(\ref{eq:wheels-H}). Due to the anisotropy the spin functions
$|S M\rangle$ are mixed strongly and $S$ loses its significance as good quantum number. This changes also the
dynamics of the N\'eel vector, defined as $\hat{\bf n} = ( \hat{\bf S}_A - \hat{\bf S}_B )/(S_A +S_B)$, which
is then not anymore that of a rotation but tunneling. In the quantum spectrum this may be seen by the fact
that the two lowest levels (ground state \& first excitation) approach each other but separate from the next
higher-lying levels, as shown exemplarily in Fig.~\ref{fig:wheels-nvt}(d): $\Delta_{01}$ becomes smaller than
e.g. $\Delta_{12}$.

A semi-classical theory provided a clear description \cite{Chi98-wheel-mqt,Mei01-wheel-mqt}. Depending on the
magnetic parameters and strength and orientation of the magnetic field, several scenarios occur; we here
focus on two: (A) $D<0$ and zero magnetic field and (B) $D>0$ and large magnetic fields with $\bf B$ along
the $x$ axis. In both cases, the N\'eel vector is strongly localized along two directions, namely ${\bf n} =
\pm {\bf z}$ (A) or $\pm {\bf y}$ (B). Classically, the ground state is then characterized by the spin
configurations with N\'eel order sketched in Fig.~\ref{fig:wheels-nvt}(a) and (b), respectively, and rotation
is hampered by an energy barrier, corresponding to a potential $V({\bf n})$ with minima at the respective
orientations [Fig.~\ref{fig:wheels-nvt}(c)]. However, quantum fluctuations allow for tunneling of the N\'eel
vector, which lifts the classical degeneracy and opens a tunneling gap $\Delta_{QTNV}$ in the energy
spectrum, which then corresponds to $\Delta_{01}$.

The N\'eel-vector dynamics is characterized by the tunneling action $S_0 / \hbar$, attempt frequency $\hbar \omega_0$,
barrier height $\Delta U$, and tunneling amplitude $\Delta_0$,
\begin{eqnarray}
\label{eq:wheels-nvtparamters}
 S_0/\hbar &=& N s \sqrt{2 |D/(2J)| },  \\
 \hbar \omega_0 &=& s \sqrt{ 8 |D (2J)| },  \\
 \Delta U &=& N s^2 |D|,  \\
 \Delta_0 &=& 8 \hbar \omega_0 \sqrt{ \frac{ S_0/\hbar }{ 2 \pi } } \exp( - S_0/\hbar ).
\end{eqnarray}
The two cases (A) and (B) need to be distinguished now, which we do through a parameter $c=2$ (A) and $c=1$
(B), respectively. The ground state energy is then given as $\frac{c}{2} \hbar \omega_0$, and the tunneling
splitting as $c \Delta_0$ [in the case (A) the tunneling gap has to our knowledge not yet been calculated,
but numerical results suggest $\Delta_{QTNV} \approx 2 \Delta_0$]. The semi-classical theory for QTNV becomes
valid for large tunneling actions $\frac{1}{2c} S_0 / \hbar \gg 1$, which is equivalent to stating that the
ground-state energy is smaller than the barrier height, $\Delta U \gg \frac{c}{2} \hbar \omega_0$, or the
N\'eel vector strongly localized, $\langle 0 | \hat{ n }_{z/y} | 1 \rangle^2 \rightarrow 1$, where the matrix
element is given as $\langle 0 | \hat{ n }_{z/y} | 1 \rangle^2 \approx 1 - c/(S_0/\hbar)$, where $z$ (A) or
$y$ (B) is the respective N\'eel vector component.

As regards applying the semi-classical theory some points are worth noticing. First, it turned out that the $J$ and $D$
values as they appear in Eq.~(\ref{eq:wheels-H}) or through $a_1^{qm}$ and $b_1^{qm}$ in Eq.~(\ref{eq:wheels-HABfull})
should not be inserted in the semi-classical formulae, but "corrected" $J$, $D$ values as they are obtained by using
$a_1^{sc} \equiv 4/N$ and $b_1^{sc} \equiv 2 N s^2 / [N s( N s + 2 ) - 3]$ \cite{Wal09-fe18} (for a detailed discussion
see \onlinecite{Kon11-review}). Not doing so leads to e.g. significantly overestimated tunneling actions, which went
unfortunately unnoticed in early works. Furthermore, the crossover from weak to strong anisotropy, or from rotation to
tunneling of the N\'eel vector, is continuous and not abrupt [Fig.~\ref{fig:wheels-nvt}(d)], and the QTNV scenario is
hence necessarily approximate \cite{San05-fe10nvt}. This introduces some ambiguity, and different criteria for when
QTNV is realized can be given, e.g., that the tunneling levels should fall below the top of the barrier or that the
tunneling splitting should be exponentially small. It is noted in passing that since the semi-classical theory as it
stands is an approximate theory, agreement with semi-classical theory is a sufficient but not necessary criterium for
QTNV.

\begin{figure}
\includegraphics[width=6.5cm]{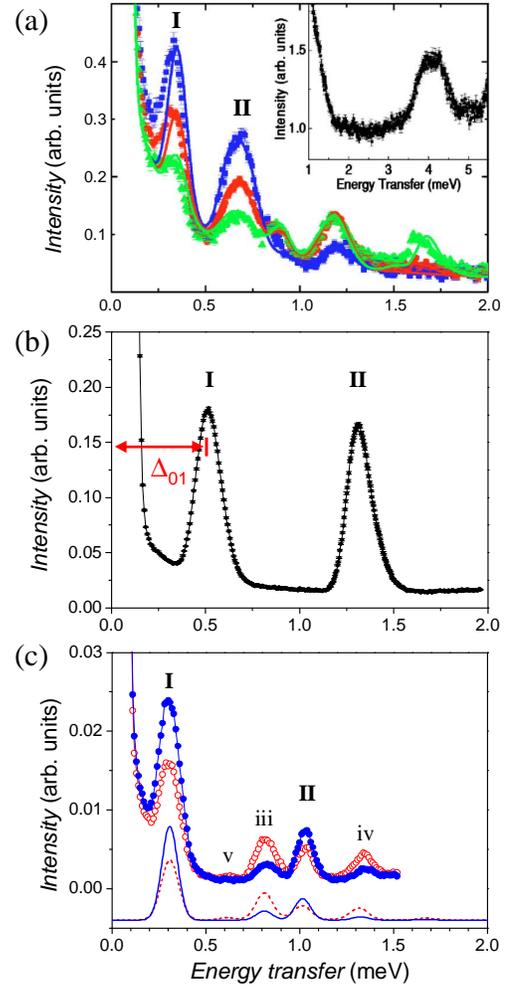}
 \caption{\label{fig:wheels-nvtins} (Color online)
Neutron-energy loss spectra for (a) Fe$_{10}$ [$T =$ 2~K (blue), 5~K (red), 10~K (green)], (b) CsFe$_8$ ($T =
2.4$~K), and (c) Fe$_{18}$ [$T =$ 1.9~K (blue), 4.2~K (red)] showing transition I, which in the QTNV regime
correspond to the N\'eel vector tunneling splitting $\Delta_{01}$. Adapted from
\onlinecite{San05-fe10nvt,Wal05-csfe8ins1,Wal09-fe18}.}
\end{figure}

The low-lying excitations in zero field were measured by INS for Fe$_{10}$, CsFe$_8$, and Fe$_{18}$
\cite{Wal05-csfe8ins1,San05-fe10nvt,Wal09-fe18}. Fe$_{10}$ and CsFe$_8$ were described before. The chemical formula of
Fe$_{18}$ is [Fe$_{18}$(pd)$_{12}$(pdH)$_{12}$(O$_2$CEt)$_6$(NO$_3$)$_6$](NO$_3$)$_6$$\cdot$$x$MeCN ($x \approx 48$);
the eighteen Fe$^{3+}$ ($s = 5/2$) ions are arranged in a cycle (Fig.~\ref{fig:wheels-wheels}) \cite{Kin06-fe18synth}.
The system crystallizes in space group R3- and the molecule exhibits crystallographic S$_6$ symmetry. One nitrate and
ca. 8MeCN solvent molecules are disordered. INS data for the three wheels are shown in Fig.~\ref{fig:wheels-nvtins}.
The lowest excitation I could clearly be detected as well as several higher excitations (for CsFe$_8$ see
Sec.~\ref{sec:lc-csfe8}), which facilitated a precise determination of the magnetic parameters. In Fe$_{10}$ a
substantial rhombic term $|E/D| = 0.21$ was found, and structural disorder had to be included in the analysis. In
Fe$_{18}$ a high-energy INS experiment evidenced a modulation of the exchange constants along the ring consistent with
the C$_3$ symmetry \cite{Umm12-Fe18}. Notably, the ratio of the excitation energies of transitions I and II decreases
in the sequence of Fe$_{10}$, CsFe$_8$, and Fe$_{18}$. For better comparison of the wheels, the magnetic parameters
reported in the original works were converted to $2j$ and $d$ of $\hat{ H}_{AB}$, Eq.~(\ref{eq:wheels-HABfull}) (in
case of Fe$_{10}$ the rhombic contribution was neglected and for Fe$_{18}$ the appropriate averaged $J$ was used). The
results are listed in Table~\ref{tab:lc-nvt}.

\begin{table}
\caption{\label{tab:lc-nvt} Experimental values and characteristic parameters of QTNV for Fe$_{18}$ ($c=1$), CsFe$_8$
($c=2$), and Fe$_{10}$ ($c=2$). Adapted from \cite{Kon11-review}.}
\begin{ruledtabular}
\begin{tabular}{cccc}
 & Fe$_{18}$ & CsFe$_8$ & Fe$_{10}$ \\
\hline
$2j$ [K] & -5.1 & -11.1 & -6.31 \\
$d$ [K] & 0.021 & -0.104 & -0.0276 \\
$\Delta_{01}$ [K] & not measured  & 5.92 & 3.83 \\
\hline
$S_0/\hbar$ & 5.90 & 4.03 & 3.42 \\
$\Delta U$ [K] & 22.2 & 22.5 & 9.25 \\
$\frac{c}{2} \hbar \omega_0$ [K] & 7.52 & 22.3 & 10.8 \\
$c \Delta_0$ [K] & 0.320 & 5.08 & 4.18 \\
\hline
$(S_0/\hbar) / c$ & 2.95 & 1.01 & 0.96 \\
$\langle 0 | \hat{ n }_{z/y} | 1 \rangle^2$ & 0.83 & 0.50 & 0.42 \\
$\Delta_{QTNV} / \Delta U$ & 0.014 & $\approx 0.23$ & $\approx 0.45$ \\
\end{tabular}
\end{ruledtabular}
\end{table}

For both CsFe$_8$ and Fe$_{10}$ one finds $d<0$ or $D<0$, while for Fe$_{18}$ $D>0$. Hence, in CsFe$_8$ and
Fe$_{10}$ [case (A)] the transition I observed by INS would directly correspond to the N\'eel vector
tunneling gap if QTNV were realized in them. The tunneling gaps estimated by the semi-classical theory
roughly agree with the observed gaps, but the other parameters compiled in Table~\ref{tab:lc-nvt} indicate
that QTNV is not well realized in Fe$_{10}$, and that CsFe$_8$ is borderline. For $D>0$ as in Fe$_{18}$ QTNV
does not occur at low fields \cite{Chi98-wheel-mqt}, but in the high-field regime, which in Fe$_{18}$ is
reached above 10.6~T [case (B)]. Here, QTNV is well realized in Fe$_{18}$ as e.g. indicated by the
exponentially small tunneling gap. The situation in the three wheels is probably most clearly demonstrated by
the energy diagrams shown in Fig.~\ref{fig:wheels-nvt}(e) - (g).

\begin{figure}
\includegraphics[width=6.5cm]{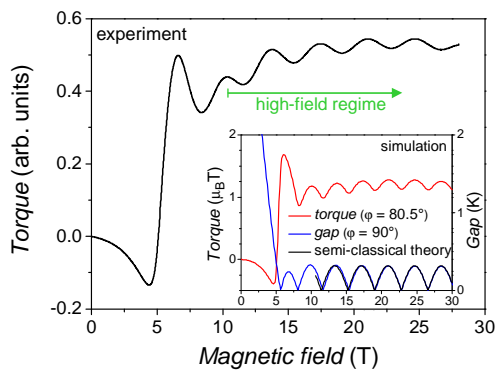}
 \caption{\label{fig:wheels-nvtfe18} (Color online)
Magnetic torque vs field for $B$ nearly perpendicular to $z$ (angle = 80.5$^\circ$, $T = 0.1$~K), showing
"wiggles" due to the oscillations of the N\'eel-vector tunneling gap. The inset shows the simulated
field-dependent torque and the tunneling gap as calculated quantum mechanically and semiclassically. Adapted
from \onlinecite{Wal09-fe18}.}
\end{figure}

The high-field N\'eel vector tunneling gap in Fe$_{18}$ was not directly observed via spectroscopic
techniques, but the field-dependent oscillations in the N\'eel vector tunneling gap due to quantum
interference were detected in low-temperature high-field magnetic torque measurements \cite{Wal09-fe18}. In
the semi-classical theory for case (B) the phase of the ground-state wave function contains a topological
term $\pi N g \mu_B B / |4 (2J)|$, which is proportional to the field. Hence, in increasing fields the phase
is repeatedly tuned through destructive and constructive interference, which gives rise to an oscillation of
the tunneling splitting according to
\begin{eqnarray}
\label{eq:wheels-nvtosc}
 \Delta(B) = \Delta_{QTNV} \left| \sin\left(\pi \frac{N g \mu_B B}{ 4 |2J| } \right) \right|.
\end{eqnarray}
Since the tunneling splitting also affects the ground-state energy, $E_0(B) = \epsilon(B) + \Delta(B)/2$,
where $\epsilon(B)$ is a smooth function, the oscillations can be detected by magnetization or torque
measurements at low temperatures. Indeed, the observed "wiggles" in the torque curve
(Fig.~\ref{fig:wheels-nvtfe18}) do directly correspond to the oscillations in the tunneling gap, which is
demonstrated e.g. by comparing the numerically calculated curves for the torque and the tunneling gap (inset
to Fig.~\ref{fig:wheels-nvtfe18}). The analysis also showed that the semi-classical theory yields highly
accurate results, which underpins the notion of QTNV in Fe$_{18}$.

It is added that wiggles in the torque as function of field can also occur due to a first-order mixing of the
$|S,M=-S\rangle$ and $|S+1,M=-S-1\rangle$ spin levels at the field-induced level-crossing points by the magnetic
anisotropy; an example for this $S$-mixing mechanism will be presented below in Sec.~\ref{sec:lc-mn3x3}. However,
theoretically QTNV cannot be described by $S$-mixing of two spin levels reflecting its different underlying physics,
and also experimentally, the two mechanisms can unambiguously be distinguished from each other. For instance, the
dependence on the angle between magnetic field and anisotropy axis $z$ allows a clear-cut decision: In the $S$-mixing
scenario the wiggles occur for both nearly parallel and perpendicular fields, in contrast to the observations in
Fe$_{18}$, where ordinary staircase-like profiles are observed for parallel fields, as predicted by the QNVT scenario.

\subsection{"Modified" antiferromagnetic molecular wheels}
\label{sec:lc-modwheels}

\begin{figure}
\includegraphics[width=6.5cm]{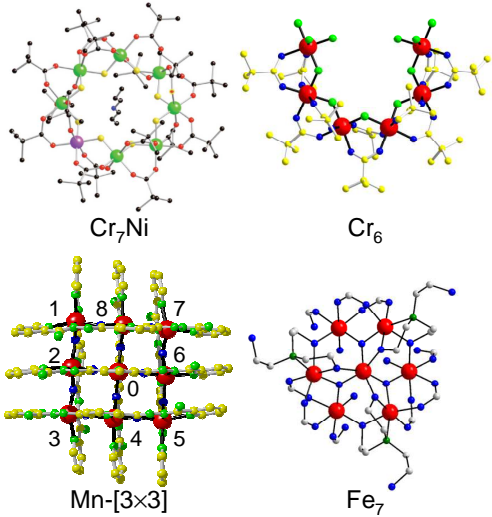}
 \caption{\label{fig:modwheels-modwheels} (Color online)
Molecular structures of the doped wheel Cr$_7$Ni: [$^n$Pr$_2$NH$_2$][Cr$_7$NiF$_8$Piv$_{16}$],
 short chain Cr$_6$: [R$_2$NH$_2$]$_3$[Cr$_6$F$_{11}$Piv$_{10}$(H$_2$O)],
 Mn-[3$\times$3] grid representing a wheel with magnetic center: [Mn$_9$(2POAP-2H)$_6$],
 and disc Fe$_7$: [Fe$_7$O$_3$tea$_3$Piv$_9$(H$_2$O)$_3$].}
\end{figure}

The ring topology considered in Sec.~\ref{sec:lc-wheels} can be varied in a number of ways, by "slight"
modifications. The topologies which will be addressed here can be put into four categories. First, one of the
magnetic metal ions in the ring, which carry spin $s$, is replaced by another magnetic metal ion with
different spin $s_0 \neq s$. These clusters shall be denoted as "doped wheels", and the foreign $s_0$ ion as
impurity. Second, the cyclic boundary conditions are changed to open boundaries, e.g., by replacing one of
the magnetic ions in the ring with a diamagnetic ion or by removing one metal center. These clusters shall be
denoted as "short chains". Third, a magnetic ion is added at the center of the ring with coupling paths such
that the lattice remains bipartite. And fourth, a magnetic ion is added at the center with the coupling paths
introducing spin frustration, then called "discs". Representative examples are Cr$_7$Ni, Cr$_6$,
Mn-[3$\times$3], and a Fe$_7$ molecule, shown in Fig.~\ref{fig:modwheels-modwheels}.

Much interest in such systems came from the theoretical suggestions that an uncompensated spin introduced
into an antiferromagnetic wheel e.g. by doping may act as a tracer spin for the quantum tunneling of the
N\'eel vector, which in this way could experimentally be observed and possibly manipulated by EPR methods,
which otherwise would not be possible \cite{Mei01-dopedwheels,Mei01-wheel-mqt}. Furthermore, the excess spin
may have $S' = 1/2$, which then suggests application of the cluster as quantum bit (qubit)
\cite{Mei03-qubit1,Mei03-qubit2,Tro05-qubit}. Using such "mesoscopic spin-1/2" clusters as "spin cluster
qubits" could provide advantages such as easier addressing. Significant progress in this direction have been
made on the Cr$_7$Ni wheel, e.g., long coherence times were observed, magnetic coupling between two Cr$_7$Ni
clusters introduced, and entanglement demonstrated \cite{Ard07,Tim09,Can10}. Reviews are available
\cite{Aff05-review,Aff06-review}. Since the physics is related mainly to the ground state, these exciting
developments will however not be discussed. We will focus here on the basic question of the impact of
topology on magnetic cluster excitations. The Mn-[3$\times$3] grid molecule will be considered first,
followed by the doped wheels and short antiferromagnetic chains. An abundance of discs have been synthesized
\cite{Hos-review}, but detailed studies of the cluster excitations were to our knowledge not reported.

\subsubsection{The Mn-[3$\times$3] grid molecule}
\label{sec:lc-mn3x3}

Molecular [n$\times$m] grids attracted considerable interest in chemistry, for the employed preprogrammed self-assembly
synthesis strategy and their physical properties. For reviews see \onlinecite{Rub04-gridreview,Daw09-gridreview}, the
magnetic properties were reviewed in \onlinecite{Wal05-gridreview}. The Mn-[3$\times$3] grid \cite{Zha00-mn3x3} is most
interesting from the perspective of cluster excitations. The molecule may be crystallized with different counter ions
and solvents, yielding e.g. [Mn$_9$(2POAP-2H)$_6$](ClO$_4$)$_6$$\cdot$3.75CH$_3$CN$\cdot$11H$_2$O (\textbf{1}) or
[Mn$_9$(2POAP-2H)$_6$](NO$_3$)$_6$$\cdot$H$_2$O (\textbf{2}). Nine Mn$^{2+}$ ($s = 5/2$) metal ions are arranged on the
vertices of a 3$\times$3 grid (Fig.~\ref{fig:modwheels-modwheels}). They crystalize in space group C2/c, and the
molecules exhibit a slightly distorted D$_{2d}$ symmetry with the S$_4$ symmetry axis (= $z$ axis) perpendicular to the
grid plane. Considering the symmetry, the appropriate spin Hamiltonian for describing the magnetism reads
\begin{eqnarray}
\label{eq:mn3x3-H}
 \hat{H} &=& -2J_R \left( \sum_{i=1}^7 \hat{\bf s}_i \cdot \hat{\bf s}_{i+1} + \hat{\bf s}_8 \cdot \hat{\bf s}_1 \right)
\nonumber \\ &&
  -2J_C \left( \hat{\bf s}_2  + \hat{\bf s}_4  + \hat{\bf s}_6 + \hat{\bf s}_8 \right) \cdot \hat{\bf s}_0
\nonumber \\ &&
   + D_R \sum_{i=1}^8 \hat{s}^2_{i,z}  + D_C \hat{s}^2_{0,z}
   + g \mu_B \hat{\bf S} \cdot {\bf B},
\end{eqnarray}
where the spins are numbered as given in Fig.~\ref{fig:modwheels-modwheels}(c). In principle, the $D$ values
for the corner and edge Mn ions could be different, but this was found to not affect the magnetism
significantly.

\begin{figure}
\includegraphics[width=8cm]{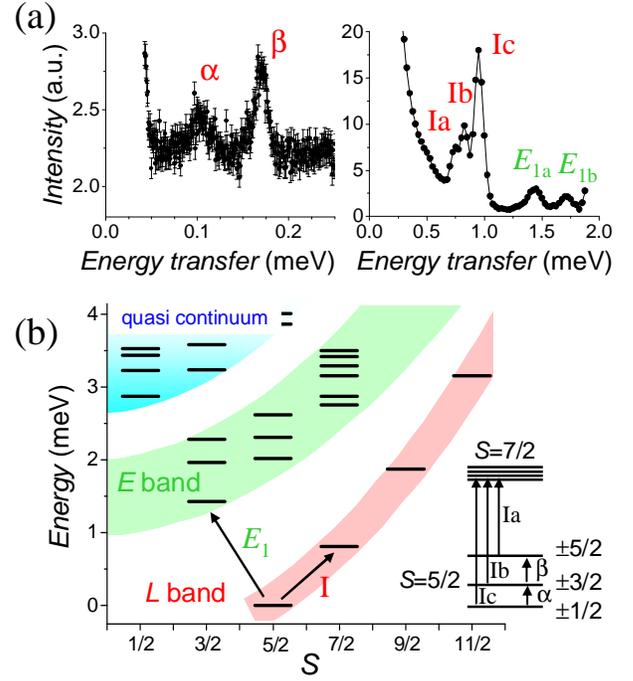}
 \caption{\label{fig:modwheels-mn3x3insspk} (Color online)
(a) Neutron-energy loss spectra of Mn[3$\times$3] (\textbf{2}) at 2.5~K in two different energy ranges. (b)
Simulated energy spectrum with anisotropy neglected. The observed INS transitions are indicated by arrows.
The inset sketches the observed transitions within the $S=5/2$ ground and first excited $S=7/2$ multiplet.
 Adapted from \onlinecite{Gui04-mn3x3}.
}
\end{figure}

From magnetization measurements, antiferromagnetic exchange interactions and a $S= 5/2$ ground state were
inferred, which can be rationalized by the classical spin configuration, where corner and central spins point
up and edge spins point down. The excitation spectrum up to energies of 4~meV was studied by INS
\cite{Gui04-mn3x3}. Some results are presented in Fig.~\ref{fig:modwheels-mn3x3insspk}(a). The dimension of
the Hilbert space is 10~077~696, and sophisticated numerical approaches had to be developed for analyzing the
data. Good agrement was obtained for $2J_R = 2J_C = -0.43$~meV and $D_R = D_C = -0.012$~meV. The higher-lying
excitations revealed a small deviation of the exchange constants from the S$_4$ symmetry assumed in
Eq.~(\ref{eq:mn3x3-H}). The calculated energy spectrum with $D_R$ and $D_C$ set to zero is presented in
Fig.~\ref{fig:modwheels-mn3x3insspk}(b). The two lowest transitions $\alpha$ and $\beta$ stem from the
zero-field splitting of the $S= 5/2$ ground state, as sketched in Fig.~\ref{fig:modwheels-mn3x3insspk}(b).
Peaks Ia, Ib, and Ic correspond to transitions from the zero-field splitting levels of the ground multiplet
to the next higher-lying $S=7/2$ multiplet, and peaks $E_{1a}$ and $E_{1b}$ go from the ground multiplet to
the lowest $S=3/2$ multiplets.

Besides that the INS data of such a large cluster as Mn-[3$\times$3] were successfully interpreted, the
inspection of the determined energy spectrum is interesting: As in antiferromagnetic wheels, the excitations
may be classified as $L$ and $E$ band, or N\'eel-vector rotation and spin waves. This observation can be
linked to the bipartite topology of the grid lattice, suggesting a "classical" spin structure.

\begin{figure}
\includegraphics[width=6.5cm]{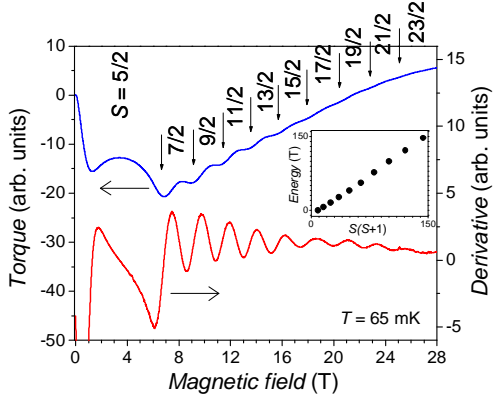}
 \caption{\label{fig:modwheels-mn3x3torque} (Color online)
Magnetic torque vs field (upper curve) and the field derivative (lower curve) for Mn-[3$\times$3]
(\textbf{1}) with $B$ nearly perpendicular to the magnetic $z$ axis (angle = 80.5$^\circ$, $T = 0.1$~K),
showing the oscillatory torque behavior. The inset shows the energies of the spin multiplets as extracted
from the level-crossing fields \cite{WalUP-mn3x3torque}.}
\end{figure}

The $L$ band has been demonstrated in magnetic torque measurements \cite{Wal04-mn3x3torque}. At very low
temperatures the torque exhibits an oscillatory field dependence, Fig.~\ref{fig:modwheels-mn3x3torque}, which
can be related to a sequence of level crossings, where the ground state changes from $S=5/2$ to $S=7/2$, $7/2
\rightarrow 9/2$, and so one, similar to the situation in the wheels sketched in
Fig.~\ref{fig:wheels-spectrum}. In the experiments all states of the $L$ band up to $S= 23/2$ were observed,
and the energies of the multiplets as determined from the level-crossing fields do indeed follow the Land\'e
pattern $S(S+1)$, as displayed in the inset to Fig.~\ref{fig:modwheels-mn3x3torque}. Hence, the $L$ band (or
tower of states or quasi-degenerate joints states), which is the precursor to long-range N\'eel ordering in
the infinite lattice \cite{Ber92-qdjs} has been experimentally demonstrated for the square-lattice topology.

These findings suggest the construction of an effective spin Hamiltonian for antiferromagnetic 3$\times$3
grids. Applying the procedure used for the antiferromagnetic wheels to the eight metal ions on the periphery
yields
\begin{eqnarray}
\label{eq:mn3x3-HAB}
 \hat{H}_{ABC} &=& -2j_R \hat{\bf S}_A \cdot \hat{\bf S}_B -2J_C \hat{\bf s}_0 \cdot \hat{\bf S}_B
\nonumber \\ &&
   + d_R \left( \hat{S}^2_{A,z} + \hat{S}^2_{B,z} \right) + D_C \hat{s}^2_{0,z}
\nonumber \\ &&
   + g \mu_B \hat{\bf S} \cdot {\bf B},
\end{eqnarray}
with the sublattices $A = \{1,3,5,7\}$ and $B = \{2,4,6,8\}$, and $j_R = 0.526 J_R$ and $d_R = 0.197 D_R$ for the
Mn-[3$\times$3] grid. $\hat{H}_{ABC}$ was indeed demonstrated to produce highly accurate results, and was found crucial
in the analysis of experimental data \cite{Dat07-m3x3smix,Wal04-mn3x3torque,Wal05-mn3x3nvt}.

The oscillations in the torque signal originate from an interesting quantum-mixing mechanism
\cite{Car03-mn3x3,Wal04-mn3x3torque}. A magnetic anisotropy is generally expected to induce mixing of spin
multiplets ($S$ mixing), which often may be treated perturbatively, implying a "small" effect (but see also
Sec.~\ref{sec:lc-nvt}). However, if two states are close in energy, i.e., essentially degenerate, then
standard (non-degenerate) perturbation theory will obviously break down, and the effect of the perturbation
not be small or mixing of the states large. The anisotropy produces thus a strong mixing of the
$|S,M=-S\rangle$ and $|S+1,M=-S-1\rangle$ states involved at a level crossing, and the ground state is
described as a superposition
\begin{eqnarray}
\label{eq:mn3x3-mixing}
 |g\rangle \propto a(B) |S \rangle + b(B) |S+1 \rangle
\end{eqnarray}
where the field-dependent $a$ and $b$ become equal at the level crossing (an obvious shorthand notation for
the states was used; $a^2 + b^2 = 1$). Since states with different total spin are mixed, the total spin will
fluctuate strongly, also called quantum oscillations of the total spin \cite{Car03-mn3x3}. This mixing is
directly related to the oscillatory torque curve observed in experiment, which hence provides evidence for
this phenomenon.

$S$ mixing should in principle also enable a direct detection of exchange splittings through EPR experiments, since the
EPR selection rule $\Delta S = 0$ would not hold exactly. However, the mixing is usually not strong enough for such EPR
transitions to gain sufficient intensity, but through this mechanism the $S=5/2 \rightarrow S=7/2$ transition could
directly be observed in Mn-[3$\times$3] as function of field in a multi-frequency single-crystal EPR experiment
\cite{Dat07-m3x3smix}.

\subsubsection{"Doped" even-membered antiferromagnetic wheels}
\label{sec:lc-dopedwheels}

A series of octa-nuclear hetero-nuclear wheels of general chemical formula [H$_2$NR$_2$][Cr$_7$M'F$_8$Piv$_{16}$], or
Cr$_7$M' in short, with e.g. M' = Cu$^{2+}$, Ni$^{2+}$, or Mn$^{2+}$ were synthesized
\cite{Lar03-cr7m,Lay05-cr7m,Aff07-cr7mreview,Bak11-cr7cu}, and their excitations studied by different techniques, among
which were low-temperature torque magnetometry \cite{Car05-cr7mtorque}, high-frequency EPR \cite{Pil09-cr7msmix}, and
INS \cite{Cac05-cr7mins,Bak11-cr7cu}. Also an analogous Fe$_7$Mn cluster was investigated using INS
\cite{Gui07-fe7mins}. As mentioned before, the molecule Cr$_7$Ni has attracted most interest, because of its potential
use in quantum information, and is focused on here.

The synthesis strategy resulting in Cr$_7$Ni is extremely flexible, and potentially many derivatives of Cr$_7$Ni exist
\cite{Aff07-cr7mreview}. The material used in INS experiments was of chemical formula
[H$_2$N(C$_2$D$_5$)$_2$][Cr$_7$NiF$_8$(O$_2$(C$_5$D$_9$)$_{16}$] and crystallizes in space group $P$4, without solvent
molecules in the crystal lattice. Seven Cr$^{3+}$ ($s = 3/2$) ions and one Ni$^{2+}$ ($s_0 = 1$) ion form a ring as
shown in Fig.~\ref{fig:modwheels-modwheels}. The material has two advantageous properties for INS; it can be deuterated
to a large extent and large single crystals can be grown. The appropriate generic spin Hamiltonian reads
\begin{eqnarray}
\label{eq:cr7ni-H}
 \hat{H} &=& -2J \sum_{i=1}^6 \hat{\bf s}_i \cdot \hat{\bf s}_{i+1}
  -2J' \left( \hat{\bf s}_1  + \hat{\bf s}_7 \right) \cdot \hat{\bf s}_0
\nonumber \\ &&
   + D \sum_{i=1}^7 \hat{s}^2_{i,z}  + D' \hat{s}^2_{0,z}
   + g \mu_B \hat{\bf S} \cdot {\bf B},
\end{eqnarray}
where the spins $\hat{\bf s}_1$ to $\hat{\bf s}_7$ refer to the Cr$^{3+}$ ions, and $\hat{\bf s}_0$ is the spin of the
Ni$^{2+}$ ion. In view of the molecular symmetry, also a rhombic term $\sum_i E_i ( \hat{s}^2_{i,x} -  \hat{s}^2_{i,y}
)$ is expected and was evidenced in EPR experiments \cite{Pil09-cr7msmix}, but was not resolved in magnetic torque and
INS experiments \cite{Cac05-cr7mins,Car05-cr7mtorque}.

\begin{figure}
\includegraphics[width=6.5cm]{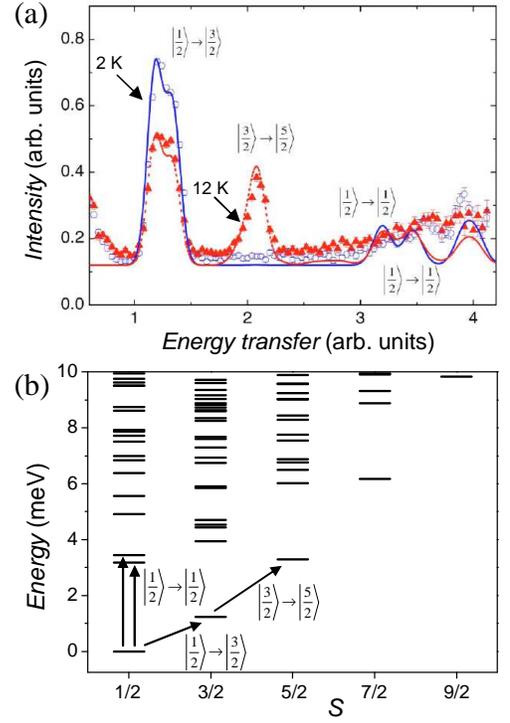}
 \caption{\label{fig:modwheels-cr7niinsspk} (Color online)
(a) Neutron-energy loss spectra on Cr$_7$Ni at $T = 2$~K and 12~K. Circles represent the experimental data;
lines the simulation result. Adapted from \onlinecite{Cac05-cr7mins}.
 (b) Simulated energy spectra with anisotropy neglected. Observed INS transitions are indicated.
}
\end{figure}

The exchange interactions in Cr$_7$Ni are antiferromagnetic and the ground state is $S= 1/2$, consistent with
the expectation from the classical spin configuration with N\'eel order. The magnetic excitations were
studied by two INS experiments, which shall be described in the following. In a first experiment the
excitations up to energies of $\sim$4~meV were detected \cite{Cac05-cr7mins}. A spectrum is presented in
Fig.~\ref{fig:modwheels-cr7niinsspk}(a). One large cold feature, with a double-peak structure, is observed at
1.27~meV, and a hot peak at 2.08~meV. At higher energies evidence for further magnetic scattering intensity
is observed. The lowest feature can be associated to the transition from the $S=1/2$ ground state to the
lowest-lying $S=3/2$ multiplet, and the double-peak structure of this feature to a zero-field splitting in
the $S=3/2$ level. The zero-field splitting (0.15~meV) is much smaller than the center of gravity of the
$S=3/2$ multiplet (1.27~meV); anisotropy is hence weak in this molecule, which is of relevance for its
potential application as qubit \cite{Tro05-qubit}. The 2.08~meV feature corresponds to an $S= 3/2 \rightarrow
5/2$ transition. The data could well be simulated using Eq.~(\ref{eq:cr7ni-H}), yielding $2J = -1.46$~meV,
$2J' = -1.69$~meV, $D^{lig} = -0.03$~meV, and ${D^{lig}}' = -0.6$~meV. The simulated energy spectrum with
anisotropy neglected is shown in Fig.~\ref{fig:modwheels-cr7niinsspk}(b). The spectrum reveals again a
$L$-band structure, which is confirmed by the INS experiment, and high-field torque magnetometry
\cite{Car05-cr7mtorque}. Also $E$-band states are detected in the INS data.

\begin{figure}
\includegraphics[width=6.5cm]{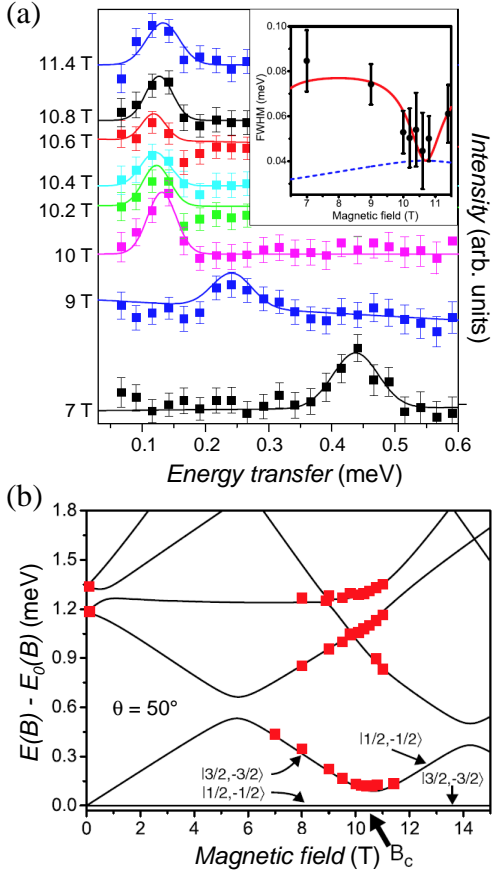}
 \caption{\label{fig:modwheels-cr7niinsfield} (Color online)
(a) High-resolution neutron-energy loss spectra on a crystal sample of Cr$_7$Ni as function of magnetic field
($T = 66$~mK, $\theta = 50^\circ$). (b) Simulated low-lying energy levels as function of magnetic field
(lines) with all observed INS transition energies included (squares). Adapted from \onlinecite{Car07-cr7nifield}. }
\end{figure}

The unique advantage of the Cr$_7$Ni system, that large deuterated single-crystals can be grown, allowed the
direct experimental observation by INS of the level-crossing behavior as function of a magnetic field
\cite{Car07-cr7nifield}. In this experiment, a crystal of 0.4~g weight was measured at $T = 66$~mK with
magnetic field applied in a range of 0 - 11.5~T and an angle of $\theta = 50^\circ$ with respect to the $z$
axis. Experimental results are shown in Fig.~\ref{fig:modwheels-cr7niinsfield}. At the level crossing at $B_c
= 10.5$~T a small gap of 0.12~meV is observed, i.e., an avoided level crossing, which originates from the $S$
mixing induced by the weak anisotropy effects. As explained in Sec.~\ref{sec:lc-mn3x3}, at the level crossing
where states become almost degenerate, the mixing effect will be strong, the wave function described by
Eq.~(\ref{eq:mn3x3-mixing}) with $|a| = |b|$, and the total spin oscillates. This INS experiment hence
provides direct evidence for the quantum oscillations of the total spin in Cr$_7$Ni at the level crossing.

\subsubsection{Short antiferromagnetic chains}

The synthesis strategy which allowed us to generate the doped antiferromagnetic wheels in
Sec.~\ref{sec:lc-dopedwheels} also afforded the generation of short antiferromagnetic chains, either by
replacing one magnetic metal ion  in the ring-like structure by a diamagnetic ion such as Zn$^{2+}$ or
Cd$^{2+}$, or modifying the synthesis method to yield structures called "horseshoes" \cite{Aff07-cr7mreview}.
The members of this family of clusters were studied, besides the thermodynamic techniques, by EPR
\cite{Pil07-cr7cdhfepr}, XMCD \cite{Ghi09-cr6cr7cvxmcd}, NMR \cite{Ami10-cr7m,Mic06-cr7cdnmr}, and INS
\cite{Cac05-cr7mins,Bak11-cr6,Och07-cr6,Och08-cr6cr7,Bia09-cr8zn}. Short antiferromagnetic chains were also
obtained through doping the Fe$_{18}$ wheel (Sec.~\ref{sec:lc-nvt}) with diamagnetic Ga$^{3+}$
\cite{Hen08-fe17ga}.

From the view point of the physics of the magnetic excitations, comparison of short antiferromagnetic chains
with even-membered antiferromagnetic wheels, or chains with open and periodic boundary conditions, should prove
interesting \cite{Och07-cr6}, as well as comparing antiferromagnetic chains with even and odd number of metal
centers \cite{Och08-cr6cr7}. A large body of literature exists on one-dimensional quantum spin chains, and also
finite chains were studied \cite{Hag90,Dit94,Fuj98,Bog04}. It appears natural that the physical pictures
developed there can also be extended to the short antiferromagnetic chains considered here. However, there are
indications that this expectation is not fulfilled and the situation in the antiferromagnetic chains much more
involved \cite{Kon11-review}. A definitive answer is not available at the moment. In the following we will
describe the molecular horseshoe Cr$_6$, which represents a short antiferromagnetic chain with length $N= 6$,
exhibiting $S=0$ ground states.

The generic spin Hamiltonian for short antiferromagnetic chains is
\begin{eqnarray}
\label{eq:shortchains-H}
 \hat{H} = -2J \sum_{i=1}^{N-1} \hat{\bf s}_i \cdot \hat{\bf s}_{i+1}
   + D \sum_{i=1}^N \hat{s}^2_{i,z}  + g \mu_B \hat{\bf S} \cdot {\bf B},
\end{eqnarray}
with $s_i = s$ for all ions. From molecular symmetry also a rhombic term $\sum_i E_i ( \hat{s}^2_{i,x} -
\hat{s}^2_{i,y} )$ and a modulation of the exchange coupling constant, in particular at the ends of the
chain, may be present. The presence of next-nearest-neighbor exchange was also suggested \cite{Bia09-cr8zn}.
However, these effects are considerably smaller than those due to the leading terms given in
Eq.~(\ref{eq:shortchains-H}).

\begin{figure}
\includegraphics[width=6.5cm]{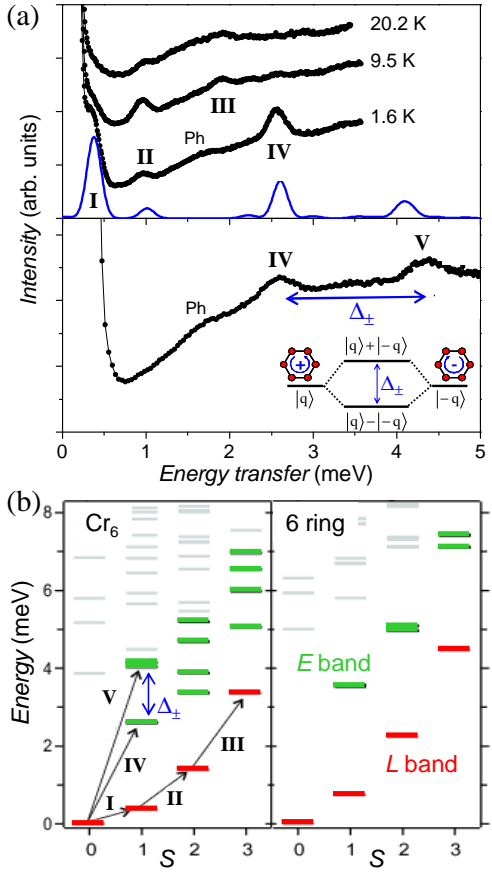}
 \caption{\label{fig:modwheels-cr6insspk} (Color online)
(a) Neutron-energy loss spectra on Cr$_6$. The solid lines represent simulations). The inset sketches the
formation of standing spin waves in a chain and the link to the gap in the spin-wave spectrum. (b) Simulated
energy spectra for Cr$_6$ and a hypothetical $N=6$, $s=3/2$ antiferromagnetic wheel. The observed INS
transitions and the gap in the $E$-band are indicated. Adapted from \onlinecite{Och07-cr6}. }
\end{figure}

Various derivatives of Cr$_6$ have been synthesized and studied by INS
\cite{Lar03-horseshoe,Bak11-cr6,Och07-cr6,Och08-cr6cr7}. Here the cluster
[NH$_2$R]$_3$[Cr$_6$F$_{11}$Piv$_{10}$(H$_2$O)] is considered \cite{Lar03-horseshoe}. It crystallizes in space group
$P$2$_1$/$c$, and the anion forms a string of six Cr$^{3+}$ ($s = 3/2$) ions, see Fig.~\ref{fig:modwheels-modwheels}.
Using INS the magnetic excitation spectrum up to energies of 5~meV was studied \cite{Och07-cr6}; results are shown in
Fig.~\ref{fig:modwheels-cr6insspk}(a). Three cold feature I, IV and V are observed, and two hot transitions II and III.
The data analysis yielded $2J = -1.27$~meV and $D=0$ in Eq.~(\ref{eq:shortchains-H}). The model was later refined to
$2J = -1.4$~meV, $2J_{edge} = -1.1$~meV, $D= -0.028$~meV and $|E| = 0.005$~meV, where $J_{edge}$ refers to the coupling
strengths of the outer coupling paths \cite{Och08-cr6cr7}.

\begin{figure}
\includegraphics[width=6.5cm]{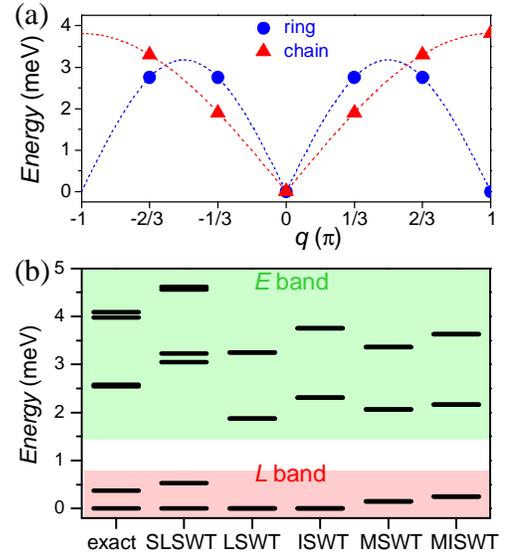}
 \caption{\label{fig:modwheels-cr6swt} (Color online)
(a) Spin-wave dispersion as obtained in linear SWT for Cr$_6$ and a hypothetical $N=6$, $s=3/2$
antiferromagnetic wheel. (b) Exact excitation energies in the $S=1$ sector of Cr$_6$ and comparison to the
results of various variants of SWT.  LSWT = linear SWT, ISWT = interacting SWT, MSWT = modified linear SWT,
MISWT = modified interacting SWT, SLSWT = spin-level SWT.  Adapted from \onlinecite{Och07-cr6} to which we
also refer for details.}
\end{figure}

In Fig.~\ref{fig:modwheels-cr6insspk}(b) are shown the simulated energy spectrum for Cr$_6$, using the
parameters of the simplified model, and for comparison the energy spectrum of a hypothetical $s= 3/2$
antiferromagnetic wheel. The $L \& E$-band picture is immediately recognized in Cr$_6$, and the observed INS
transitions demonstrate the $L$- and $E$-band states. In comparison to the wheel, however, the $E$ band
consists of two sub groups with an energy gap $\Delta_\pm$ in the $S=1$ sector. This splitting can be
associated to the formation of standing spin waves in the chain, as opposed to running waves in the wheel.
The basic argument is simple, and familiar from textbooks. In the hexa-nuclear wheel the spin waves with
shift quantum number $q$ and $-q$, corresponding to left and right running waves, are degenerate because of
cyclic symmetry. However, the open boundary in the chain or missing link acts as a perturbation lifting the
degeneracy, resulting in the formation of the symmetric and anti-symmetric linear combinations of the wave
function $|q\rangle$ and $|-q\rangle$, or standing waves indeed. The mechanism is sketched in
Fig.~\ref{fig:modwheels-cr6insspk}(a). Hence the splitting of the $E$-band transitions into peaks IV and V
observed in experiment directly demonstrates the standing spin waves in the Cr$_6$ antiferromagnetic chain.

These conclusions are supported by linear SWT \cite{And52-swt}, which can approximate the energies of the $E$-band
states in the $S=1$ sector \cite{Wal02-spindyn,Och07-cr6}. The result for the Cr$_6$ horseshoe and the same
hypothetical $N=6$, $s=3/2$ wheel as before is drawn in Fig.~\ref{fig:modwheels-cr6swt}(a), clearly demonstrating the
gap in the $E$-band spectrum because of the open-chain topology. This motivated an analysis of the exact quantum
spectrum in Cr$_6$ by different variants of SWTs and a newly suggested spin-level SWT, which adds quantum-corrections
to Eq.~(\ref{eq:wheels-HAB}) in first-order \cite{Och07-cr6}. Obviously interacting SWT does best in reproducing the
exact energies, but the accuracy is a modest 8\%. The results of a similar analysis for the CsFe$_8$ and Fe$_{18}$
antiferromagnetic wheels in Sec.~\ref{sec:lc-csfe8} are recalled \cite{Dre10-csfe8ins3,Umm12-Fe18}.

However, as mentioned before, the physics of the magnetic excitations in short antiferromagnetic chains,
although displaying a $L \& E$-band structure in the energy spectrum, presents some subtleties which are
difficult to understand \cite{Kon11-review}. In the eight membered short chain Cr$_8$Zn, for instance, a
detailed analysis of the wave functions indicated a significant mixing of the $L$- and $E$-band states
\cite{Bia09-cr8zn}.

\subsection{Spin frustration in antiferromagnetic molecular clusters}
\label{sec:lc-fe30}

In the previous sections clusters with antiferromagnetic HDVV interactions on a bipartite lattice were
discussed, but studying quantum spin-frustration effects in large clusters is obviously also of high
interest. In fact, since the possible geometrical arrangements of metal ions and ligand linkages are not
restricted by the constraint of translational invariance in "zero-dimensional" clusters, competing
interaction paths are almost always present in polynuclear magnetic molecules, and bipartite magnetic
molecules are rather the exception than the rule. However, the research on the magnetic excitations in
spin-frustrated systems concentrated on few model systems.

Regular antiferromagnetic spin triangles and lattices incorporating triangular units are most often
considered in this context. The HDVV Hamiltonian of a triangle has been given in
Eq.~(\ref{eq:sc-trimerHlin}), where for a regular triangle $s_i = s$ for all ions and $J' = J$. For $s =
1/2$, the energy spectrum consists of a doublet of two $S=1/2$ spin multiplets in the ground state, which
transforms according to the irrep $E$ of the $D_3$ symmetry group, and a higher-lying $S=3/2$ multiplet at
energy $\frac{3}{2}|2J|$. The degeneracy of the two $S=1/2$ multiplets is (often) considered as criterium for
spin frustration. However, small deviations such as distortions of the triangle leading to $J' \neq J$ or
Dzyaloshinski-Moriya interactions lift the degeneracy in the ground state opening a gap $\Delta$
(Fig.~\ref{fig:frust-v15}). The ratio $\Delta/|2J|$ may here be considered as a figure of merit.

This structure of low-lying energy levels has attracted huge interest for a variety of reasons. Spin
frustration is one of them, but corresponding clusters are also attractive models for studying the
Landau-Zener-St\"uckelberg (LZS) transitions or dissipation and decoherence in general
\cite{Lan32-LZS,Zen32-LZS,Stu32-LZS,Leg87-disspation,Dob00-decoherence,Chi00-v15buttefly}, or for
applications as qubits \cite{Wer04-v15resonant}, and Rabi oscillations were indeed observed
\cite{Ber08-v15rabi}. Many molecular tri-nuclear clusters are available and were studied, also by INS, for
their excitations. However, usually distortions are strong; the realization of regular triangles in bounded
clusters is apparently not preferred by nature. The V$_{15}$ molecule is one of the best realizations of a
regular triangle, according to its $\Delta/|2J|$, and its magnetic excitations are discussed in the next
section.

Topologies with higher nuclearity, which also incorporate triangular units and are of high symmetry, such as
the cubeoctahedron or icosidodecahedron, have also been intensely studied, however, mostly by theory
\cite{Cep05-swtfe30,Sch05-metamagnetic,Rou08-fe30,Kon09-frust,Sch01-frustreview}. Only few experimental
spectroscopic investigations of the cluster excitations are available. We will here focus on the magnetic
Keplerate molecule Fe$_{30}$, which has become a main representative in this field of research, also because
of its relationship to the Kagom\'e lattice ("Kagom\'e on a sphere")
\cite{Sch01-frustreview,Sch01-fe30rbmodel,Rou08-fe30}.

Many more molecules such as tetra-nuclear "butterfly" and hepta-nuclear disc-like molecules or odd-membered
antiferromagnetic wheels and cycles have been synthesized and studied for spin frustration effects.

\subsubsection{The V$_{15}$ molecule}
\label{sec:lc-v15}

\begin{figure}
\includegraphics[width=6.5cm]{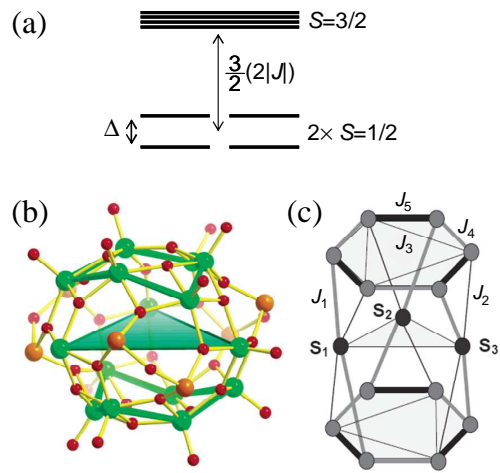}
 \caption{\label{fig:frust-v15} (Color online)
(a) Sketch of the energy level scheme of an antiferromagnetic spin-1/2 triangle. (b) Structure of the anion
of V$_{15}$. The thick green lines indicate the two hexagons, and the triangular area the spin triangle. (c)
Sketch of the exchange interactions in the V$_{15}$ molecule. Adapted from \onlinecite{Tar07-v15mag} and
\onlinecite{Cha04-v15ins2}. }
\end{figure}

The cluster [V$_{15}$As$_6$O$_{42}$(H$_2$0)]K$_6$$\cdot$8H$_2$O, or V$_{15}$ in short, contains 15 V$^{4+}$
ions ($s = 1/2$) whose arrangement can be described as two hexagons sandwiching a triangle, see
Fig.~\ref{fig:frust-v15}(a) and (b) \cite{Mue88-v15synth,Gat91-v15}. The system crystallizes in space group
$R\bar{3}c$, and the molecule exhibits a nominal crystallographic $D_3$ symmetry, however, a water molecule
sits at the center of V$_{15}$. Magnetic susceptibility demonstrated that the exchange interactions in
V$_{15}$ indicated in Fig.~\ref{fig:frust-v15}(b) are antiferromagnetic and strong [$2J_1 \approx 2J_4
\approx -13$~meV, $2J_2 \approx 2J_3 \approx -26$~meV, $2J_5 \approx -70$~meV], such that the hexagons are in
a singlet state at temperatures below ca. 100~K and do not contribute magnetic moment. At low temperatures
the magnetism can hence be described as that of a regular spin triangle with an effective antiferromagnetic
interaction of $2J = -0.211(2)$~meV \cite{Cha02-v15ins1}.

This picture was confirmed by detailed magnetization measurements at low temperatures \cite{Tar07-v15mag}.
However, a small gap in the $S=1/2$ ground-state doublet was detected. Magnetization data indicated $\Delta =
7(2)$~$\mu$eV \cite{Bar02-v15lzs} and low-frequency EPR $\Delta = 3$~$\mu$eV \cite{Kaj07-v15epr}, while from
INS a gap of $\Delta = 35(2)$~$\mu$eV was concluded \cite{Cha02-v15ins1}. The origin of the gap has been
discussed controversially, most often it has been associated to Dzyaloshinski-Moriya interactions, but also
to a distortion by e.g. the central water molecule.

\begin{figure}
\includegraphics[width=8cm]{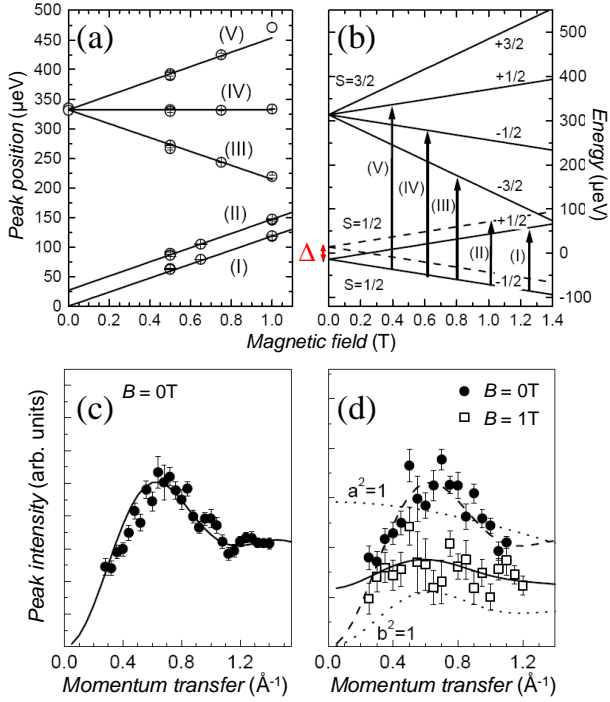}
 \caption{\label{fig:frust-v15ins}
(a) Field dependence of the INS peak energies observed in V$_{15}$ at $T = 45$~mK. Lines are fits to the
data. (b) Derived energy spectrum and assignment of INS transitions. (c) $Q$ dependence of the intensities of
peaks III + IV + V in zero field. (d) $Q$ dependence of the intensity of transitions III + IV + V at 0~T
(circles) and transition I at 1~T (squares). Lines are fits to the data. Adapted from \onlinecite{Cha04-v15ins2}. }
\end{figure}

Evidence for the latter came from an INS experiment performed on a fully deuterated polycrystalline sample of
V$_{15}$ at ultra-low temperatures and with magnetic fields applied \cite{Cha04-v15ins2}. The field
dependence of the observed INS peaks and their assignment are presented in Figs.~\ref{fig:frust-v15ins}(a)
and (b), respectively, confirming spectroscopically the expected energy spectrum. The INS energies could
excellently be fit and yielded $2J = -0.212(2)$~meV and $\Delta = 27(3)$~$\mu$eV. A careful analysis of the
magnetic field and $Q$ dependence of the INS peak intensities, shown in Figs.~\ref{fig:frust-v15ins}(c) and
(d), allowed discriminating between Dzyaloshinski-Moriya interactions and distortions. It has been concluded
that the gap $\Delta$ comes from a slightly distorted triangle, with exchange interactions $2J_{12} =
-0.21$~meV, $2J_{23} = -0.23$~meV, and $2J_{13} = -0.20$~meV.

\subsubsection{The Fe$_{30}$ Keplerate molecule}

\begin{figure}
\includegraphics[width=8cm]{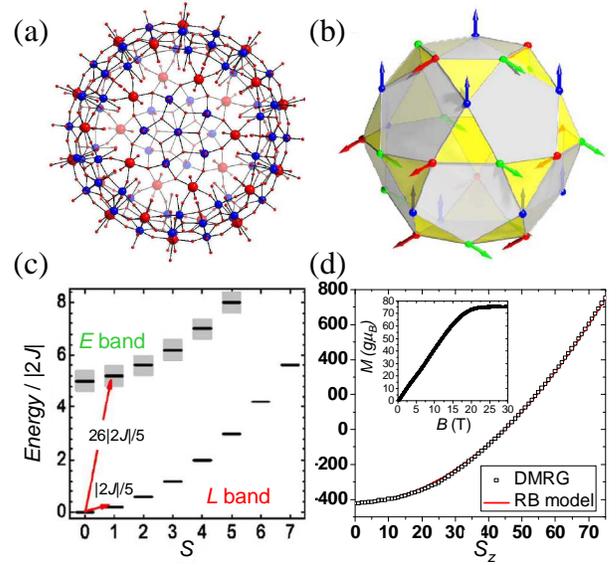}
 \caption{\label{fig:frust-fe30} (Color online)
(a) Molecular structure of Fe$_{30}$. (b) Fe$^{3+}$ core forming an icosidodecahedron. The spin orientations
in the classical ground state are represented by the arrows, the colors red, green and blue refer to the
sublattices $A$, $B$, and $C$, respectively. (c) Low-lying energy spectrum as function of total spin $S$ as predicted
in the rotational-band model Eq.~(\ref{eq:fe30-Heff}), showing the $L$ and $E$ band. (d) Lowest energy in each sector
$S_z$ as calculated by DMRG (black squares) and comparison to the $L$ band in the rotational-band model (red line). Both curves
essentially superimpose confirming the $S(S+1)$ energy dependence of the lowest eigenstates in Fe$_{30}$. The inset
shows the experimental magnetization curve. Pictures: (a,b,d) courtesy of J. Schnack, (c) adapted from
\onlinecite{Gar06-fe30ins}. }
\end{figure}

The Fe$_{30}$ molecule
[Mo$_{72}$Fe$_{30}$O$_{252}$(Mo$_2$O$_7$(H$_2$O))$_2$-(Mo$_2$O$_8$H$_2$(H$_2$O))(CH$_3$COO)$_{12}$(H$_2$O)$_{91}$]$\cdot$150H$_2$O
consists of 30 antiferromagnetically coupled Fe$^{3+}$ ions ($s = 5/2$) which are located at the vertices of
an icosidodecahedron \cite{Mue99-fe30synth}. The molecular structure is displayed in
Fig.~\ref{fig:frust-fe30}(a) and the iron metal core in Fig.~\ref{fig:frust-fe30}(b). The spin arrangement
clearly supports pronounced quantum-spin frustration effects, however, the large $s = 5/2$ spins also suggest
classical or semi-classical approaches \cite{Mue01-review}. The dimension of the Hilbert space in this
molecule is a staggering 2.2$\times$10$^{23}$ and understanding its magnetic excitation spectrum is obviously
challenging. Fe$_{30}$ has in fact become an ideal test ground for developing theoretical schemes and
physical concepts.

Classically, the antiferromagnetic ground state is characterized by three sublattices $A$, $B$, and $C$, and is highly
degenerate \cite{Axe01-fe30class}. Figure~\ref{fig:frust-fe30}(b) depicts one of the possible classical ground-state
spin configurations. For the quantum spectrum a rotational-band model with a three sublattice structure was conjectured
\cite{Sch01-fe30rbmodel}, and based on ideas as those in Sec.~\ref{sec:lc-effH} a Hamiltonian was derived,
\begin{eqnarray}
\label{eq:fe30-Heff}
 \hat{H}_{ABC} = -\frac{2J}{5} \left( \hat{\bf S}_A \cdot \hat{\bf S}_B + \hat{\bf S}_B \cdot \hat{\bf S}_C + \hat{\bf S}_C \cdot \hat{\bf S}_A
 \right),
\end{eqnarray}
where $\hat{\bf S}_A$, $\hat{\bf S}_B$, $\hat{\bf S}_C$ describe the three sublattices, with spin lengths $S_A = S_B =
S_C = 25$. The predicted low-lying energy spectrum is presented in Fig.~\ref{fig:frust-fe30}(c), and shows an $L \&
E$-band structure, similar as in the antiferromagnetic wheels (Sec.~\ref{sec:lc-cr8}). However, the degeneracies and
spatial symmetry labels are of course different. For the existence of the $L$-band in Fe$_{30}$ solid evidence came
from the experimental magnetization curve, which increases linearly with magnetic field up to saturation at a critical
field $B_c = 17.7$~T consistent with a $S(S+1)$ energy dependence of the lowest states in each spin sector [inset to
Fig.~\ref{fig:frust-fe30}(d)]. From the critical field the interaction strength has be estimated to $2J = -0.134$~meV.
The $L$ band was also produced in a density matrix renormalization group (DMRG) calculation,
Fig.~\ref{fig:frust-fe30}(d) \cite{Exl03-fe30dmrg}.

\begin{figure}
\includegraphics[width=6.5cm]{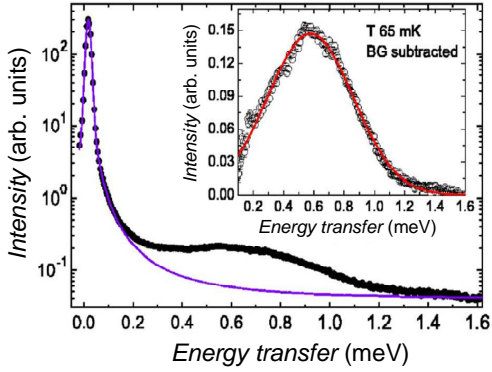}
 \caption{\label{fig:frust-fe30ins} (Color online)
Neutron-energy loss spectrum od Fe$_{30}$ at $T = 65$~mK. The solid line represents the estimated background,
and the such determined magnetic scattering is displayed in the inset. Adapted from
\onlinecite{Gar06-fe30ins}. }
\end{figure}

The higher-lying magnetic excitations in Fe$_{30}$ were probed by INS experiments on a deuterated
polycrystalline sample. An experimental spectrum is displayed in Fig.~\ref{fig:frust-fe30ins}, showing a
broad magnetic feature in the energy range 0.3-1.1~meV. The INS data could qualitatively be interpreted using
Eq.~(\ref{eq:fe30-Heff}) and associated to the $E$ band. Its excitation energy is predicted to $(26/5)|2J| =
0.67$~meV [Fig.~\ref{fig:frust-fe30}(c)] in rough agreement with the maximum in the neutron scattering
intensity (inset to Fig.~\ref{fig:frust-fe30ins}). The detailed analysis yielded $2J = -0.108$~meV. The width
of the magnetic feature however remained unexplained.

\begin{figure}
\includegraphics[width=8cm]{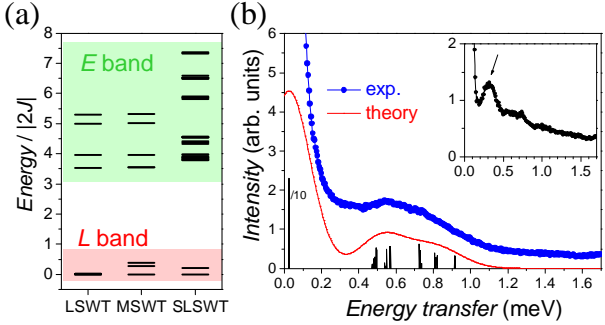}
 \caption{\label{fig:frust-fe30swt} (Color online)
(a) Low-temperature excitation spectrum of Fe$_{30}$ as predicted by linear SWT (LSWT), modified linear SWT
(MSWT), and spin-level SWT (SLSWT). (b) Experimental INS spectrum of Fe$_{30}$ (points) and simulated
spectrum using SLSWT (line). Inset: Experimental minus simulated spectrum, providing evidence for magnetic
scattering at ca. 0.3~meV which the SLSWT model does not account for. Adapted from
\onlinecite{Wal07-swtfe30}. }
\end{figure}

In order to obtain a better insight the magnetic excitation spectrum was also calculated using modified
linear SWT \cite{Cep05-swtfe30}. The predicted spectrum is shown in Fig.~\ref{fig:frust-fe30swt}(a). This
theory suggests magnetic scattering in the energy range of $3.5|2J|$-$5.5|2J|$ and hence accounts at least
partially for the observed broadening. A novel spin-level SWT, which implements a first-order quantum
correction to Eq.~(\ref{eq:fe30-Heff}), yielded an excitation spectrum in the range of $3.8|2J|-7.4|2J$
[Fig.~\ref{fig:frust-fe30swt}(a)], and allowed to reproduce most of the observed magnetic scattering using
$2J = -0.125$~meV [Fig.~\ref{fig:frust-fe30swt}(b)], consistent to within 7\% with the finding from the
magnetic data \cite{Wal07-swtfe30}. However, a peak-like magnetic scattering at ca. 0.3~meV is not
reproduced.

It was pointed out that these SWT techniques, when applied to three- (and more) partite systems, have
conceptual drawbacks \cite{Wal07-swtfe30}. The spin-level SWT was later extended to higher-orders
\cite{Sch09-approxeigenvalues}, further emphasizing this point. A fully satisfying understanding of the
magnetic excitations in Fe$_{30}$ is still lacking \cite{Gar06-fe30ins}.

%% file: chapter5_smms_v2012_10_02.tex

\section{Single-molecule magnets}
\label{sec:smm}

\subsection{Introduction}
\label{sec:smm-intro}

\begin{figure}
\includegraphics[width=6.5cm]{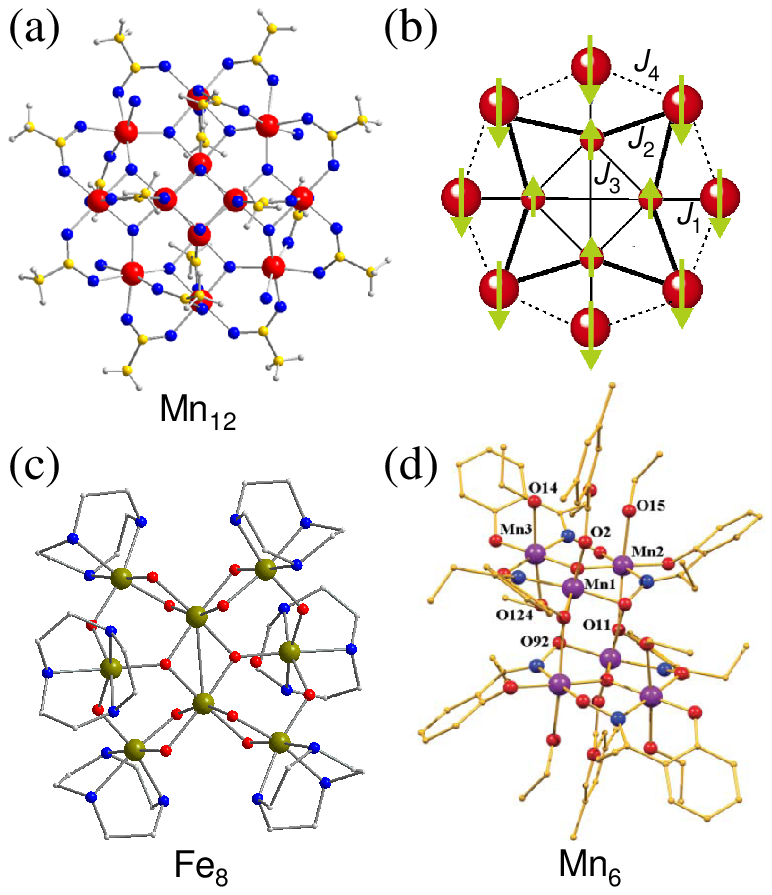}
 \caption{\label{fig:smm-smms} (Color online)
Molecular structures of three prominent SMMs,
  (a) Mn$_{12}$: [Mn$_{12}$O$_{12}$(CH$_3$COO)$_{16}$(H$_2$O)$_4$],
  (c) Fe$_8$: [Fe$_8$O$_2$(OH)$_{12}$(tacn)$_6$]Br$_8$$\cdot$9H$_2$O,
  (d) Mn$_6$: [Mn$_6$O$_2$(Et-sao)$_6$(O$_2$CPh(Me)$_2$)$_2$(EtOH)$_6$].
Panel (b) shows the metal core in the Mn$_{12}$ molecule, the exchange coupling paths, and the classical ground state
spin configuration.
}
\end{figure}

The single-molecule magnets (SMMs) are a special subclass of the molecular nanomagnets. Some prominent
examples are shown in Fig.~\ref{fig:smm-smms}. The SMMs are distinguished by exhibiting slow relaxation of
the magnetization or magnetic hysteresis at low temperatures, below a blocking temperature $T_B$. This
phenomenon is not due to a long-range ordered magnetic ground state as in conventional magnets, but arises
from an energy barrier for spin reversal at the molecular level. Furthermore, quantum phenomena such as
quantum tunneling of the magnetization are observed as characteristic steps in the magnetic hysteresis
curves. These unique magnetic properties discovered about 15 years ago stimulated enormous research and
allowed addressing fundamental questions in quantum mechanics as well as suggested applications in
information technology as classical or quantum bits or in molecular spintronics. Many authoritative reviews
and books were written, some of them are
\onlinecite{Chr00-smmreview,Chu98-QTMbook,Gat03-review,Gat06-book,Fri00-smmreview,Bog08-smmspintronics,Bar05-mn12review,Bar99-smmreview,Leu03-nvtreview}.

The magnetic spectrum in the SMMs is in principle also described by the microscopic spin Hamiltonian
Eq.~(\ref{eq:lc-microscopic}). However, the most fascinating SMM phenomena such as quantum tunneling of the
magnetization are related to the ground-state multiplet, and the anisotropy splittings in it. The effective
spin Hamiltonian, which is known as "rigid-spin" or "giant-spin" or "single-spin" model, reads then
\begin{eqnarray}
\label{eq:smm-giantH}
\hat{H} &=& \hat{H}_z + \hat{H}_{\perp}
\\&=&
 D \hat{S}_z^2 + B^0_4 \hat{ O }^0_4(S)
 \nonumber \\ &&
 + E \left( \hat{S}_x^2 - \hat{S}_y^2 \right) + B^2_4 \hat{ O }^2_4(S) + B^4_4 \hat{ O }^4_4(S),
\end{eqnarray}
where $S$ is the total spin of the ground-state multiplet. The terms were distinguished into $\hat{H}_z$ and
$\hat{H}_{\perp}$ according to whether they commute with $\hat{S}_z$ or not, and only the usually most relevant
higher-order terms were listed [terms up to 6$^{th}$ order were demonstrated in high-precision experiments
\cite{Bar07-mn12transverse}, compare also with Sec.~\ref{sec:sc-higherorders}]. In a magnetic field the Zeeman terms
$\mu_B g_z \hat{S}_z B_z$ and $\mu_B ( g_x \hat{S}_x B_x +  g_y \hat{S}_y B_y)$ have to be added to $\hat{H}_z$ and
$\hat{H}_{\perp}$, respectively. In principle the giant-spin model can be derived perturbatively from
Eq.~(\ref{eq:lc-microscopic}).

\begin{figure}
\includegraphics[width=6.5cm]{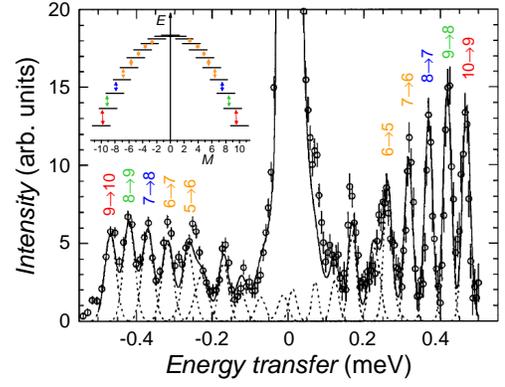}
 \caption{\label{fig:smm-fe8ins} (Color online)
INS spectrum of Fe$_8$ at 9.6~K. Circles represent the experimental data, dashed lines Gaussian fits, and
the solid line a simulation. The transitions are labeled according to $|\pm M\rangle \rightarrow |\pm M'\rangle$. The inset
shows the parabolic energy spectrum with the allowed INS transitions indicated by arrows.
 Adapted from \onlinecite{Cac98-fe8ins}
}
\end{figure}

In a SMM the uniaxial anisotropy, which is the dominant zero-field splitting term in
Eq.~(\ref{eq:smm-giantH}), has to be of easy-axis type or $D<0$. The energy levels $|M\rangle$ of the
ground-state spin multiplet organize then into a parabolic band according to $D M^2$, with the $M = \pm S$
levels being lowest and separated by an energy barrier of height $U = |D|S^2$ for integer and $U = |D|(S^2-
1/4)$ for half-integer $S$ (see inset to Fig~\ref{fig:smm-fe8ins}). This energy barrier is responsible for
the slow relaxation of the magnetization. The non-commuting terms contained in $\hat{H}_{\perp}$ induce a
mixing between the $|M\rangle$ levels, or quantum tunneling indeed.

As regards spectroscopy, the major goal is the precise determination of the parameters in the giant-spin Hamiltonian,
where in particular the non-commuting terms in $\hat{H}_{\perp}$ are of most interest, because of their relation to the
quantum tunneling rates. As the relevant transitions are intra-multiplet with $\Delta S = 0$ many experimental
techniques can accomplish the task. EPR is one of them, and high-frequency high-field EPR has indeed played a most
important role \cite{Gat06-smmeprreview}. INS has also been very valuable, adding the advantage of a zero-field
spectroscopy.

Figure~\ref{fig:smm-fe8ins} shows an INS spectrum recorded on a non-deuterated sample of the compound
[Fe$_8$O$_2$-(OH)$_{12}$(tacn)$_6$]Br$_8$$\cdot$9H$_2$O, or Fe$_8$ henceforth \cite{Cac98-fe8ins}. The
molecule contains eight Fe$^{3+}$ ions ($s = 5/2$) in a butterfly-like arrangement,
Fig.~\ref{fig:smm-smms}(c). It crystallizes in space group P1 and exhibits approximate D$_2$ molecular
symmetry \cite{Wie84-fe8synth}. The frustrated AFM interactions in the molecule result in a $S= 10$ ground
state. In the experimental INS data in Fig.~\ref{fig:smm-fe8ins} the allowed $\Delta M = \pm1$ transitions
are observed, demonstrating the zero-field splitting in this spin multiplet. The "picket-fence" INS spectrum,
which is typical for SMMs, results from the fact that the transition energies given by the dominant $D$ term
in Eq.~(\ref{eq:smm-giantH}) vary as $|D|( 2M -1)$ or linearly in $M$. Slight deviations from the regular
pattern are however seen, which relate to the other terms in Eq.~(\ref{eq:smm-giantH}), and precise values
for its five parameters could be determined \cite{Cac98-fe8ins}. Besides the scientific impact of this work,
it also showed that excellent INS data can be recorded on non-deuterated samples of molecular nanomagnets,
which established the basis for many of the subsequent INS works in this area.

Most of the SMMs which have been synthesized so far are based on 3d metal ions, and several of them were studied by
INS, but mostly for the transitions within the ground-state multiplet (for reviews see
\onlinecite{Bas03-insreview,Bir06-mn12review}). However, being related to anisotropy and not the exchange splittings,
these studies shall not be further discussed here; Fe$_8$ may serve as a representative example. The potential of INS
to also detect exchange splittings has been applied to only very few SMMs, such as Fe$_4$ \cite{Car04-fe4highins},
Fe$_8$ \cite{Car06-fe8highins}, Mn$_{12}$, and Mn$_6$, and derivates of them. In the following the situation in
Mn$_{12}$ and Mn$_6$ will be presented.

It is added that in the last few years the focus shifted from 3d-based SMMs to 3d-4f or 4f SMMs (also 5d ions became of
interest). Since 4f ions are type J and bring in significant anisotropy, while exchange interactions are weak, the
analysis of experimental data is typically more involved than in 3d-based type S/Q clusters. INS work on lanthanide
containing SMMs has just begun \cite{Klo09-cu2tb2ins,Dre12-CrDyCr}, but promises exiting results in the future.

\subsection{The Mn$_{12}$ cluster}
\label{sec:smm-mn12}

The compound [Mn$_{12}$O$_{12}$(CH$_3$COO)$_{16}$(H$_2$O)$_4$], or Mn$_{12}$ in short, was the first molecule
for which SMM behavior and quantum tunneling of the magnetization was observed below a blocking temperature
of ca. 3.5~K \cite{Gat06-book}, and became the prototype SMM. The system crystallizes in space group $I_4$,
and the molecule exhibits $S_4$ symmetry \cite{Lis80-mn12synth}. It contains an inner tetrahedral core of
four Mn$^{4+}$ ions ($s = 3/2$) and an outer ring of eight Mn$^{3+}$ ions ($s = 2$),
Fig.~\ref{fig:smm-smms}(a). The nearest-neighbor Heisenberg exchange interactions in the cluster result in a
total spin $S = 10$ ground state, whose classical spin structure can be depicted as displayed in
Fig.~\ref{fig:smm-smms}(b). The molecule's properties in its $S = 10$ ground-state multiplet have been
heavily investigated \cite{Gat06-book}, also by INS
\cite{Hen97-mn12ins,Mir99-mn12ins,Zho99-mn12ins,Bir04-mn12ins,Sie05-mn12inspressure,Wal06-mn12instimeresolved}.

However, understanding how the competing Heisenberg interactions in the molecule give rise to the $S = 10$ ground state
is also of high interest, since through the $S$-mixing effects induced by the anisotropy the higher-lying spin
multiplets can also significantly affect the $S = 10$ ground-state multiplet and tunneling rates
\cite{Car04-smmsmix,Bar07-mn12transverse}. Several attempts were made to infer the values of the exchange constant
$J_1$, $J_2$, $J_3$ and $J_4$ [see Fig.~\ref{fig:smm-smms}(b)] with controversial conclusions
\cite{Ses93-mn12,Har96-mn12couplings,Reg02-mn12couplings,Rag01-mn12couplings}.

\begin{figure}
\includegraphics[width=6.5cm]{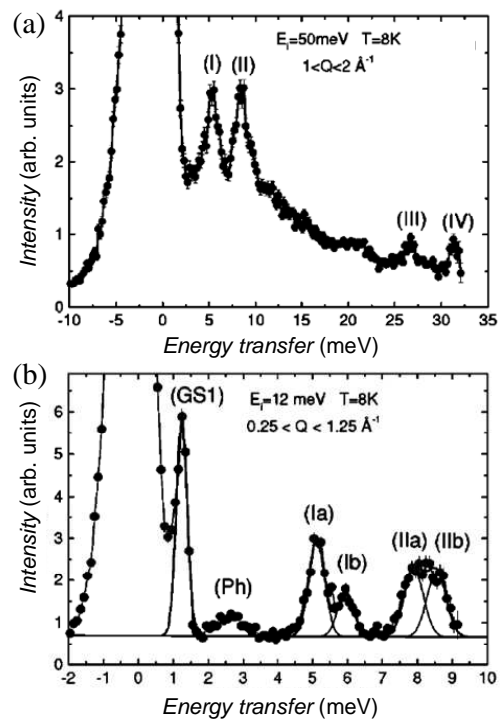}
 \caption{\label{fig:smm-mn12ins}
Neutron-energy loss spectra of Mn$_{12}$ recorded in two energy regimes.
  Data were recorded at the direct TOF spectrometer MARI at Rutherford Appleton Laboratory ISIS.
  Adapted from \onlinecite{Cha04-mn12ins}.
}
\end{figure}

\begin{figure}
\includegraphics[width=6.5cm]{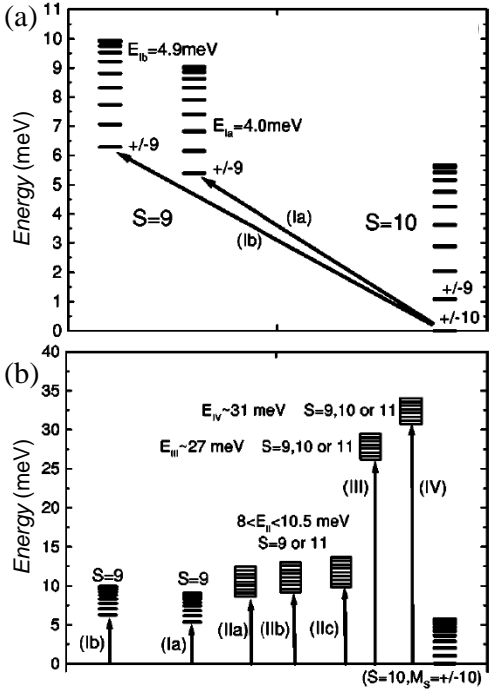}
 \caption{\label{fig:smm-mn12spk}
 Energy level scheme as derived from the experimental INS data. Panel (a) shows a zoom into the low-energy spectrum.
  Adapted from \onlinecite{Cha04-mn12ins}.
}
\end{figure}

The issue was targeted by high-energy INS experiments on a deuterated sample of Mn$_{12}$ \cite{Cha04-mn12ins}. INS
spectra are shown in Fig.~\ref{fig:smm-mn12ins}. A peak at ca. 1.2~meV is found, which originates from the
$|\pm10\rangle \rightarrow |\pm9\rangle$ transition within the $S=10$ ground-state multiplet [$D = -0.057$~meV
\cite{Bir04-mn12ins}]. Besides that, several inter-multiplet transitions reflecting the exchange splittings were
observed in the energy range up 35~meV. From a detailed analysis of the experimental data the energy level scheme
presented in Fig.~\ref{fig:smm-mn12spk} was derived. Interestingly, the anisotropy splittings in Mn$_{12}$ are not
significantly smaller than the exchange splittings, as expected in the strong-exchange limit, which can be seen in
Fig.~\ref{fig:smm-mn12spk}(a) from the fact, that the lowest level of the lowest $S=9$ multiplet falls below the top of
the $S=10$ ground-state multiplet. Hence, $S$ mixing plays a relevant role, which was later confirmed in a
high-precision EPR experiment on the $S=10$ ground state \cite{Bar07-mn12transverse}. Using numerical large-scale
calculations, the exchange coupling constants were determined to $2J_1 = -5.79$~meV, $2J_2 = -5.33$~meV, $2J_3 =
-0.67$~meV and $2J_4 = - 0.48$~meV.

The exchange couplings $J_3$ and $J_4$ are one order of magnitude smaller than $J_1$ and $J_2$, and spin frustration is
hence small in the Mn$_{12}$ molecule, which justifies the interpretation of the $S=10$ ground state in terms of the
classical spin configuration shown in Fig.~\ref{fig:smm-smms}(b) and suggests a topological relation to AFM wheels
\cite{Wal05-gridreview}. In the context of Sec.~\ref{sec:lc-cr8}, this implies that the $L \& E$-band concept is also
realized in the Mn$_{12}$, which was underlined by a successful linear SWT analysis of the INS excitation spectrum
\cite{Cha04-mn12ins}.

\subsection{The Mn$_{6}$ clusters}
\label{sec:smm-mn6}

The SMM [Mn$_6$O$_2$(Etsao)$_6$(O$_2$CPh(Me)$_2$)$_2$(EtOH)$_6$], or Mn$_6$ henceforth, which has been synthesized 14
years after the initial discovery of SMM behavior in Mn$_{12}$ \cite{Ses93-mn12}, was yet the first molecule found to
exhibit a higher anisotropy barrier than that in Mn$_{12}$ \cite{Mil07-mn6recordaniso}. It contains six Mn$^{3+}$ ions
($s = 2$), which are arranged into two triangles linked by oxygen atoms, see Fig.~\ref{fig:smm-smms}(d). It
crystallizes in space group P2$_1$/n and the molecule exhibits C$_{2h}$ symmetry \cite{Mil07-mn6recordaniso}. Several
derivates of this molecule have been synthesized and a magneto-structural correlation for the exchange couplings
established \cite{Mil07-mn6magnetostruct}.

Magnetic measurements demonstrated a $S = 12$ ground state due to overall FM exchange interactions in the
cluster. The zero-field splitting parameter in the $S=12$ ground state was estimated to $D \approx
-0.054$~meV, corresponding to an energy barrier of $|D|S^2 \approx 7.7$~meV. The exchange couplings were
however determined to $2J \approx 0.4$~meV, hence suggesting that in Mn$_6$ $S$ mixing is very strong and the
strong-exchange limit or giant-spin model break down.

\begin{figure}
\includegraphics[width=6.5cm]{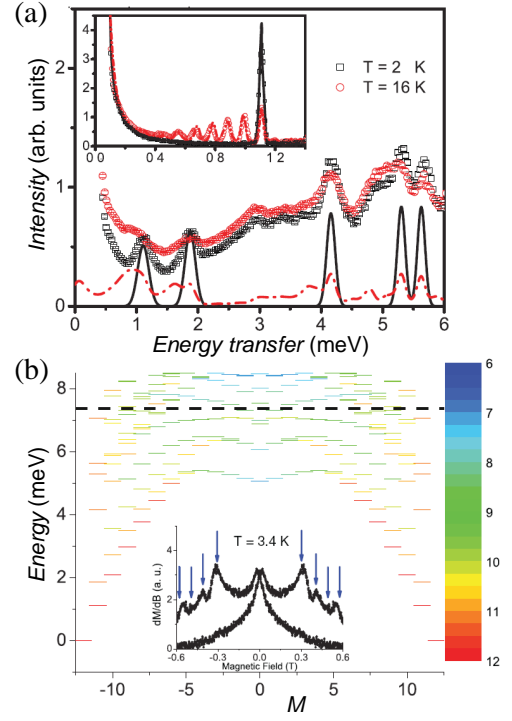}
 \caption{\label{fig:smm-mn6isnspk} (Color online)
(a) Neutron-energy loss spectra in Mn$_6$ recorded for two energy regimes up to 6~meV at the indicated temperatures
(symbols). The lines represent simulations using Eq.~(\ref{eq:smm-mn6H}) and the parameters given in
the text. (b) Calculated energy spectrum as function of $M$. The color represents the value of the
expectation value $\langle \hat{\bf S}^2 \rangle$.
  Adapted from \onlinecite{Car08-mn6ins}.
}
\end{figure}

This has been indeed convincingly confirmed in an INS experiment on a non-deuterated sample of Mn$_6$
\cite{Car08-mn6ins}. Experimental INS data recorded in two energy windows are presented in
Fig.~\ref{fig:smm-mn6isnspk}(a). At low energies, at 1.1~meV and below, the "picket-fence" spectrum characteristic for
the transitions within the ground-state multiplet of a SMM is observed, see inset to Fig.~\ref{fig:smm-mn6isnspk}(a).
However, at already slightly higher energies (1.9~meV) the first inter-multiplet transition appear, followed by several
more in the experimental energy range of 6~meV. This directly points to the fact, that several higher-lying spin
multiplets are partially nested with the ground-state multiplet.

\begin{figure}
\includegraphics[width=6.5cm]{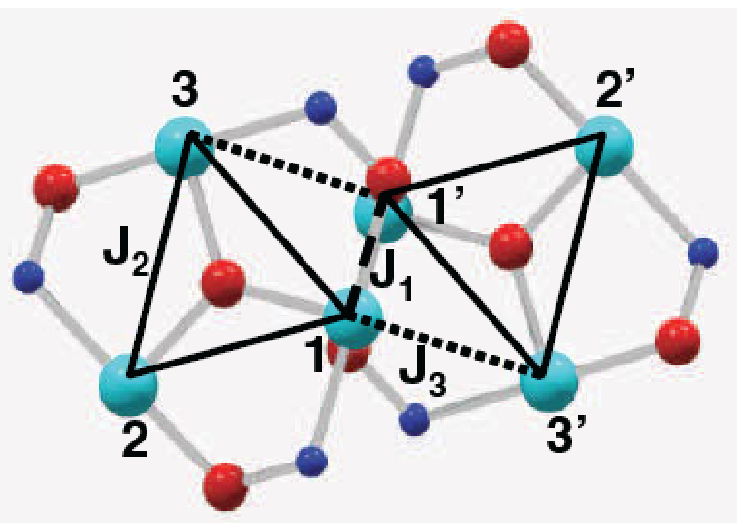}
 \caption{\label{fig:smm-mn6core} (Color online)
The core of Mn$_6$ with the labeling of the Mn$^{3+}$ sites and the assumed coupling paths indicated.
  Adapted from \onlinecite{Car08-mn6ins}.
}
\end{figure}

The data analysis was based on the Hamiltonian
\begin{eqnarray}
\label{eq:smm-mn6H}
 \hat{H} &=& -2 \sum_{i<j} J_{ij} \hat{\bf s}_i \cdot \hat{\bf s}_j
   + \sum_i  \left[ D_i \hat{s}_{iz}^2 + B^0_{4i} \hat{ O }^0_4(s_i) \right],\quad
\end{eqnarray}
where three exchange coupling paths $J_1$, $J_2$, $J_3$ were assumed as displayed in Fig.~\ref{fig:smm-mn6core}. The
ligand-field parameters for sites 1 and 1', 2 and 2', and 3 and 3' are identical by symmetry, those of sites 2, 3, 2',
3' were assumed to be equal, and $B^0_{4,1}/B^0_{4,2} = D_1/D_2$ employed. The experimental data could be excellently
reproduced with the best-fit parameters $2J_1 = 0.84(5)$~meV, $2J_2 = 0.59(3)$~meV, $2J_3 = -0.01(1)$~meV, $D_1 =
-0.20(1)$~meV, $D_2 = -0.76(2)$~meV, and $B^0_{4,1} = -0.0010(3)$~meV. The resulting simulated energy spectrum is shown
in Fig.~\ref{fig:smm-mn6isnspk}(b), which demonstrates that the energies of the $|M\rangle$ states of the $S=12$
ground-state multiplet do not follow the generic parabolic $M^2$ curve but deviate strongly from it at higher energies,
and that indeed several higher-lying $S=11$ multiplets fall below the top of the $S=12$ multiplet. In
\onlinecite{Car08-mn6ins} it was further demonstrated that this unique structure of the energy spectrum has significant
effects on the magnetic relaxation rates.

%% file: chapter6_quantumspins_v2012_10_02.tex

\section{Quantum spin systems}
\label{sec:qss}

\subsection{Introduction}
\label{sec:qss-intro}

Quantum spin systems have been attracting much attention due to numerous magnetic features which cannot be interpreted
by conventional spin models \cite{Sac08}. In particular, classical magnetic phases such as ferromagnetism and N\'eel
antiferromagnetism are prevented by (strong) quantum fluctuations, which are present in magnetic compounds if some of
the following conditions are fulfilled:

(i) The spin quantum number of the magnetic ions is low, i.e., $s_i = 1/2$ or 1.

(ii) The dimensionality of the magnetic system is low, i.e., $d = 1$ or 2.

(iii) The connectivity of the network of magnetic ions (i.e., the number of spins to which each spin is coupled) is
low.

(iv) The interactions between the magnetic ions are geometrically frustrated.

Representatives of the categories (i) and (ii) have been the subject of detailed investigations for a long time. They
include one-dimensional magnets such as KCuF$_3$ \cite{Lak05,Hut79} and two-dimensional magnets such as La$_2$CuO$_4$
\cite{Hea10,Col01,Vak87}, the latter being of tremendous interest due to the observation of doping-induced
superconductivity \cite{Bed86}. More recently, novel materials have been synthesized which are formed by two and three
magnetically coupled spin chains, e.g., two-leg spin ladders such as SrCu$_2$O$_3$ \cite{Azu94} and
(C$_5$H$_{12}$N)$_2$CuBr$_4$ \cite{Rue08,Thi09} as well as three-leg spin ladders such as Sr$_2$Cu$_3$O$_5$
\cite{Azu94}. Typical realizations of the category (iii) are weakly coupled dimer-based compounds including KCuCl$_3$
\cite{Kat98,Cav99}, TlCuCl$_3$ \cite{Oos02,Rue03}, NH$_4$CuCl$_3$ \cite{Kur99,Rue04a}, BaCuSi$_2$O$_6$
\cite{Jai04,Seb05,Rue07}, SrCu$_2$(BO$_3$)$_2$ \cite{Kod02,Kag00}, Cs$_3$Cr$_2$Br$_9$ \cite{Leu85,Gre04},
Sr$_3$Cr$_2$O$_8$ \cite{Qui10,Wan11}, and Ba$_3$Mn$_2$O$_8$ \cite{Uch02,Sto08}. Trimer-based compounds such as
La$_4$Cu$_3$MoO$_{12}$ \cite{Qiu05} as well as the compound SrCu$_2$(BO$_3$)$_2$ (characterized by an array of mutually
perpendicular dimers) fulfill in addition the criterion (iv).

According to the topic of the present work, we will consider here only compounds for which the presence of magnetic
clusters is the most important ingredient to understand their quantum spin properties. The following sections will
therefore focus firstly on weakly interacting AFM dimer systems, for which quantum phase transitions involving gapless
excitations and Bose-Einstein condensation have been observed most convincingly \cite{Gia08}. Secondly, the phenomenon
of a spin-Peierls transition will be described that occurs in quasi one-dimensional (1D) antiferromagnets due to the
formation of spin pairs as a result of dimerization of the regular array of magnetic ions \cite{Bra75}. Thirdly, the
formation of magnetic polarons observed in transition-metal perovskites will be discussed which evolve upon hole doping
and behave like magnetic clusters embedded in a non-magnetic matrix \cite{Phe06}.

\subsection{Dimer-based antiferromagnets}
\label{sec:qss-dimers}

\begin{figure}
\includegraphics[width=8cm]{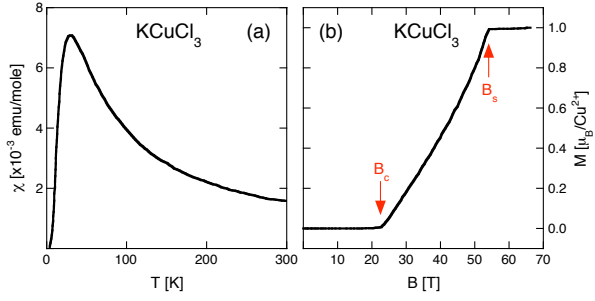}
  \caption{\label{fig:qss-fig1}
(a) Magnetic susceptibility measured for KCuCl$_3$ with $B=0.5$~T oriented perpendicular to the cleavage plane
(1,0,-2). Adapted from \onlinecite{Tan97}. (b) Magnetization measured for KCuCl$_3$ at $T=1.7$~K with $B$ perpendicular to the cleavage plane. The
saturation field $B_s \approx 60$~T was outside the available field range.
 Adapted from \onlinecite{Oos02}.
}
\end{figure}

\begin{figure}
\includegraphics[width=8cm]{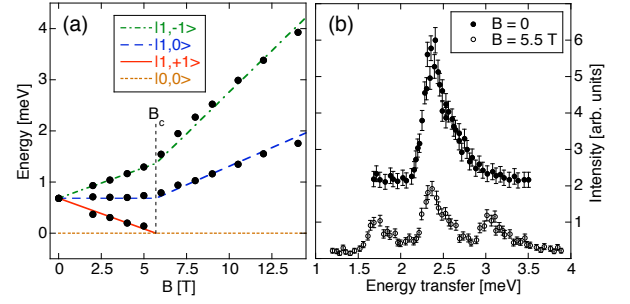}
  \caption{\label{fig:qss-fig2} (Color online)
Selected results obtained from INS experiments on TlCuCl$_3$. (a) Field dependence of the magnetic
excitation energies measured at ${\bf Q}=(0,4,0)$. The solid lines reflect a linear Zeeman model. The critical field is
$B_c=5.7$~T.  Adapted from \onlinecite{Rue03}. (b) Splitting of the singlet-triplet excitation measured at ${\bf Q}=(-0.5,0,2)$ and $T=1.5$~K. The
asymmetric line shapes are typical resolution effects.
 Adapted from \onlinecite{Cav02}.
}
\end{figure}

In dimer-based compounds the two magnetic ions are antiferromagnetically coupled according to the spin Hamiltonian
Eq.~(\ref{eq:sc-dimerH}). In contrast to isolated dimer systems discussed in Sec.~\ref{sec:sc}, the coupling between
the dimers cannot be neglected, but the intra-dimer exchange coupling $J_0$ is larger than the coupling between the
dimers $J_1 =J_0/\gamma$. As long as $\gamma \gg 1$, the thermodynamic magnetic properties resemble those of isolated
dimer systems, i.e., the dimer ground state $|S,M\rangle$ is a singlet $|0,0\rangle$, and so preserves full rotational
invariance, unlike the N\'eel state of a square lattice antiferromagnet with $\gamma =1$. The singlet state is
separated from the excited triplet states $|1,+1\rangle$, $|1,0\rangle$ and  $|1,-1\rangle$ by an energy gap $\Delta =2
|J_0|$. This picture is confirmed by the magnetic susceptibility measured for the dimer compound KCuCl$_3$ \cite{Tan97}
as shown in Fig.~\ref{fig:qss-fig1}(a), which readily allows the determination of the gap energy, similar to the
analysis of Fig.~\ref{fig:sc-crocrchi}. However, the high-field magnetization data \cite{Oos02} displayed in
Fig.~\ref{fig:qss-fig1}(b) are basically different from the case observed for isolated cluster systems. The latter
exhibit sharp step-like enhancements of the magnetization (see Figs.~\ref{fig:sc-ni4mag} and \ref{fig:wheels-fe10mag}),
which are absent for KCuCl$_3$. Instead, the magnetization increases rather slowly above the critical field $B_c
\approx 23$~T, and the saturation moment of 1~$\mu_B$/Cu$^{2+}$ is reached at the saturation field $B_s \approx 53$~T
far above $B_c$. This is due to the inter-dimer coupling $J_1$ giving rise to energy dispersion of the singlet-triplet
excitations which may be called triplons. The triplon dispersion can be calculated in the random phase approximation
\cite{Jen91}, since perturbation theory in $1/\gamma$ is well defined. For the case of dimers forming a square lattice
we find
\begin{eqnarray}
\label{eq:qss-eq1}
E_M( {\bf k} ) &=& -2 J_0 - 2 J_1 \left[ \cos(k_x d) + \cos( k_y d) \right] + g \mu_B B M,
\nonumber \\ &&
\end{eqnarray}
where ${\bf k} = ( k_x, k_y)$ is the quasiparticle wave vector, $d$ the lattice constant, $B$ an external magnetic
field, and $M = 0, \pm1$. In zero field the three triplons are degenerate. With increasing field the triplons are split
into three branches with $M = +1, 0, -1$ as shown in Fig.~\ref{fig:qss-fig2}(a). At a critical magnetic field $B_c$,
the energy of the lowest triplet component $|1,+1\rangle$ intersects the ground-state singlet $|0,0\rangle$ and the
ground state changes, thus $B_c$ is a quantum critical point separating a gapped spin-liquid state ($B < B_c$) from a
field-induced magnetically ordered state ($B > B_c$). The triplet components $|1,+1\rangle$ can be regarded as bosons
with hard-core repulsion, thus Bose-Einstein condensation occurs at the quantum critical point $B_c$, i.e., the gas of
triplet bosons undergoes a phase transition into a novel condensate state with macroscopic occupation of the
single-particle ground-state.

The state of an individual dimer at position ${\bf r}$ is well approximated by a linear combination of the singlet
$|0,0\rangle$ and the triplet $|1,+1\rangle$,
\begin{eqnarray}
\label{eq:qss-eq2}
|\Psi\rangle = u |0,0\rangle - v \exp[i({\bf k} \cdot {\bf r} - E_{+1} t)] |1,+1\rangle,
\end{eqnarray}
where $E_{+1}$ is defined by Eq.~(\ref{eq:qss-eq1}), and the amplitudes $u$ and $v$ depend on the magnetic field
\cite{Mat02,Mat04}. The expectation values of the spin operator components of the dimer in the state
Eq.~(\ref{eq:qss-eq2}) are
\begin{eqnarray}
\label{eq:qss-eq3}
\langle \hat{s}_{1,x} \rangle &=& -\langle \hat{s}_{2,y} \rangle \propto  \frac{u v}{2} \cos({\bf k} \cdot {\bf r} - E_{+1} t), \\
\langle \hat{s}_{1,y} \rangle &=& -\langle \hat{s}_{2,x} \rangle \propto \frac{u v}{2} \sin({\bf k} \cdot {\bf r} - E_{+1} t), \\
\langle \hat{s}_{1,z} \rangle &=& -\langle \hat{s}_{2,z} \rangle \propto \frac{v^2 }{2}.
\end{eqnarray}
The condensate at $B>B_c$ can be associated with the transverse order parameters $\langle \hat{s}_{i,x} \rangle$ and
$\langle \hat{s}_{i,y} \rangle$ which are oppositely aligned at the dimer sites 1 and 2 due to the AFM dimer coupling
$J_0$. The rotational symmetry $O(2)$ of the underlying spin Hamiltonian is spontaneously broken for $B>B_c$, giving
rise to a dramatic change in the nature of the magnetic excitation spectrum. According to Eq.~(\ref{eq:qss-eq1}), the
mode associated with the lowest triplet $|1,+1\rangle$ exhibits a quadratic dispersion around the wave vector ${\bf
k}_0= (\pi/a,\pi/a)$ and becomes gapless at the critical field $B_c$, but for $B>B_c$ it transforms in a fully
isotropic system into a gapless Goldstone mode with a sound-like, linear dispersion
\begin{eqnarray}
\label{eq:qss-eq4}
E( {\bf k} ) &=& \hbar s | {\bf k}- {\bf k}_0|
\end{eqnarray}
where $s$ is a velocity \cite{Mat02,Mat04}.

\begin{figure}
\includegraphics[width=8cm]{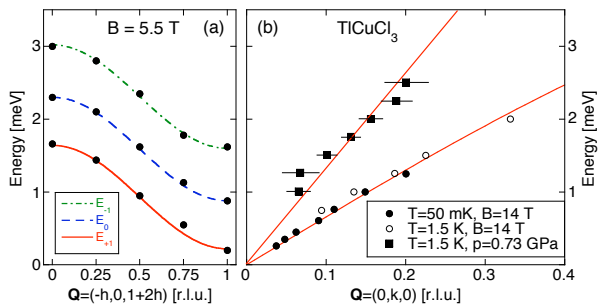}
  \caption{\label{fig:qss-fig3} (Color online)
(a) Energy dispersion of the triplons observed in TlCuCl$_3$ at $T=1.5$~K and $B=5.5$~T. The lines correspond to model
expectations based on a three-dimensional extension of Eq.~(\ref{eq:qss-eq1}). Adapted from \onlinecite{Cav02}.
(b) Energy dispersion of the low-lying
magnetic excitations observed in TlCuCl$_3$ at different temperatures, fields and pressures. The lines denote the
linear behavior of the Goldstone mode according to Eq.~(\ref{eq:qss-eq4}).
 Adapted from \onlinecite{Rue03,Rue04a}.
}
\end{figure}

One of the most widely studied dimer-based antiferromagnets is the monoclinic compound TlCuCl$_3$ in which the
Cu$^{2+}$ ions are arranged in centro-symmetric pairs. The intra-dimer coupling $J_0=-2.7$~meV dominates the
inter-dimer couplings $|J_1| < 1$~meV. INS experiments confirmed the singlet-triplet nature of the magnetic excitations
by applying an external magnetic field as visualized in Fig.~\ref{fig:qss-fig2} \cite{Rue03}. For $B>0$ the
singlet-triplet excitation is clearly split into three lines due to the Zeeman effect, with the central  $\Delta M=0$
line being twice as intense as the $\Delta M= \pm1$ side lines as predicted by Eq.~(\ref{eq:basics-ins-S}). The
triplons are well characterized by a three-dimensional extension of Eq.~(\ref{eq:qss-eq1}) with a quadratic dispersion
around ${\bf Q}=(0,0,1)$ as shown in Fig.~\ref{fig:qss-fig3}(a). The spin gap has a minimum value $\Delta \approx
0.70$~meV at the zone center $(0,0,1)$. With increasing field strength, the lowest triplet state $|1,+1\rangle$ reduces
its energy continuously and overcomes the spin gap for a critical field $B_c=5.7$~T, see Fig.~\ref{fig:qss-fig2}(a).
Since a finite number of triplet states $|1,+1\rangle$ is created at $B_c$, the system undergoes a phase transition to
a magnetically ordered state. This condensation of triplet states at $T \approx 0$ is therefore a prototype of a
quantum phase transition.

What is the experimental proof that TlCuCl$_3$ is Bose-Einstein condensed at $B_c$? Important evidence is provided by
the critical exponent $\phi$ in the field dependence of the critical temperature, $T_c(B) \propto (B_c - B)^\phi$. The
theory for a three-dimensional Bose gas predicts a universal value $\phi =2/3$ (see, e.g. \onlinecite{Gia99}), which
was experimentally confirmed, e.g., for the dimer compounds BaCuSi$_2$O$_6$ \cite{Jai04} and TlCuCl$_3$ \cite{Tan07} in
some temperature range. The ultimate proof for the Bose-Einstein condensation in TlCuCl$_3$, however, is offered by the
properties of the magnetic excitation spectrum above $B_c$ according to Eq.~(\ref{eq:qss-eq4}). This theoretical
prediction was observed for the first time by INS experiments in TlCuCl$_3$ as shown in Fig.~\ref{fig:qss-fig3}(b)
\cite{Rue03}, thus the presence of a spin-wavelike mode with a linear dispersion is a convincing signal for the
existence of the Bose-Einstein condensate.

The singlet-triplet gap of TlCuCl$_3$ can also be closed by the application of hydrostatic pressure (thereby modifying
the parameter $\gamma$) which occurs at a critical pressure $p_c \approx 0.1$~GPa. In contrast to the field induced
case, all the triplet components can condense into the singlet ground state at the quantum critical point $p_c$. The
magnetic excitation spectrum has again the nature of a gapless Goldstone mode \cite{Mat04} which was experimentally
confirmed in the pressure-induced ordered phase as shown in Fig.~\ref{fig:qss-fig3}(b) \cite{Rue04b}. In later INS
experiments, it was demonstrated that only the longitudinal and one transverse triplet component soften at $p_c$,
whereas the other transverse triplet component retains a finite gap at $p_c$ \cite{Rue08}. For $p>p_c$, the gap
energies of both transverse components remain constant, whereas that of the longitudinal component gradually increases.
The data could consistently be interpreted by using a mainly linear pressure dependence of the exchange parameters as
well as a tiny exchange anisotropy. The gap energy of the longitudinal mode increases with the ordered magnetic moment
above $p_c$, with a fundamental ratio of $\sqrt 2$ between the gaps in the ordered and disordered states, thereby
providing a non-trivial experimental test of the $S_\varphi$ field theory \cite{Sac11}. Such an amplitude mode is not
present in a classical description of ordered magnets, but is a direct consequence of the underlying quantum
criticality.

Up to the present, the compound TlCuCl$_3$ has remained a prototype of a quantum antiferromagnet in which evidence for
the Bose-Einstein condensation was given by many techniques. The concept of a Bose-Einstein condensation has been
applied to some other dimer based compounds mentioned in Sec.~\ref{sec:qss-intro} as well, but often factors such as
the large values of the exchange parameters as well as the presence of anisotropies violating the rotational symmetry
prevent the complete softening of the lowest triplet component $|1,+1\rangle$ at $B_c$. The influence of anisotropies,
which can be determined rather precisely by EPR techniques (see, e.g., \onlinecite{Kol04}), was discussed elsewhere
\cite{Gia08}.

\begin{figure}
\includegraphics[width=8cm]{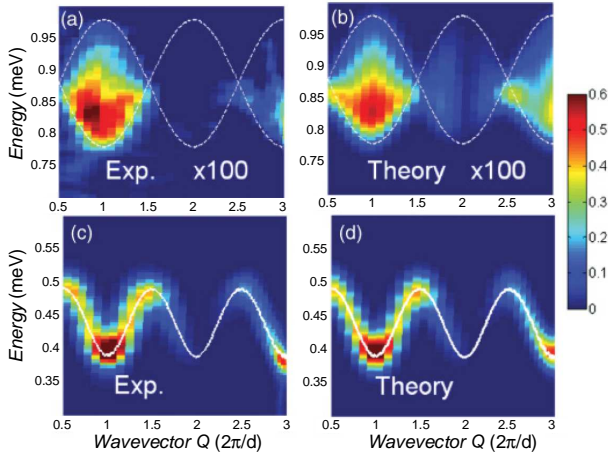}
  \caption{\label{fig:qss-fig10} (Color online)
Excitation spectra of copper nitrate at $T=0.12$~K. Panels (a) and (c) show background-subtracted two- and one-magnon
INS data, respectively, while panels (b) and (d) show the simulated $T=0$ spectra. Adapted from \onlinecite{Ten12}. }
\end{figure}

Dimerized antiferromagnetic chain systems are closely related to the ACuCl$_3$ (A = K, Tl) compounds discussed above. A
well studied example is copper nitrate, Cu(NO$_3$)$_2$$\cdot$2.5(D$_2$O), in which the spin-1/2 Cu$^{2+}$ ions are
arranged as chains of copper pairs. Each pair has a singlet ground state separated from a triplet at $|2 J_0| =
0.44$~meV. The weak interdimer coupling $J_1=J_0/\gamma$ with $\gamma \approx 4$ yields triplet excitations that
propagate coherently along the chain \cite{Xu00}. The experimental data displayed in Fig.~\ref{fig:qss-fig10} are well
described by the one-dimensional variant of Eq.~(\ref{eq:qss-eq1}) with ${\bf k} =( k_x , 0 )$ where $x$ denotes the
chain direction. Above the singlet-triplet gap there is a second gap to the multimagnon continuum. Two-magnon bound
states with $S=1$ are visible in INS experiments \cite{Ten03,Ten12} and have a dispersion \cite{Uhr96,Sch03}
\begin{eqnarray}
\label{eq:qss-eq4a}
E_{BS}({\bf k}) = -2 J_0 - \frac{J_1}{2} [ 1 + 4 \cos^2(k_x d)],
\end{eqnarray}
as shown in Fig.~\ref{fig:qss-fig10}. These $S=1$ bound states exist only over the range $|n \pi - k_x d| \leq \pi /3$,
where $n$ is an odd integer.

\subsection{Spin-Peierls dimerization}
\label{sec:qss-spinpeierls}

For a compound built up by identical atoms the elastic energy is lowest if the atoms are equally spaced. However, there
are compounds where the atoms move from an equally-spaced crystal to one in which the spacing alternates, i.e., the
atoms form pairs. This is called dimerization (first proposed for one-dimensional systems by Peierls in the 1930s in a
textbook entitled "Quantum theory of solids"), which is made possible through lowering of the free energy of the
electronic subsystem by this maneuver. The early examples mainly included polymer-like organic materials, characterized
by antiferromagnetic Heisenberg chains, being first discovered in the compound TTF-CuS$_4$C$_4$(CF$_3$)$_4$
\cite{Bra75} and afterwards in MEM-(TCNQ)$_2$ \cite{Hui79} and in DEM-(TCNQ)$_2$ \cite{Sch82}. Here we exemplify the
phenomenon of a spin-Peierls transition for the linear-chain compound CuGeO$_3$ \cite{Nis94}, followed by a discussion
of the three-dimensional compound CeRu$_2$Al$_{10}$, in which recently a dimerization has been suggested to occur
\cite{Rob10}.

\begin{figure}
\includegraphics[width=8cm]{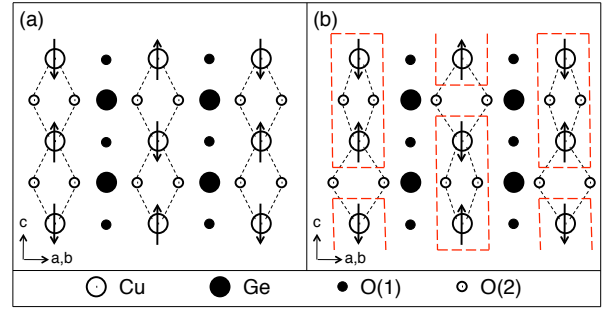}
  \caption{\label{fig:qss-fig4} (Color online)
Schematic representation of the structure of CuGeO$_3$. The arrows denote the AFM spin alignment of the Cu$^{2+}$ ions.
The dashed lines mark the Cu-O(2)-Cu superexchange bonds. (a) $T>T_c=14$~K. (b) Copper dimerization for $T<T_c$
indicated by the dashed boxes.
}
\end{figure}

CuGeO$_3$ is a linear Cu$^{2+}$ ($s_i=1/2$) chain compound crystallizing in the orthorhombic space group $Pbmm$ with
$a=4.81$~{\AA}, $b=8.47$~{\AA}, and $c=2.941$~{\AA} at room temperature \cite{Vol67}. The coupling of the Cu$^{2+}$
ions positioned at (1/2,0,0) is strong along the $c$ axis giving rise to physical properties typical of a
one-dimensional magnet. The magnetic susceptibility rapidly drops to zero below $T_c=14$~K \cite{Has93}, which points
to the opening of a finite energy gap associated with a singlet ground state. This can be understood in terms of
lattice dimerization, i.e., the formation of Cu$^{2+}$ pairs resulting in a spin-Peierls ground state. Neutron
diffraction experiments indeed gave evidence for Cu-Cu dimerization along the c axis below $T_c$ with an interatomic
separation of 2.926~{\AA}, compared to 2.941~{\AA} in the high-temperature structure \cite{Hir94}, accompanied by
shifts in the position of the O(2) ions in the $(a,b)$ plane as illustrated in Fig.~\ref{fig:qss-fig4}. Since the
coupling between two Cu$^{2+}$ ions in the $c$ direction is mainly due to a superexchange interaction through O(2), the
dimerization is clearly driven by the O(2) shifts, which results in two unequal and alternating exchange parameters
along the $c$ direction:
\begin{eqnarray}
\label{eq:qss-eq5}
J_{1,2}(T) = J_c \left[ 1 \pm \delta(T) \right].
\end{eqnarray}

\begin{figure}
\includegraphics[width=8cm]{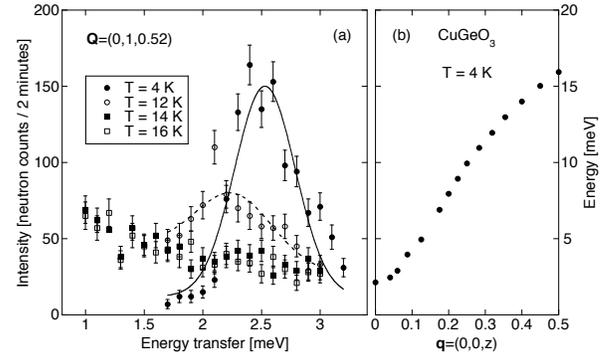}
  \caption{\label{fig:qss-fig5}
(a) Energy scan profiles of the magnetic excitation observed for CuGeO$_3$ at ${\bf Q} =(0,1,0.52)$ for various
temperatures. The full and dashed lines denote Gaussian fits to the data obtained at $T=4$ and 12~K,
respectively. (b) Dispersion of the magnetic excitation observed in CuGeO$_3$ along the $c$ direction at $T=4$~K.
 Adapted from \onlinecite{Nis94}.}
\end{figure}

The existence of the spin-Peierls gap was verified by INS experiments as shown in Fig.~\ref{fig:qss-fig5}(a)
\cite{Nis94}. At $T=4$~K a sharp peak corresponding to the singlet-triplet transition appears at an energy
transfer of 2.5~meV for ${\bf Q}=(0,1,0.52)$ which is close to the Brillouin zone center $(0,1,0.5)$. With
increasing temperature the peak moves to lower energies and broadens ($T=12$~K), while at $T=T_c=14$~K the
energy gap vanishes as expected. The dispersion of the singlet-triplet transition is shown in
Fig.~\ref{fig:qss-fig5}(b). The energy at the zone boundary ($z=1/2$) yields directly the intra-chain
exchange parameter $J_c=10.4$~meV from the  formula $\pi J_c/2 = 16.3$~meV derived by Des Cloizeaux and
Pearson \cite{Clo62}. This information is useful to derive the value of $\delta(T)$ in
Eq.~(\ref{eq:qss-eq5}). The mean-field theory gives
\begin{eqnarray}
\label{eq:qss-eq6}
 \delta(T) = \frac{ \Delta(T) }{ p J_c },
\end{eqnarray}
with $p=1.637$ \cite{Bra75}. Substituting $\Delta(0)=2.1$~meV by extrapolation of the gap energy $\Delta(4K)=2.5$~meV
[see Fig.~\ref{fig:qss-fig5}(a)] to zero temperature yields $\delta(0)=0.12$ and $J_1(0)/J_2(0)=1.27$ from
Eq.~(\ref{eq:qss-eq5}).

\begin{figure}
\includegraphics[width=5cm]{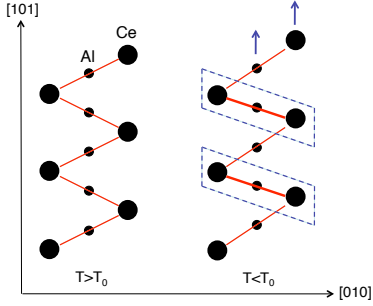}
  \caption{\label{fig:qss-fig6} (Color online)
Schematic representation of the Ce-Al zig-zag chains in CeRu$_2$Al$_{10}$. The Ce$^{3+}$ dimerization for $T<T_0$ is
indicated by the dashed boxes.
}
\end{figure}

Recently the ternary compound CeRu$_2$Al$_{10}$ has attracted much attention because of a phase transition
taking place at $T_0=27$~K whose origin remained unclear initially \cite{Str09}. CeRu$_2$Al$_{10}$
crystallizes in the orthorhombic space group $Cmcm$ in which the Ce$^{3+}$ ions are separated from each other
by an exceptionally large distance of 5.2~{\AA}, so that the interpretation of $T_0$ as a magnetic phase
transition has to be discarded. Alternative mechanisms such as a charge-ordered state as well as
spin-density-wave formation also have serious shortcomings \cite{Nis09,Mat09}. \cite{Tan10a,Tan10b} suggested
the formation of Ce dimers within the $(a,c)$ plane, bearing some similarities to a spin-Peierls transition.
Neutron scattering experiments confirmed the latter interpretation \cite{Rob10}. Neutron diffraction
measurements gave evidence for a displacement of the Al atoms below $T_0$, resulting in a spin-Peierls
transition along the one-dimensional Ce-Al zig-zag chains accompanied by a dimerization of the Ce$^{3+}$ ions
as sketched in Fig.~\ref{fig:qss-fig6}.

\begin{figure}
\includegraphics[width=8cm]{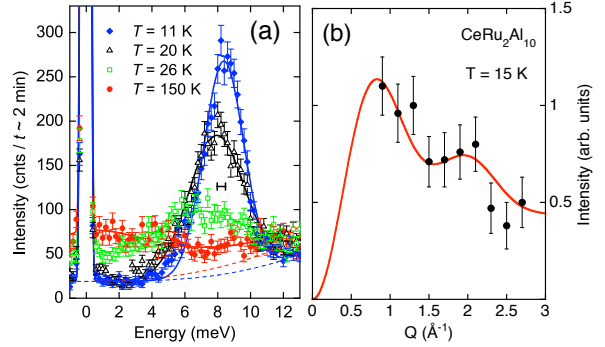}
  \caption{\label{fig:qss-fig7} (Color online)
(a) Temperature dependence of the energy spectra of neutrons scattered from CeRu$_2$Al$_{10}$ for $Q=1.5$~{\AA}$^{-1}$.
The lines denote least-squares fits. (b) $Q$ dependence of the intensity of the peak at
8~meV. The line corresponds to the dimer cross-section Eq. (3.9) with $R=5.2$~{\AA}.
 Adapted from \onlinecite{Rob10}.
}
\end{figure}

The final proof of the Ce$^{3+}$ dimerization is provided by the results of INS experiments performed for
CeRu$_2$Al$_{10}$ as shown in Fig.~\ref{fig:qss-fig7}(a) \cite{Rob10}. At temperatures $T<T_0$ there is no spectral
weight at low energies. There is a well defined peak at 8~meV corresponding to the gap energy. With increasing
temperature the gap shifts to lower energies. The peak at 8~meV can therefore be identified as the singlet-triplet
excitation associated with the formation of Ce$^{3+}$ dimers below $T_0$ which is supported by the oscillatory $Q$
dependence of the intensities according to the dimer cross-section Eq. (3.9) as illustrated in
Fig.~\ref{fig:qss-fig7}(b). From the peak position and Eq. (3.3) the effective intra-dimer exchange coupling results as
$J_{eff} = -4$~meV, which was considered to be unrealistically large \cite{Rob10}. However, this value is based on a
truncated basis with $s = 1/2$. The true exchange coupling between the Ce$^{3+}$ ions results from scaling with the de
Gennes factor $\xi = (g-1)^2 j(j+1)$, where $j$ is the total angular momentum. For Ce$^{3+}$ with $g=6/7$ and $j=5/2$
we have $\xi =0.18$, and for an $s=1/2$ system ($g=2$, $j=1/2$) we find $\xi =0.75$, thus the correction by the de
Gennes factor reduces the true exchange coupling to $J_{Ce-Ce} \approx -1$~meV.

\subsection{Polarons}
\label{sec:qss-polarons}

In complex materials, competing interactions can lead to the spontaneous formation of nano-sized regions of a different
phase. If an additional charge is introduced into the material by doping, a fermionic quasiparticle called polaron can
be formed. The resulting lattice polarization and deformation acts as a potential well that decreases the mobility of
the charge. Polarons have spin, though two close-by polarons are spinless. The latter is called a bipolaron whose
existence was the driving idea behind the discovery of high-temperature superconductivity in the copper-oxide
perovskites \cite{Bed86}. In the meantime the existence of charge-ordered stripes was postulated \cite{Zaa89} and
experimentally verified \cite{Luc03}. The stripes are superconducting regions, separated by AFM regions which act as
Josephson junctions by the proximity effect.

The existence of polarons is the key to understand the rich phase diagrams of the giant-magnetoresistive manganese and
cobalt perovskites upon doping \cite{Sal01,Phe06}. This is demonstrated here for the hole-doped lanthanum cobaltates of
type La$_{1-x}$Sr$_x$CoO$_3$. The ground state of the parent compound LaCoO$_3$ is nonmagnetic, corresponding to a
low-spin state of Co$^{3+}$ ions with $s_i=0$. It was widely believed that the addition of each hole into pristine
LaCoO$_3$ through the substitution of a Sr$^{2+}$ ion for the La$^{3+}$ ion creates a Co$^{4+}$ ion in the lattice
which has a nonzero value of $s_i$ in any spin-state configuration, thereby inducing a magnetic moment in the system.
However, already lightly doped cobaltates with $x \approx 0.002$ give rise to an order of magnitude larger magnetic
susceptibility than expected \cite{Yam96}. It was proposed that the holes introduced by Sr doping do not remain
localized at the nearby Co site, but each hole is distributed among several neighboring Co sites, leaving the latter in
the intermediate Co$^{3+}$ state (with $s_i=1$) and thereby forming a multi-site magnetic polaron. Such spin-state
polarons behave like ferromagnetic nanoparticles with a very large total spin in an insulating nonmagnetic matrix.

\begin{figure}
\includegraphics[width=6.5cm]{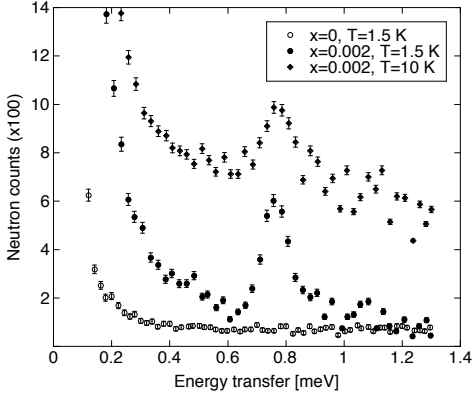}
  \caption{\label{fig:qss-fig8}
Energy spectra of neutrons scattered from La$_{0.998}$Sr$_{0.002}$CoO$_3$ at $T=1.5$ and 10~K. The open
circles correspond to the nonmagnetic reference compound LaCoO$_3$. For clarity, the intensities of the $T=10$~K data
are shifted by 400 neutron counts.
 Adapted from \onlinecite{Pod08}.
}
\end{figure}

\begin{figure}
\includegraphics[width=6.5cm]{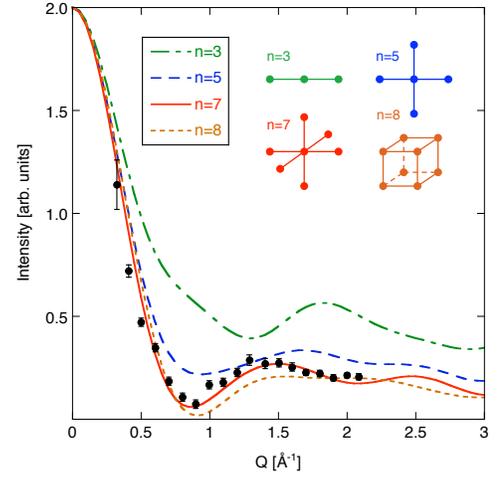}
  \caption{\label{fig:qss-fig9} (Color online)
$Q$ dependence of the intensity of the transition observed at 0.75~meV in La$_{0.998}$Sr$_{0.002}$CoO$_3$.
The insert sketches different types of Co multimers. The lines are the result of structure factor calculations based on
Eq.~(\ref{eq:qss-eq7}).
 Adapted from \onlinecite{Pod08}.
}
\end{figure}

The existence of spin-state polarons was confirmed by inelastic neutron scattering experiments as shown in
Fig.~\ref{fig:qss-fig8}. A magnetic excitation is observed for La$_{0.998}$Sr$_{0.002}$CoO$_3$ at an energy
transfer of 0.75~meV, which is absent for the undoped parent compound LaCoO$_3$ \cite{Pod08}. The ground
state of Co$^{3+}$ in the intermediate spin state is an orbitally degenerate triplet which is split by a
small trigonal ligand field into a singlet and a doublet. The transition between these two levels is the
source of the peak observed at 0.75~meV. The peak intensity diminishes with increasing temperature, in
agreement with the Boltzmann population factor Eq.~(\ref{eq:basics-Zp}) for a singlet-doublet transition. The
$Q$ dependence of the intensity of the observed excitation exhibits a clear oscillatory behavior as shown in
Fig.~\ref{fig:qss-fig9}, which reflects the size as well as the shape of the polaron through the structure
factor. The neutron cross-section of a cluster comprising $N$ magnetic ions can be approximated for
polycrystalline material by an extension of Eq.~(\ref{eq:basics-ins-sigma-powder0}):
\begin{eqnarray}
\label{eq:qss-eq7}
\frac{d^2 \sigma }{d \Omega d \omega} &\propto& F^2(Q) \sum_{i<j=1}^N
[ \langle S || \hat{T}^{(1)}(s_i) || S' \rangle^2
+ 2 \frac{ \sin(Q R_{ij}) }{ Q R_{ij} }
\nonumber \\ && \times
\langle S || \hat{T}^{(1)}(s_i) || S' \rangle \langle S' || \hat{T}^{(1)}(s_j) || S \rangle ] ,
\end{eqnarray}
For the special case of a  $\Delta S = 0$ transition (which is relevant in the present context), the reduced matrix
elements can be factorized and set to 1. The lines in Fig.~\ref{fig:qss-fig9} correspond to calculated cross sections
for different Co clusters sketched in the insert. We clearly see that the $Q$ dependence of the cross section is an
unambiguous fingerprint of the geometry of the multimers; in particular, the data observed for the 0.75~meV transition
in La$_{0.998}$Sr$_{0.002}$CoO$_3$ are perfectly explained by the scattering from an octahedrally shaped Co heptamer.
The total moment of this heptamer, consisting formally of one central Co$^{4+}$ ion ($s_i=1/2$) and six surrounding
Co$^{3+}$ ions ($s_i=1$), is 13~$\mu_B$, in good agreement with the magnetic susceptibility data \cite{Yam96}.

The result of the above experiment gives a clear microscopic explanation why hole doping of as little as 0.2\%
dramatically affects the overall magnetic properties of the entire system, i.e., the magnetic susceptibility is an
order of magnitude larger than expected. Additional charge carriers increase the number of such spin-state polarons,
which form a percolative network resulting in a metallic state with long-range ferromagnetic order at the critical Sr
concentration $x_c=0.18$ \cite{Phe06}. The formation of spin-state polarons may be a common mechanism present in other
Mn- and Co-based oxide perovskites as well.

%% file: chapter7_conclusions_v2012_10_02.tex
\section{Conclusions}
\label{sec:conclusion}

In this review we have attempted to provide an impression of the physics which one encounters in magnetic spin
clusters, and their relevance to a variety of different physical systems. The presentation had obviously been limited
to examples whose magnetic excitations have already been very well or comparatively well understood. Many of the
important scientific questions which are under current research or will possibly become of relevance in future have
however not been addressed, though some of them were indicated at few places in the text. We wish to conclude here by
picking them up again.

The small magnetic clusters have been demonstrated to be ideal experimental systems for studying the basic mechanisms
of the magnetic interactions between spins and the underlying physical principles. The research on them started at
least six decades ago and many fundamental questions could be addressed (Sec.~\ref{sec:sc}). However, despite this long
history, the research field is not yet exhausted, but continuously increases due to the ongoing improvements of the
experimental equipment. For instance, for many fascinating extended magnetic compounds the corrections to the HDVV
model such as the weaker anisotropic and/or higher-order exchange interactions have to be known with accuracy as they
can be crucial for understanding their phase diagrams. However, the higher-order interactions are intrinsically hidden
in the commonly applied analyzes of spin-wave dispersion relations, but become accessible by studying the cluster
excitation spectra in the related diluted materials; an example was given in Sec.~\ref{sec:IIIB6}.

For most magnetic systems the conventional HDVV model, possibly including the weaker corrections to it, is perfectly
appropriate, which is not surprising for magnetic ions of type S and type Q. The basic assumptions underlying the HDVV
picture are phenomenological, but there was little clear-cut experimental evidence in the past for an eventual failure
of this model. The situation is quite different for clusters with magnetic type L and type J ions, and the exchange in
general has to be described by more involved interaction terms. An obvious extension of the HDVV coupling is to
introduce multipole interactions based on standard tensor operator techniques or to replace the exchange parameter
$J_{ij}$ by an exchange tensor $J(m_i,m_j;m'_i,m'_j)$, with $m_i$ and $m'_i$ being the spin quantum numbers of the
initial and final ionic states, respectively; the latter formalism was verified in the dimeric Ho$^{3+}$ compound
Cs$_3$Ho$_2$Br$_9$ \cite{Fur90}. This topic will certainly become more important in the future, e.g., magnetic
molecules incorporating 4f metal ions are currently intensely studied \cite{Klo09}.

The observed existence of three-spin interactions in magnetic systems discussed in Sec.~\ref{sec:IIIB5} is of
particular relevance in the general context of many-body interactions, which already fascinated Kopernikus (1473-1543)
and Kepler (1571-1630) in their investigations of the mutual gravitational interaction between three planets. There has
been ample experimental evidence for many-body interactions in the past, notably in studies of ultra-cold gases of
alkaline atoms \cite{Buc07}, the stability of molecules like ozone \cite{Zha07}, four-atom exchange in $^3$He
\cite{McM75}, chirality in magnetic compounds \cite{Gri05}, ring-exchange \cite{Col01} and three-body correlations of
vortex states \cite{Men06} in high-temperature superconductors, or domain-wall fluctuations in (anti-) ferromagnets
\cite{Shp07}. Novel quantum phases with intriguing physical properties can arise from many-body interactions which are
usually described on the basis of pair potentials $G({\bf r},t)$. However, it would be highly desirable to extend the
analytic tools beyond van Hove's theory \cite{Vho54} to include higher-order correlation functions in both space and
time, for which approximate solutions exist, e.g., for $G({\bf r}, {\bf r}',t)$ \cite{Hu07}, $G({\bf r},t,t')$
\cite{Wor81} and $G({\bf r}, {\bf r}',t,t')$ \cite{Vzo01}. Adequate experimental techniques should be developed as well
which provide a direct access to higher-order correlation functions. Indeed, novel neutron scattering techniques were
suggested for this purpose which include neutron interferometry \cite{Rau98} and spin-echo techniques \cite{Gri04}.

Besides the small magnetic clusters the ground state and excitation spectrum in the large magnetic clusters, as they
have been called in this review, has emerged as an attractive research field in the last 15 years. The scientific
questions encountered in them are ultimately related to the possibility of many-body (quantum) effects, with links to
the field of quantum spin systems. However, as compared to the system sizes considered in the latter area, in which
one-, two-, and three-dimensional lattices of interacting quantum spins are considered, the sizes of the "large
magnetic clusters" have to be considered small as they are (far) away from the finite-size scaling regime, in which the
magnetic properties start to resemble those in the thermodynamic limit $N \rightarrow \infty$ (exceptions are spin-1/2
clusters because of the short correlation length for $s_i = 1/2$, see Sec.~\ref{sec:lc-wheels} and
Fig.~\ref{fig:wheels-cr8theory2}). In this sense the large magnetic clusters should be called zero dimensional and
belong to the class of mesoscopic systems.

A typical consequence of that could be described as the "loss" of the wave vector as a good quantum number, since
translational symmetry or a finite-size version of it such as the cyclic symmetry in wheels is generally not present in
the large clusters. On the one hand this implies that concepts which are developed for extended lattices may have to be
adapted or interpreted in novel ways if applied to the large clusters. The application of spin-wave theory to the
antiferromagnetic wheels and short chains (Secs.~\ref{sec:lc-cr8} and \ref{sec:lc-dopedwheels}) represents a showcase.

For the antiferromagnetic Heisenberg rings or molecular wheels the cyclic symmetry allows introducing a shift quantum
number $q$ which emerges into a wave vector in the thermodynamic limit. Therefore the available literature results for
the spin-wave dispersion relation in the antiferromagnetic chain, as they have been derived from the various spin-wave
theories \cite{Iva04-swtreview}, can be taken over directly with the wave vector replaced by the discrete values of
$q$. As was demonstrated in Sec.~\ref{sec:lc-csfe8} for the CsFe$_8$ molecular wheel this yields the energies in the
$E$ band with some accuracy. However, besides this success, fundamental issues are eminent. For instance, spin-wave
theories do \emph{a priori} violate the spin rotational invariance of the HDVV Hamiltonian, and their applicability to
large clusters with their disordered ground states is hence fundamentally flawed, yet they can produce reasonable
energies and one may ask why. Also, since the wave vector becomes discretized, magnetic excitations with long wave
lengths do not exist in molecular wheels (and large magnetic clusters in general), which raises the question of whether
the excitations in the $E$ band, albeit their energies can be derived by spin-wave theory, should actually be
interpreted as spin waves. This is a sensible question, and the notion of "cluster spin waves" may be introduced
\cite{Stu11-mn10}. We here adopted a pragmatic view and called any excitation which is obviously related to a spin-wave
energy a spin wave.

The situation becomes even more involvled in the short antiferromagnetic chains. At the level of linear spin-wave
theory the literature result for the finite antiferromagnetic chain can again directly be carried over, but this does
not work for the more sophisticated spin-wave theories such as interacting spin-wave theory since the open boundaries
result in site dependent corrections. That is, these theories have to be formulated in real space and not momentum
space. Furthermore, a description of the excitations in classical terms, which is at the heart of the $L \& E$-band
concept and motivated applying spin-wave theories, is actually not obvious in the short chains as the wave functions do
not significantly overlap with the semiclassical configurations. Surprisingly, their ground state is in fact neither in
the classical nor the quantum regime and the intuitively clear distinction between these two regimes becomes blurred
\cite{Kon11-review}.

The loss of the wave vector on the other hand is also indicative of the possibility of lattice topologies in large
clusters which are not possible in extended systems. In fact, most magnetic molecules which were synthesized have a
complex topology which cannot systematically be expanded into an infinite lattice. The V$_{15}$ or Mn$_{12}$ molecules
presented in Secs.~\ref{sec:lc-v15} and \ref{sec:smm-mn12} are examples. However, for obvious reasons, the research on
the many-body aspects of magnetic cluster excitations concentrated on lattice topologies or magnetic molecules with an
"appealing" symmetry, but the most interesting magnetic phenomena may be overlooked this way. In this sense only the
simplest systems were studied so far, yet the understanding of their excitations can be challenging and the case of
Fe$_{30}$ establishes a dramatic example. The large number of lattice topologies available through the molecular
nanomagnets, which thanks to the productivity of chemists will certainly grow further, presents obviously a wide area
for future research.

Except of two cases \cite{Wal06-mn12instimeresolved,Car07-cr7nifield} (see also Sec.~\ref{sec:lc-v15}) the INS works on
molecular nanomagnets were using powder/polycrystalline samples. Experiments on single crystals would offer new
possibilities to unravel the many-body physics in the cluster excitations, e.g., the spin-pair correlation function can
be mapped out more directly than possible with powder samples. Single-crystal INS experiments have become state of the
art in the last decade for inorganic compounds. The current drastic improvements in the INS technique make such
experiments possible now also for molecular nanomagnets (which are challenging because of the small number of magnetic
centers as compared to the many non-magnetic ligand atoms). Single-crystal INS work on molecular nanomagnets will
certainly been seen more often in the next years.

Magnetic cluster systems have also been considered as promising units for quantum computing. Much effort went into
"mesoscopic spin-1/2 clusters" or "antiferromagnetic cluster qubits", i.e., large clusters which exhibit a total spin
$S = 1/2$ ground state and may be used as qubits at low temperatures, where only the ground state is thermally
populated \cite{Mei03-qubit2,Mei03-qubit1}. Such clusters may present advantages over atomic scale qubits, such as
easier addressing and readout. Significant progress were made, e.g., coherence times long enough for quantum
computations could be achieved in the Cr$_7$Ni "doped" antiferromagnetic wheel \cite{Ard07} and two molecular "qubits"
could be linked such as to provide entanglement between them suggesting the possibility of two-qubit operations
\cite{Can10,Aff05-review,Aff06-review}. However, the multi-level structure of the low-lying energies as it can be
provided only by magnetic clusters has also been explored. The particular structure of the ground-state spin multiplet
in the single-molecule magnets was theoretically shown to allow for an implementation of Grover's search algorithm or
to build dense and efficient memory devices \cite{Leu01}. Furthermore, by accessing also higher-lying levels nearby to
the ground state Rabi oscillations could be observed in the V$_{15}$ molecule and the Fe$_4$ single-molecule magnet
\cite{Ber08-v15rabi,Sch08}. Lastly, the spin frustration or orbital degeneracy in the ground state of regular spin
triangles has been theoretically shown to allow for a coupling of the spin degree of freedom to electric fields and
currents (spin-electric currents), such that the spin qubit can be manipulated through currents supplied to a cluster
by e.g. scanning tunneling microscope (STM) techniques \cite{Tri08,Leh07,Geo10}. These examples indicate that the
cluster excitations as present in magnetic clusters may provide novel quantum computation schemes, and exciting results
in this direction can be expected in the future. For instance, the many-body nature of the large magnetic clusters
suggests the possibility of decoherence free subspaces in them, which however has to our knowledge not yet been
explored.

Magnetic clusters play also a role in the emerging field of molecular spintronics
\cite{Bog08-smmspintronics,San06,Roc05}. Significant success has first been obtained with molecules containing one
metal ion, which have been incorporated into break junctions or contact leads produced by electro migration. For
instance, the Kondo effect could be observed in the molecule [Co(terpy(CH$_2$)$_5$-SH)$_2$]$^{2+}$ this way
\cite{Par02}. However, also the current-voltage characteristics of magnetic clusters of two metal centers or even the
single-molecule magnet Mn$_{12}$ have been measured in experiment \cite{Hee06}. Many further schemes are currently
under investigation \cite{Bog08-smmspintronics}. As in the context of quantum computation it can also be envisioned for
molecular spintronics that the unique energy level structures or complex many-body states provided by magnetic cluster
systems will open novel opportunities and allow for functionalities not currently considered.

The molecular nanomagnets represent obviously a valuable resource of (large) magnetic clusters, but in the recent years
not only the chemical route towards large clusters has been advanced, but several beautiful examples of artificially
engineered spin clusters also emerged \cite{Jam01}. Here ensembles of magnetic metal ions experiencing
nearest-neighboring exchange interactions were fabricated directly on surfaces using STM techniques, and their magnetic
excitations measured through recording the current-voltage curves. For instance, short antiferromagnetic chains of
Mn$^{2+}$ metal ions with chain lengths of $N=1 - 10$ were produced and the magnetic ground state as well as first
excited excitation observed this way \cite{Hir06}. Apparently in both areas, that of the molecular nanomagnets and the
artificially engineered spin clusters, one faces very similar scientific questions, yet the technological challenges in
putting them forward in real-world applications are complementary. In comparison with the artificially formed clusters,
the molecular nanomagnets are available in an abundance of different lattice topologies assuming complex many-body
states and their excitations can be studied in any detail using powerful experimental techniques such as inelastic
neutron scattering. However, the artificial nanostructures are available directly on surfaces and have already been
proven to maintain their magnetic properties and their function to be addressable on the surface. The molecular
nanomagnets and the artificially engineered spin systems are hence complementary in the sense that the advantages of
each class may help the other to overcome its problems \cite{Kon11-review}.

Furthermore, magnetic cluster excitations are fundamental in a variety of other systems. For instance, the clusters may
be linked together by a network of weak magnetic \emph{inter}-cluster exchange interactions, such that the cluster
excitations may "travel" through the network and become dispersive. This can lead to fascinating novel quantum phases,
and the Bose-Einstein condensation in the dimer-based compounds, which was presented in Sec.~\ref{sec:qss-dimers}, is a
striking example. However, such networks may obviously not only be built from dimers, but also trimers or tetramers,
which can additionally introduce spin frustration and possess unprecedented behavior. At this point a further link
between molecular nanomagnets and quantum spin systems emerges, as not only small clusters may be incorporated into the
network but also large clusters or molecular nanomagnets, which should be expected to result in an interesting
interplay between the complex quantum states realized within a cluster and the complex phases generated in extended
lattices. Indeed, the synthesis of extended exchange-coupled networks of, e.g., Cu$_3$ molecular units or
single-molecule magnets have already been reported \cite{Iva10,Mor09,Miy06}. All in all, the competition between the
quantum states in a magnetic cluster and the cooperativity introduced through an extended network of
\emph{inter}-cluster interactions should continue to be an attractive playground for research.

Lastly, clusters of magnetic ions are relevant in biology as they constitute the active sites in many metalloproteins.
They are ubiquitous in living matter and contain sites with one to eight metal atoms, sometimes with multiple
occurrence of the smaller clusters in the same protein molecule. The primary function of, e.g., the iron-sulfur
clusters lies in mediating one-electron redox processes and as such they are integral components of respiratory and
photosynthetic electron transfer chains. So far the understanding of the electronic ground- and excited-state
properties relied on magnetic susceptibility, magnetic circular dichroism, EPR, M\"ossbauer and resonance Raman
scattering measurements, which provided information on the anisotropy of the $g$ factor, the valence state of the iron
atoms, the total spin quantum number of the ground state, and sometimes the \emph{intra}-cluster exchange interactions
\cite{Bei97}. The latter are often determined from experiments on model systems as demonstrated, e.g., by magnetic
susceptibility measurements of a cubane-type Fe$_4$S$_4$ cluster, giving $J = -18$~meV for the exchange parameter of
the two semi-independent (Fe$_2$S$_2$)$^{2+}$ dimers \cite{Yoo97}. To our knowledge, the very informative INS technique
has not yet been applied to the study of metalloproteins, since prohibitively large deuterated samples with volumes of
typically 1~cm$^3$ would be needed. However, with the advent of third-generation neutron sources and novel
beam-focusing techniques, reducing the requested sample size by several orders of magnitude, such experiments will
become feasible in the near future.

As so often, improvements in the experimental techniques are a main driving force for scientific progress, and it is
finally mentioned that the development of the instrumentation used for studying magnetic cluster excitations has been
impressive in the last decade, and should be expected to be so also in the future. Neutron scattering can serve as a
representative example. Neutron facilities such as the high-flux reactor HFR at the Institute Laue-Langevin (ILL) in
Grenoble (France) and the spallation neutron source ISIS at the Rutherford Appleton Laboratory in Didcot (U.K.) which
both presently undergo significant upgrades as well as the successful commissioning or planing of new third-generation
neutron sources such as the Spallation Neutron Source (SNS) at the Oak Ridge National Laboratory (USA), the spallation
neutron source at J-PARC in Tokai (Japan), and the European Spallation Source (ESS) in Lund (Sweden) open truly
exciting perspectives to unravel the many unsolved questions in the area of magnetic cluster excitations.